\documentclass[namedreferences,fleqn]{SolarPhysics}
\usepackage{epsfig}

\newcommand{\be}{\begin{equation}}
\newcommand{\ee}{\end{equation}}
\newcommand{\beq}{\begin{eqnarray}}
\newcommand{\eeq}{\end{eqnarray}}

\begin{document}
\begin{article}
\begin{opening}

\title{\bf Sensitivity of a Babcock-Leighton Flux-Transport Dynamo
to Magnetic Diffusivity Profiles }

\author{E.J. \surname{ZITA$^1$}}


\institute{The Evergreen St. College, Olympia WA 98505
           \email{email: zita@evergreen.edu} 
             }


\runningauthor{E.J.Zita}
\runningtitle{Dynamo Sensitivity to Diffusivity Profiles}


\begin{abstract}
We study the influence of various magnetic diffusivity profiles on the 
evolution of the poloidal and toroidal magnetic fields 
in a kinematic flux transport dynamo model for the Sun.
The diffusivity is a poorly understood ingredient in solar dynamo models.
We mathematically construct various theoretical profiles of
the depth-dependent diffusivity, based on constraints from mixing length
theory and turbulence, and on comparisons of poloidal field evolution
on the Sun with that from
the flux-transport dynamo model.
We then study the effect of each diffusivity profile in the cyclic evolution of the
magnetic fields in the Sun, by solving the mean-field dynamo equations.
We investigate effects on 
the solar cycle periods, the maximum tachocline field strengths, and the
evolution of the toroidal and poloidal field structures inside the convection
zone, 
due to different diffusivity profiles.

We conduct three experiments:  
(I) comparing very different magnetic diffusivity profiles;  
(II) comparing different $locations$ of diffusivity gradient near the 
tachocline for the optimal profile; and (III) comparing different
$slopes$ of diffusivity gradient for an optimal profile.

Based on these
simulations, we discuss which aspects of depth-dependent diffusivity 
profiles may be most relevant for magnetic flux evolution in the Sun, 
and how certain observations could help improve knowledge of
this dynamo ingredient.
\end{abstract}
\end{opening}
\section{Introduction}
\label{Introduction} 

Starting from the work of Wang \& Sheeley (1991), flux-transport type 
solar dynamos have been of considerable interest 
for modeling solar cycles, primarily because this class of models has 
been successful in reproducing many large-scale cyclic features 
(Dikpati \& Choudhuri 1994; Choudhuri, Sch\"ussler \& Dikpati 1995; Durney 1995; Dikpati 
\& Charbonneau 1999; Nandy \& Choudhuri 2001;
Bonanno et al. 2002; Guerrero \& Munoz 2004; Rempel 2006).
(For a more complete discussion of Babcock-Leighton dynamo models, 
see Charbonneau 2007). A predictive tool based on the dynamo model 
which we use here
was calibrated by reproducing solar cycles 16-23 using sunspot data 
from previous cycles (Dikpati \& Gilman 2006; 
Dikpati 2008).
While this code is generally
periodic, it also responds appropriately nonlinearly by
adjusting its period, field strengths, and other outputs
to data inputs such as different diffusivity profiles.
The authors of our model (Dikpati, deToma, and Gilman 2006,
Dikpati \& Gilman 2007) and a number of other groups (e.g.
Choudhuri et al. 2007, Svalgaard \& Schatten 2008, and more),
made predictions of the amplitude and duration of Cycle 24.
All have been surprised by the recent ``weird solar minimum." 
This attests to the complexity of solar dynamo forecasting.

The main ingredients of our flux-transport solar dynamo model are: (i) differential rotation,
(ii) meridional circulation, (iii) a Babcock-Leighton type surface poloidal 
source, and (iv) the depth-dependent magnetic diffusivity. 
In a kinematic flux-transport dynamo model, the dynamo ingredients are generally specified by 
mathematical formulae, and the evolution equations for the 
magnetic fields are solved (Sec.2.1). 
The fact that the shape (and the initial period) of the plasma flows are specified, instead of evolving in direct response to field evolution (as in MHD models), is one intrinsic limitation of such a ``kinematic model."
Observational and theoretical knowledge
helps solar dynamo modelers prescribe suitable mathematical
forms to construct the first three solar-like ingredients, but our knowledge
about the fourth ingredient is poor. Nevertheless, Dikpati et al. (DDGAW 2004) have shown that their model can rather closely reproduce observed synoptic maps, or butterfly
diagrams (in their Figs. 8 and 9).  
Their model was tested with seven others (Jouve et al. 2008) 
and agrees well with
the common benchmarks established.  While no solar dynamo
model is complete, and we will note other limitations in 
the course of this paper as necessary, this flux transport
dynamo model is well suited for our investigations.

Appearing in the induction equation, 
${\partial \bf{B} \over \partial t} = 
\nabla \times (\bf{u} \times \bf{B}) + \eta \nabla^2 \bf{B} $,
the magnetic diffusivity $\eta$ is
defined as the ratio of the electrical resistivity to the 
magnetic permeability.  Magnetic diffusivity and plasma flows $\bf{u}$
control the rate of redistribution of magnetic flux in the dynamic
plasma.  Higher magnetic diffusivity permits more rapid changes of 
magnetic flux, while lower magnetic diffusivity can more strongly ``freeze" 
flux to the local plasma. Low diffusivity at the base of the convection 
zone, for example, permits storage of flux near the tachocline,
and inhibits
penetration of  oscillatory fields into the radiative
tachocline -- a ``skin depth effect"
(Garaud 2001; Dikpati, Gilman, and MacGregor 2006). Variations 
in the diffusivity profile
contribute to changes in the radial distribution of solar magnetic flux.
Therefore, better understanding of the magnetic diffusivity 
in the convection zone and tachocline is necessary
for better understanding of the solar dynamo cycle.

One significant uncertainty in solar dynamo models is the form of the 
radial dependence of the magnetic diffusivity, $\eta(r)$.  
Boundary values for diffusivity can be 
estimated from theoretical considerations. 
We have some idea about reasonable
magnetic diffusivity values at the solar surface (Mosher 1977), 
Below the base of the 
convection zone, under the tachocline region, classical (or ``molecular") scale turbulence (Spitzer 1962, Styx p.253)
is expected to dominate. Near the boundary of the radiative interior, 
the molecular diffusivity is estimated at
$\eta_{core} \approx {10^{3-5}} \mathrm{cm^{2} s^{-1}}$,
and higher values are typically used for numerical reasons
\cite{DGM2006}.
While penetrative convection may carry magnetic flux into an overshoot 
region beneath the tachocline, 
this effect is generally small in our model.
Overshoot can lengthen the dynamo cycle period (Sec.3),
extend the magnetic memory effect (discussed below and in Sec.3.1), or, in
extreme cases, quench the dynamo by trapping and freezing
too much flux in highly conducting deep plasma (Sec.3.2).
In the convection zone itself, turbulence can 
 enhance resistivity and diffusivity
(Leighton 1964).

At the photosphere, mixing length theory estimates for
diffusivity are $\eta_{surface} \approx {10^{11-13}} \mathrm{cm^{2} s^{-1}}$
(Simon, Title, and Weiss 1995; Schrijver 2001; Wang, Sheeley, and Lean 2002).
Dikpati et al. (2005) and Charbonneau (2007) have established practical constraints of 
$\eta_{surface} \approx {10^{11-12}} \mathrm{cm^{2} s^{-1}}$
in the outer regions of the convection zone.
Dikpati et al. (2005) have shown that 
kinematic flux-transport dynamo simulations
will not operate in the advection-dominated regime if the
diffusivity value inside the bulk of the convection zone exceeds a
certain value ($\sim 5\times 10^{11} \thinspace {\rm cm}^2 {\rm s}^{-1}$). 
While the meridional circulation
plays a crucial role in advective flux transport, 
diffusion also contributes to radial transport. 
K\"uker et al. (2001) showed earlier that alpha-omega dynamos and
advection-dominated dynamos are separated from each other by a
magnetic Reynolds number of about 10, so that a meridional flow
of about 1 m/s yields an upper limit of 
$\eta \approx 2 \times {10^{11}} \mathrm{cm^{2} s^{-1}}$.
Without a complete theory of convection zone turbulence, 
or detailed observations throughout the convection zone,
we can only conjecture the detailed diffusivity profile in this important region.

The effect of various plausible diffusivity profiles 
has been explored over a wide range of boundary values
(Zita et al. 2005);
however those simulations
were performed to study the evolution of the Sun's dynamo-generated
poloidal fields only.  
They found that certain profiles yielded flux evolution patterns that more
closely matched observations;
we use these cna others in the current study.
Zita et al. (2005) and Dikpati et al. (2006b) also found limits for physically reasonable
and computationally feasible
magnetic diffusivity values at the photosphere and tachocline, which we
use to optimize our current choice of boundary values.

In this paper, we solve the Babcock-Leighton flux-transport dynamo 
equations for $both$ the toroidal and poloidal field evolution, and we study the 
sensitivity of the model to various depth-dependent diffusivity profiles.
Our investigation is guided by such questions as: 
(i) How do the magnetic fields
evolve inside the convection zone, for different diffusivity profiles? 
(ii) How do different profiles change the dynamo period and the 
maximum toroidal field strength at the tachocline?
(iii) How is the magnetic memory influenced
by various diffusivity profiles? 
Schatten et al. (1978) suggested a ``magnetic memory" effect
in which past solar cycles could influence later solar cycles.
This was investigated numerically by
Charbonneau \& Dikpati (2000) and Dikpati et al. (2004).
Their experiments showed that shear-layer toroidal fields of cycle $n$ 
correlate most strongly with those of cycle ($n$-2), 
and Hathaway (2003) verified this observationally. 

We note that other forms of solar dynamo models 
(e.g. interface and distributed dynamos, and more)
and diverse dynamo drivers 
(e.g. near the photosphere, near the tachocline, and in combination)
are actively under investigation
(K\"uker et al. 2001, Bonnano et al. 2002, Bushby 2005, Covas et al. 2005, Hubbard \& Brandenburg 2009, and more).
Significant progress has recently been made in 3D MHD solar
dynamo modeling
(Ghizaru, Charbonneau, and Smolarkiewicz 2010, Miesch \& Toomre 2009).
Additional effects which
may be 
directly or indirectly
related to magnetic diffusivity, 
such as transport coefficients, turbulence, alpha
effects, and quenching, 
are or should be under investigation.  
With this in mind, we 
seek to contribute some insight into one aspect of the problem
at hand - the effects of static profiles of the magnetic diffusivity 
on the evolution of magnetic flux and in a kinematic flux-transport solar dynamo.
The solution of the fully dynamic problem of diffusivity interaction 
and co-evolution with magnetic fields is beyond
the scope of this paper.

After describing the model and our various diffusivity profiles
in Section 2,
we present our results in Section 3. We discuss implications
in Section 4.
\section{Method}
\label{model}

The following picture of the Babcock-Leighton flux-transport dynamo which
has emerged in the past decade or so, informed by solar physics data and
by models such as Dikpati's,
has become increasingly familiar (DDGAW 2004).
The poloidal and toroidal magnetic flux evolve by both diffusion and advection.
Differential rotation induces the toroidal field by shearing the
pre-existing fields.  New poloidal fields are generated by lifting and
twisting the toroidal fields, followed by the decay and diffusion of emerging
flux near the surface.  The surface poloidal fields are then transported toward
the poles by meridional flow, 
where they can cancel older flux from the previous magnetic cycle,
causing polar reversal.  Fields from leading polarity sunspots can diffuse
toward the equator, overcoming the poleward meridional flow there.  Part of
the poloidal flux in high latitudes is ``recycled" into the interior by
meridional circulation.  The poloidal flux which reaches the tachocline is 
then carried by meridional flow down toward the equator and sheared again 
by strong differential rotation, generating new toroidal field of the
opposite sign from that of the previous cycle.
\subsection{Mathematical formulation}
\label{math}

Our solution method and boundary conditions are generally as described in 
Dikpati \& Charbonneau (1999). 
For a Babcock-Leighton flux-transport dynamo with a surface poloidal source,
the axisymmetric mean-field dynamo equations are:

$$
\hspace{-4cm}
{\partial A \over \partial t} + {1 \over r\sin\theta}({\bf u} \cdot \nabla)
(r\sin\theta A) = \eta({\nabla}^2 - {1 \over r^2 \sin^2 \theta}) A
$$
$$
\hspace{0cm} +S_{\rm BL}(r,\theta) {\left[1+{\left({B_{\phi}|_{0.7R}(\theta,t) 
\over B_0}\right)}^2 \right]}^{-1} \thinspace B_{\phi}|_{0.7R}(\theta,t), 
\eqno(1a)$$
%
%
%
$${\partial B_{\phi} \over \partial t} + {1 \over r} \left[{\partial \over
\partial r}(r u_r B_{\phi}) + {\partial \over \partial \theta}(u_{\theta}
B_{\phi}) \right]= r\sin\theta ({\bf B}_p \cdot \nabla)\Omega
-{\bf{\hat e}}_{\phi}\thinspace \cdot \thinspace \left[
\nabla\eta\times\nabla\times B_{\phi}{\bf{\hat e}}_{\phi}\right]
$$
$$+\eta({\nabla}^2
-{1 \over r^2 \sin^2 \theta})B_{\phi},\eqno(1b)$$
\noindent
where
${\bf B}_p(r,\theta) = \nabla \times (A {\bf{\hat e}}_{\phi})$ denotes the poloidal field,
$B_{\phi}$ the toroidal field, and 
$B_0$ is the quenching amplitude of the poloidal field.
The dynamo model operates with four ingredients:
(i) solar differential rotation $\Omega$, observed by helioseismic methods
(Brown et al. 1989; Goode et al. 1991;
Tomczyk et al. 1995;
Charbonneau et al. 1997;
Corbard et al. 1998),
(ii) solar-like meridional circulation ($u_r$, $u_{\theta}$),
observed to flow from the equator toward the poles
along the photosphere
(Giles et al. 1997; Basu \& Antia 1998; Gonzalez-Hernandez et al. 2000),
and inferred, by mass conservation, to return equatorward near the tachocline,
(iii) a Babcock-Leighton-type surface poloidal field source
$S_{\rm BL}(r,\theta)$ derived from
observations of sunspot decay, and
(iv) a depth-dependent magnetic diffusivity $\eta(r)$, 
the poorly-understood ingredient which we study here.

The poloidal field is calculated at each point by curling the vector potential $\bf A $.
This scheme allows for more efficient numerical advancement than by direct calculation of the fields.
We track the evolution of magnetic flux in the meridional plane,
an $r-\theta$ cross-section of the solar sphere.
Each complete iteration of our simulations represents a full solar cycle
of roughly 22 years, when the poles return to their initial magnetic polarity.
The exact periodicity is free to vary depending on the simulation's dynamical response
to input parameter.

While we directly adopt the differential rotation from Dikpati \&
Charbonneau (1999), we use slightly modified forms for the second and third
dynamo ingredients.
The meridional circulation $\bf u$, dynamo ingredient (ii), is as in Dikpati \& Charbonneau (1999), 
except that the flow speed is set to zero at the center of the tachocline to prevent deeper penetration
(see Gilman \& Miesch 2004; R\"udiger,
Kitchatinov, and Arlt 2005). 
Our simply cycling meridional circulation is admittedly an approximation, as
helioseismic observations suggest that the flow may sometimes develop perturbations such as multiple lobes (Dikpati et al. 2004), and flow speeds at a given point may vary
within a solar cycle.
Our density stratification is slightly different from the stellar atmosphere 
of Dikpati \& Charbonneau (1999), who used
$m=0.25$ in the formula
$\rho(r)=\rho_c {[(R/r) -1]}^m$, where $\rho_c$ denotes the density
at the base of the convection zone. Their purpose 
was to align the maximum return flow speed with the center of the
tachocline. 
We slightly shift the shape of the density stratification
by using $m=1.5$. 
We use the same method as Dikpati \& Charbonneau to control the spreading and
shrinking of the streamlines near the polar and equatorial regions.
However we now use a value of 1.0 instead of 0.5 for the parameter $p$ that
governs the streamline spreading near the surface and the bottom.
The flow pattern looks similar to that used in 
Dikpati et al. (2004).

The Babcock-Leighton dynamo source term $S_{\rm BL}$ is a 
kinematic alpha effect 
near the solar surface. Instead of simply using a $\sin\theta \cos\theta$ profile as in earlier work,
we now confine the driver
closer to sunspot latitudes by multiplying it with a Gaussian.

The boundary conditions on the induction equation are chosen to satisfy
physical and simulation constraints;
boundary conditions on the poloidal field are written in terms of the vector potential.
At the poles, the vector potential must vanish: 
$A|_{\theta=0,\pi} = 0$.
At the equator, field lines must match across the two hemispheres:
${\partial A \over \partial \theta}|_{\theta=(\pi/2)} = 0$.
At the solar surface, field lines must satisfy Laplace's equation:
$\left(\nabla^2 - {1 \over r^2\sin^2\theta}\right)A|_{r=R} = 0$.
The toroidal field vanishes on all boundaries, including the equator, 
since $B_{\phi}$ is antisymmetric about the equator.
These boundary conditions do allow fields more complex
than dipolar to evolve, as long as they match across
the hemispheres.

Diffusivity gradients $(\partial \eta / \partial r)$ are taken when the vector potential is curled.  
We discuss in Section 2.2 the physical constraints on the magnetic diffusivity and our 
mathematical constructions of plausible profiles.

\subsection{Diffusivity profile experiments}
\label{experiments}

In general the magnetic diffusivity is a tensor which parametrizes turbulent processes, and turbulence can also generate alpha effects.  
We make a common approximation by considering diffusivity independently from alpha effects, 
and treating the diffusivity as a scalar
(neglecting the possible effects that an additional poloidal source may have on our dynamics). 
We model several different magnetic diffusivity profiles $\eta(r)$ in the convection zone.
High temperatures near the radiative zone reduce resistivity and diffusivity,
while turbulence near the photosphere enhances resistivity and diffusivity
(Spitzer 1962, Stix 1989).
Therefore, we generally model diffusivity increasing outward from the tachocline to the photosphere. 

\begin{figure}[ht]
\centering
\mbox{
      \psfig{file=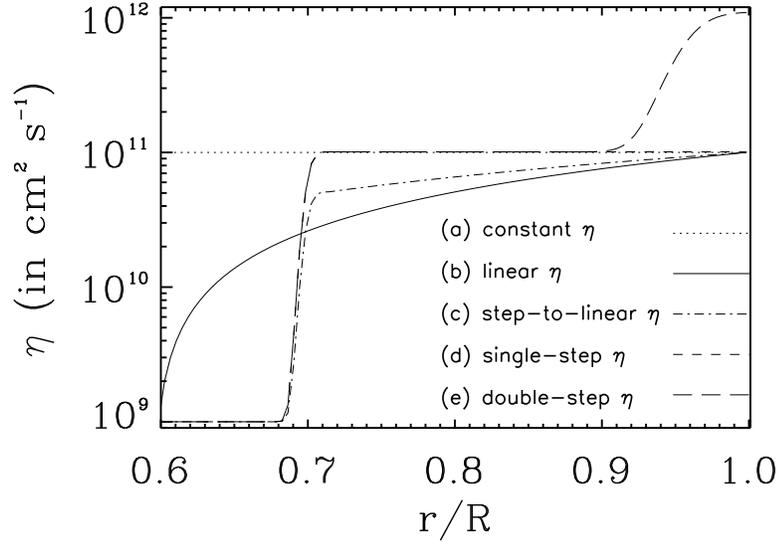,height=8.0cm}
     }
\caption{Various profiles of magnetic diffusivity 
versus radius used in model simulations in
Experiment I.}
\label{etafive}
\end{figure}

Our approach is to perform three sets of numerical experiments to explore
three outstanding questions regarding the role of diffusivity variations on 
magnetic field diffusion and solar dynamo evolution.
\begin{enumerate}
\item How does the general \emph{shape} of the diffusivity profile and its gradients
affect the evolution of the magnetic flux and the dynamo?  
To address this question, we explore the
five different diffusivity profiles in Fig. \ref{etafive} and analyze their
resultant dynamo simulations.
\item For a given diffusivity profile, with fixed boundary and gradient values,
how does the \emph{location} of the diffusivity gradient with respect to the
tachocline 
affect the evolution of the magnetic flux and the dynamo?  
To address this question, we explore four diffusivity profiles identical except 
that their inner gradients are slightly offset from the tachocline
(Fig. \ref{etapar}).
\item For diffusivity profiles with gradients centered identically with respect
to the tachocline, how does the \emph{slope} of the diffusivity gradient 
affect the evolution of the magnetic flux and the dynamo?  
To address this question, we compare dynamo simulations for four profiles
with the same boundary values, 
with inner gradients centered at the same point near the tachocline, 
with magnetic diffusivity gradients ranging from
very steep to very broad 
(Fig. \ref{etarcross}).
\end{enumerate}
For each of these three experiments we will tabulate the
cycle times and maximum tachocline field strengths,
and study timeseries of poloidal and toroidal
field contours for insight into effects on dynamo evolution.
In the absence of detailed knowledge of turbulent velocity fields deep in the
convection zone, numerical experiments such as these are good tools
for gaining better insight into magnetic diffusivity and its role in
the evolution of dynamo fields.
%
\input{psfig}
%
%
\subsubsection{Description of Experiment I: Various diffusivity profile shapes}
\label{ExI}
\vspace{0.2cm}
Our magnetic diffusivity profiles in Fig. \ref{etafive} include
a constant-$\eta$ profile (a);
%
a linearly increasing diffusivity profile (b, solid line);
a profile with one step near the tachocline, e.g. due to turbulent resistivity
enhancement throughout the convection zone 
(d, dashed line); 
a single-step profile which increases linearly beyond the tachocline 
(c, dash-dotted line);
%
and a profile with a second step near the photosphere, 
e.g. due to a combination of
granular and
supergranular turbulence (e, long-dashed line).

We set the (lower) radiative boundary at $r_{core}=0.6 R$, 
the tachocline at 
$r_{tach}=0.7 R$ ,
and the photosphere at $r_{surf}=R$.
The diffusivity is set to $\eta_{core}=10^{9} cm^2 s^{-1}$ at the radiative boundary,
$\eta_0=10^{11} cm^2 s^{-1}$ in the bulk of the convective zone and
at the photosphere in most cases, and
$\eta_{surf}=10^{12} cm^2 s^{-1}$ at the photosphere for the double-step case.
We use the double-step case (e) 
in all three experiment sets to follow.


(a) Constant diffusivity profile:  
$\eta(r) = \eta_0$

(b) Linear profile:
$$\eta_{linear}(r) = \eta_{core} + 
 \frac{\eta_{0} - \eta_{core}}{R - r_{core}} \times ({r-r_{core}})
$$

(c) Step-to-linear profile:
$$\eta(r) = \left\{ \begin{array}{ll}
\eta_{single}(r) & \mbox{where $r \leq r_{tach}$, and  
               $\frac{d \eta_{single}(r)}{dr} \neq \frac{d \eta_{linear}(r)}{dr}$ 
	       } \\
\eta_{linear}(r) & \mbox{otherwise}
\end{array} \right \} 
$$

(d) Single-step profile:
$$\eta_{single}(r) = \eta_{core} + \frac{\eta_0}{2} \left \{
1 + erf \left( \frac{r - r_{tach}}{\Delta r_{tach}} \right) 
\right \} 
$$
\vspace{0.1cm}
where $r_{tach}=0.7 R$ is the approximate center of the step,
$\Delta r_{tach}=0.04 R$ is the width of the step-up region.
The error function (${erf}$) shapes a smooth step,
and yields boundary values within about 1\% of the nominal values.

(e) Double-step profile:
$$\eta(r) = \eta_{core} + \frac{\eta_0}{2} \left \{
1 + erf \left( \frac{r - r_{tach}}{\Delta r_{tach}} \right) 
\right \} 
+ \frac{\eta_{surf}}{2} \left \{
1 + erf \left( \frac{r - r_2}{\Delta r_2} \right) 
\right \} \, \eqno(2)$$ 
%
\vspace{0.1cm}
where $r_2=0.956 R$ and 
$\Delta r_2=0.08 R$ are the approximate center and width, respectively, 
of the second step-up region (and the first step is the same as in case d).


Shapes of diffusivity profiles were chosen based on results of earlier tests
(Zita et al. 2005, Dikpati et al. 2005).
The linear profile (b) 
appears curved on the log-linear plot
of Fig. \ref{etafive}. 
Profile (c) 
is a new hybrid of (b) and (d),
transitioning from single-step to linear after the tachocline,
where the slopes of the two functions most closely match.  A five-point
smoothing is performed at the transition.
The constant diffusivity profile, (a), 
is useful for contrasting the effect of diffusivity with competing
influences such as advection, as discussed in earlier works.  
Boundary values were chosen based on the results of Zita et al. (2005) and
Dikpati et al. (2006b).
\subsubsection{Description of Experiment II: Variable locations of a fixed 
gradient}
\label{ExII}
\begin{figure}[ht]
\centering
\mbox{
      \psfig{file=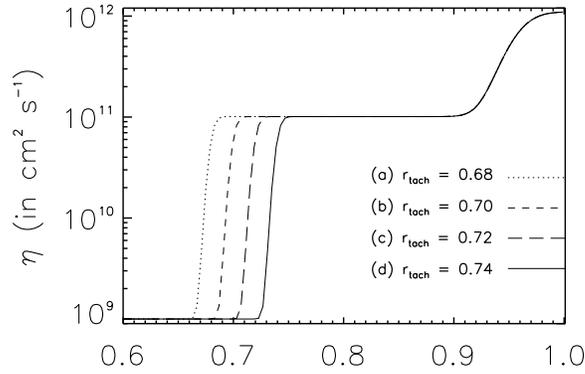,height=6.0cm}
     }
\caption{Parallel magnetic diffusivity profiles
used in model simulations of Experiment II.
(Fig.2.b = Fig.1.e)}
\label{etapar}
\end{figure}

\vspace{0.2cm}
In this experiment (Fig.\ref{etapar}) 
we use the double-step diffusivity profile of
Fig.1e and shift the location of its inner gradient 
with respect to the tachocline, i.e. the
location at which the meridional circulation closes. 
We do not change the magnitude of the gradient.
We change the location of the lower step, or inner gradient, 
simply by specifying $r_{tach}$ in equation (2).
As seen in Fig.\ref{etapar}, this diffusivity gradient
is moved $0.02 R$ inside the tachocline for (a),
is kept at the original location for (b),
as in Fig.1.3, 
and is moved $0.02 R$ and $0.04 R$ further out toward
the photosphere for (c) and (d).  
We also performed simulations with gradients further outward 
(cases 11 and 12 in Table 0), but found little change from gradients (c) and (d)
(cases 9 and 10 in Table 0), so we will present only representative cases.

The location of the gradient is mathematically specified by equation (2).
Fig.2 shows that the error function ($erf$) formulation for the diffusivity profile sets
the peak of the gradient nearer to $r_{tach}$ than the center of the gradient.
%

\subsubsection{Description of Experiment III: Gradients of various slope,
centered at the tachocline}
\label{ExIII}
\begin{figure}[ht]
\centering
\mbox{
      \psfig{file=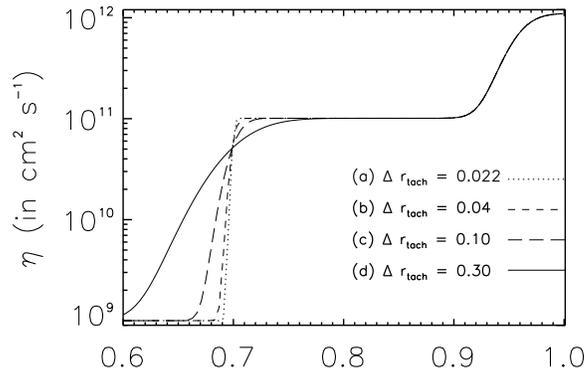,height=6.0cm}
     }
\caption{Profiles of various tachocline-centered gradients
used in model simulations of Experiment III.
(Fig.3.b = Fig.2.b)}
\label{etarcross}
\end{figure}

\vspace{0.2cm}
This experiment (Fig.\ref{etarcross}) also uses variations on our 
double-step case with
$r_{tach} = 0.70$ and $\Delta r_{tach} = 0.04$, which appears as
Fig.1.e, Fig.2.b, and Fig.3.b (and is tabulated in Table 0, Case 8).
The steepest slope numerically feasible with our code's resolution has
$\Delta r_{tach} = 0.022$, shown in Fig.3.a.  We simulated a range of 
slightly varying slopes between these two; as we found smoothly varying behavior
throughout the range (see Table 0, Cases
1-6), we also chose 
an extremely broad slope 
$\Delta r_{tach} = 0.30$ and an intermediate slope, 
$\Delta r_{tach} = 0.10$.  
Varying details of the inner step of the double-step profile
has negligible impacts on the boundary values, except for a slight rise in $\eta$ at the core for the very broadest slope,
as to be expected from Fig.3.
All of these cases have their inner gradients 
mathematically centered
on $r_{tach} = 0.70$.
Due to the specific form of the error function ($erf$), 
as noted above, 
their actual crossing points
are seen in Fig.3 to be slightly inside the tachocline.  
The results of Experiment II will show that
this location provides good sampling
of the tachocline region for representative dynamo simulations,
while gradients centered further outside the tachocline can experience
compromised dynamo performance.
%
\input{psfig}
%
%
\input{psfig}
\clearpage
\section{Results}
\label{results}

As a preview to the results of our three Experiments, we briefly discuss 
the two
primary sets of
diagnostics presented in each of the following Results subsections.  
One is
presented in tabular form to compare simulation outputs, 
and the other is graphical,
to show the evolution of poloidal and toroidal fields.


\begin{tabular}{lccccc} \\ \hline

Case & $\Delta r_{tach}$ & $r_{tach}$ & $max(B_{\phi})$ & ($T$) & Figure\\ 

\hline
(1)& 0.022    & 0.70       & 82 kG  & 14.9 years & 3.a \\ 
(2)& 0.023    & 0.70      & 67 kG   & 15.4 years & \\ 
(3)& 0.024   & 0.70      & 60 kG    & 15.9 years & \\ 
(4)& 0.025    & 0.70      & 56 kG   & 16.4 years & \\ 
(5)& 0.030    & 0.70      & 55.5 kG & 18.1 years & \\ 
(6)& 0.035    & 0.70      & 55 kG   & 18.6 years & \\ 
(7)& 0.04     &  0.68     & 20 kG   &  20.2 years & 2.a \\ 
(8)& 0.04     &  0.70     & 50 kG  &  19.0 years & 2.b, 3.b \\ 
(9)& 0.04     &  0.72     & 79 kG   & 15.9 years & 2.c \\ 
(10)& 0.04    & 0.74      & 266 kG  & 19.4 years & 2.d \\ 
(11)& 0.04    & 0.76      & 285.5 kG & 15.5 years & \\ 
(12)& 0.04    & 0.78      & 287.5 kG & 16.9 years & \\ 
(13)& 0.10    & 0.70      & 25.5 kG  & 17.9 years & 3.c \\ 
(14)& 0.20    & 0.70      & 14.5 kG  & 16.4 years & \\ 
(15)& 0.30    & 0.70      & 11.8 kG  & 15.8 years & 3.d \\ 
(16)& 0.40    & 0.70      & 11.4 kG  & 15.5 years & \\ 

\hline
\end{tabular}

Table 0. Dynamo cycle period ($T$) and maximum toroidal field strength at
the tachocline, tabulated for steepness and location of diffusivity gradients,
for all runs in Experiments II and III.
\vspace{0.2cm}

We can gain broad insight by
tabulating all the double-step runs by the approximate center ($r_{tach}$) and 
inverse slope
($\Delta r_{tach}$) 
of their inner diffusivity gradients (Table 0).
Note that a lower $\Delta r_{tach}$ corresponds to a steeper diffusivity gradient.
Overall, we find that as the slope of the diffusivity gradient decreases
at a fixed location,
the maximum tachocline field strength decreases.
That is, steeper gradients yield stronger tachocline fields
for a given gradient location $r_{tach}$.
For very steep gradients (cases 1-6 and 8),
we find that the cycle time decreases as slope increases.
For broad gradients (cases 13-16), 
the cycle time increases with slope.
In Experiment II, we find that for a constant gradient (Fig.2) at different locations 
(cases 7-12), the field strength increases asymptotically as 
the gradient is moved toward the photosphere, 
and that the cycle time 
does not have a direct relationship to the gradient location.

We examined butterfly diagrams (or time-latitude plots, or synoptic maps)
as another potential diagnostic for
testing magnetic diffusivity profiles. 
We found little variation in butterfly diagrams for our various runs.
This limits their use as a diagnostic tool in assessing the
relative physicality of different diffusivity profiles.

Finally, plots of meridional cuts of poloidal and toroidal fields
in the convection zone provide significant insight into evaluating
magnetic diffusivity profiles in this investigation,
especially when we track their time evolution.  
This is discussed in more depth starting in Sec.3.1.2.

\input{psfig}
\begin{figure}[ht]
\centering
\mbox{(a)
      \psfig{file=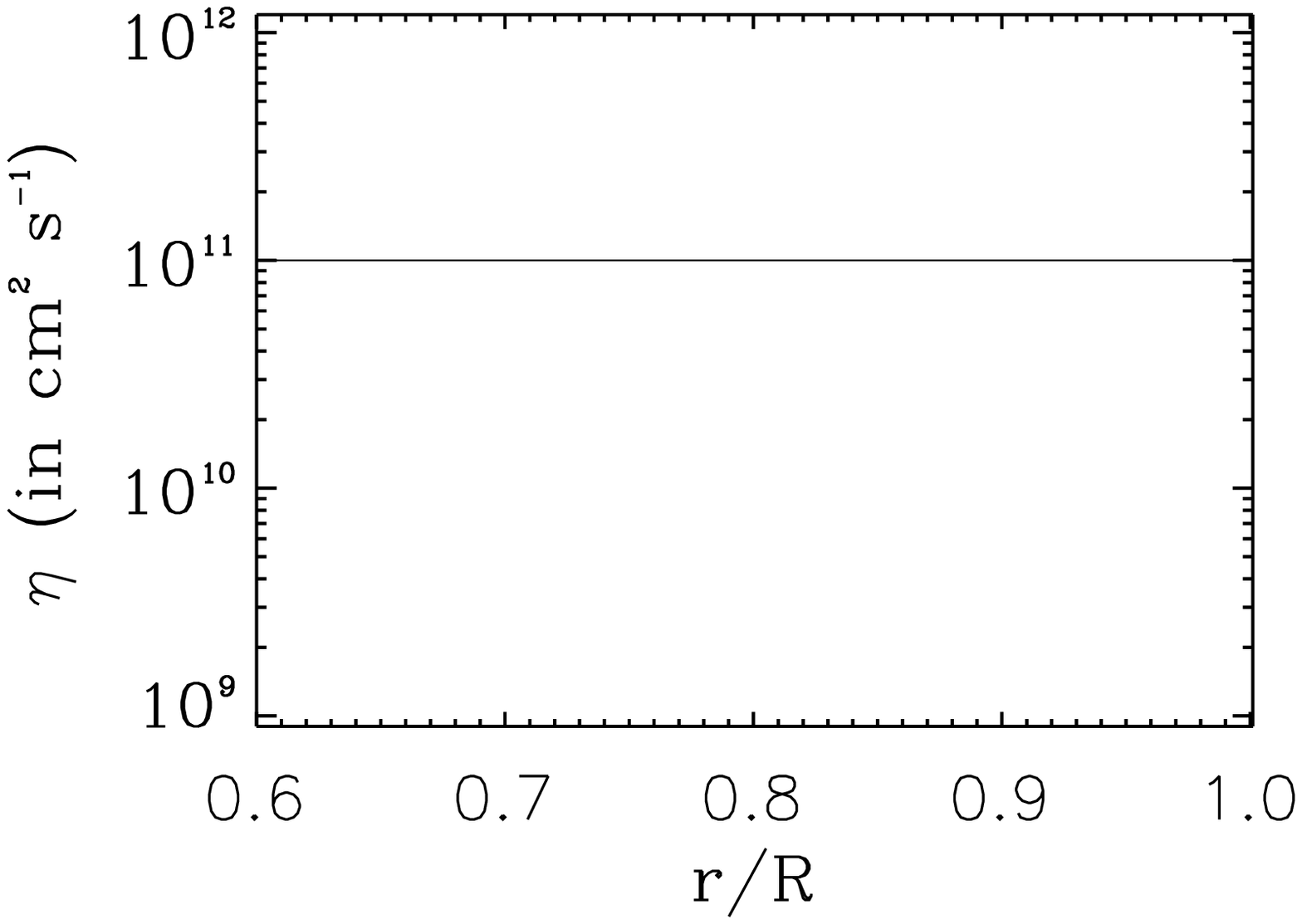,height=4.0cm}
      \psfig{file=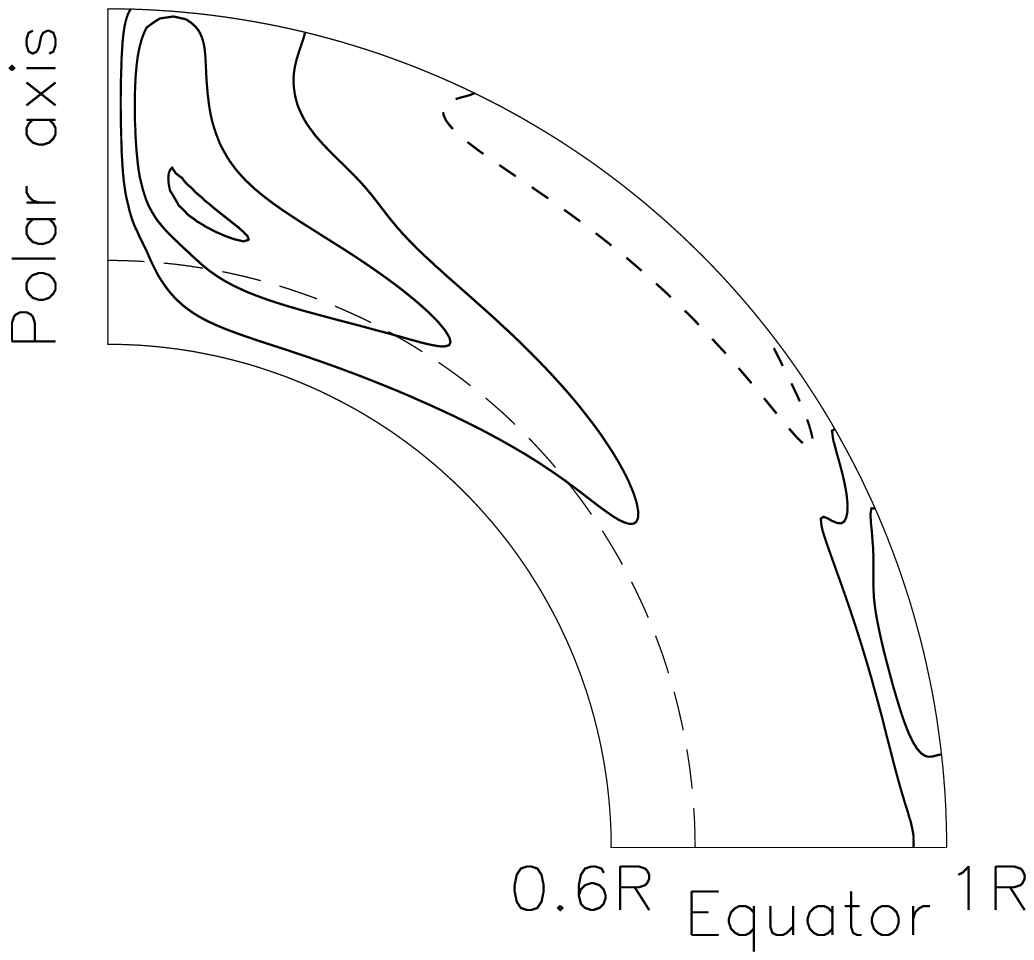,height=4.0cm}
      \psfig{file=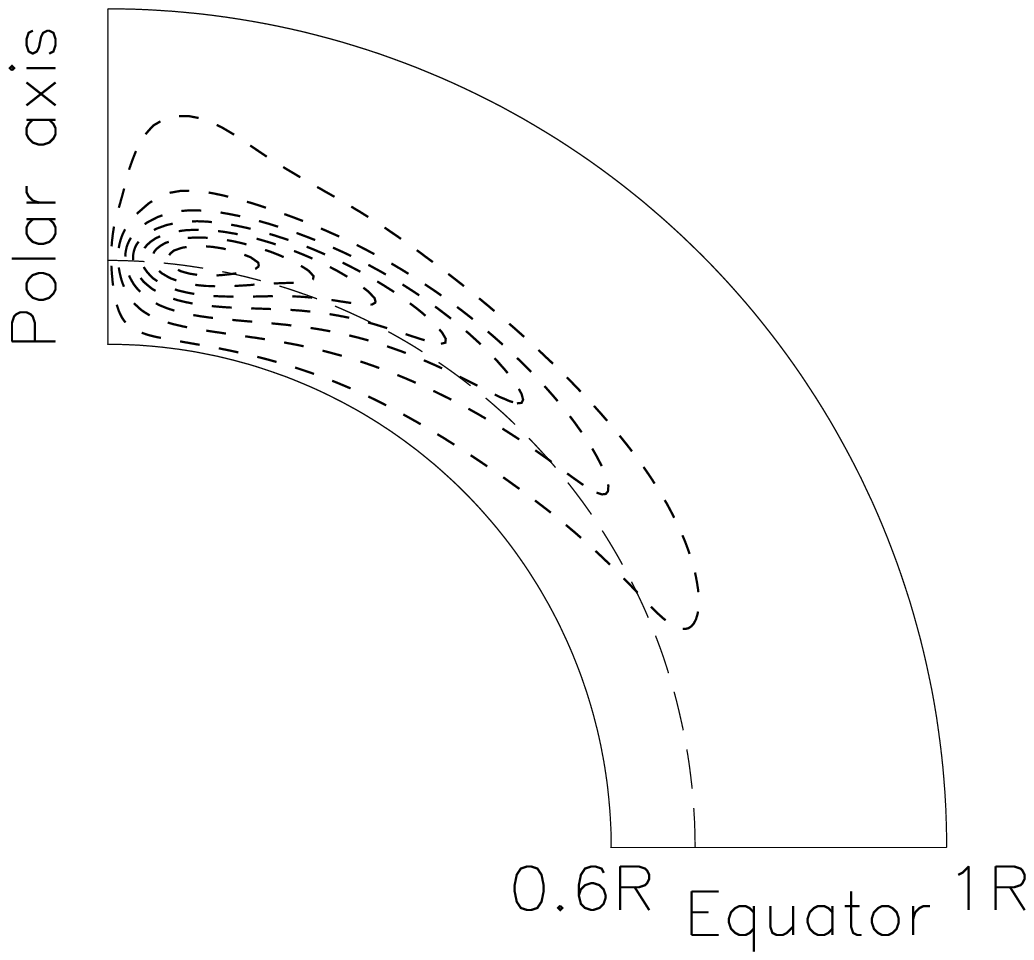,height=4.0cm}
     }
\mbox{(b)
      \psfig{file=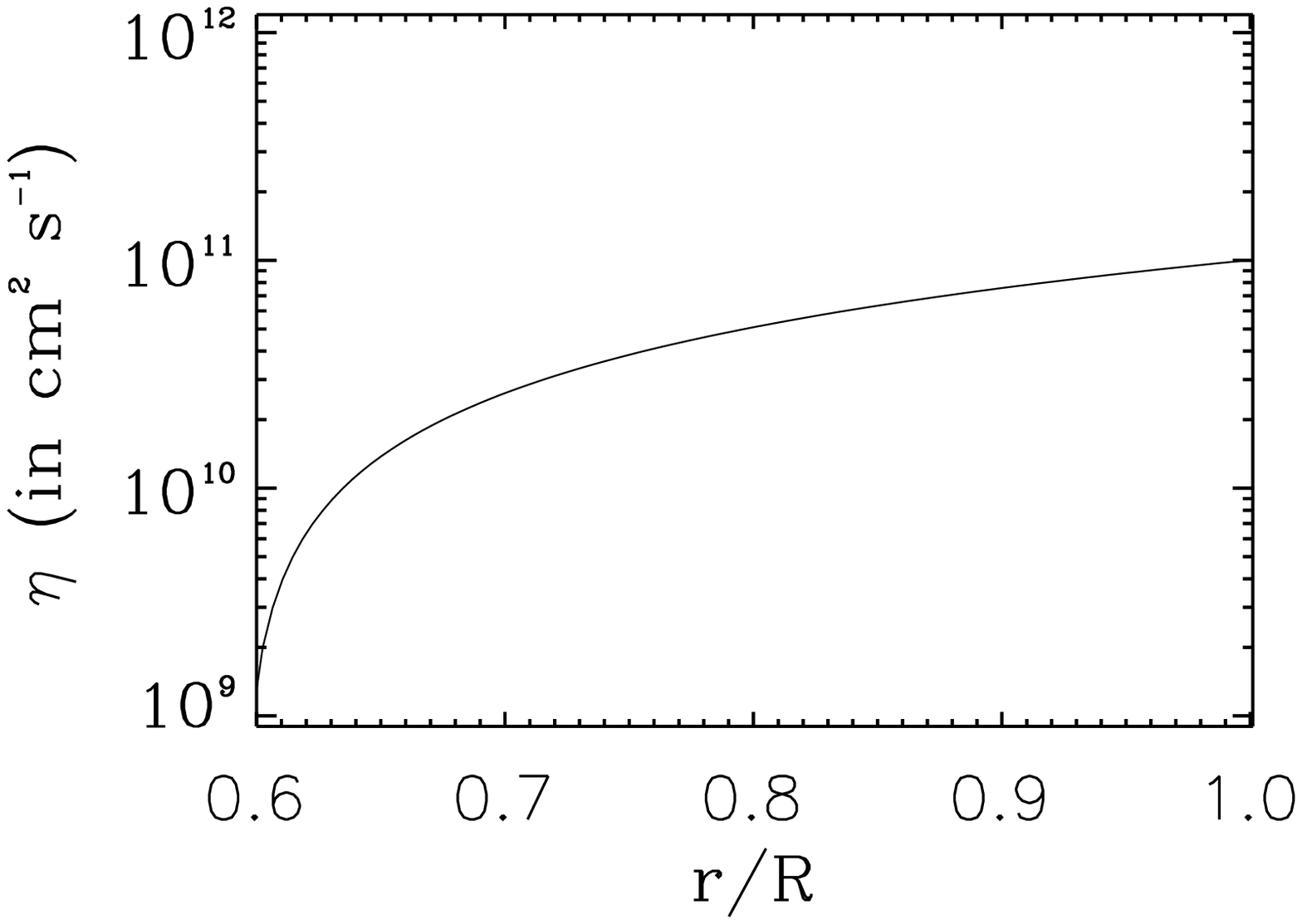,height=4.0cm}
      \psfig{file=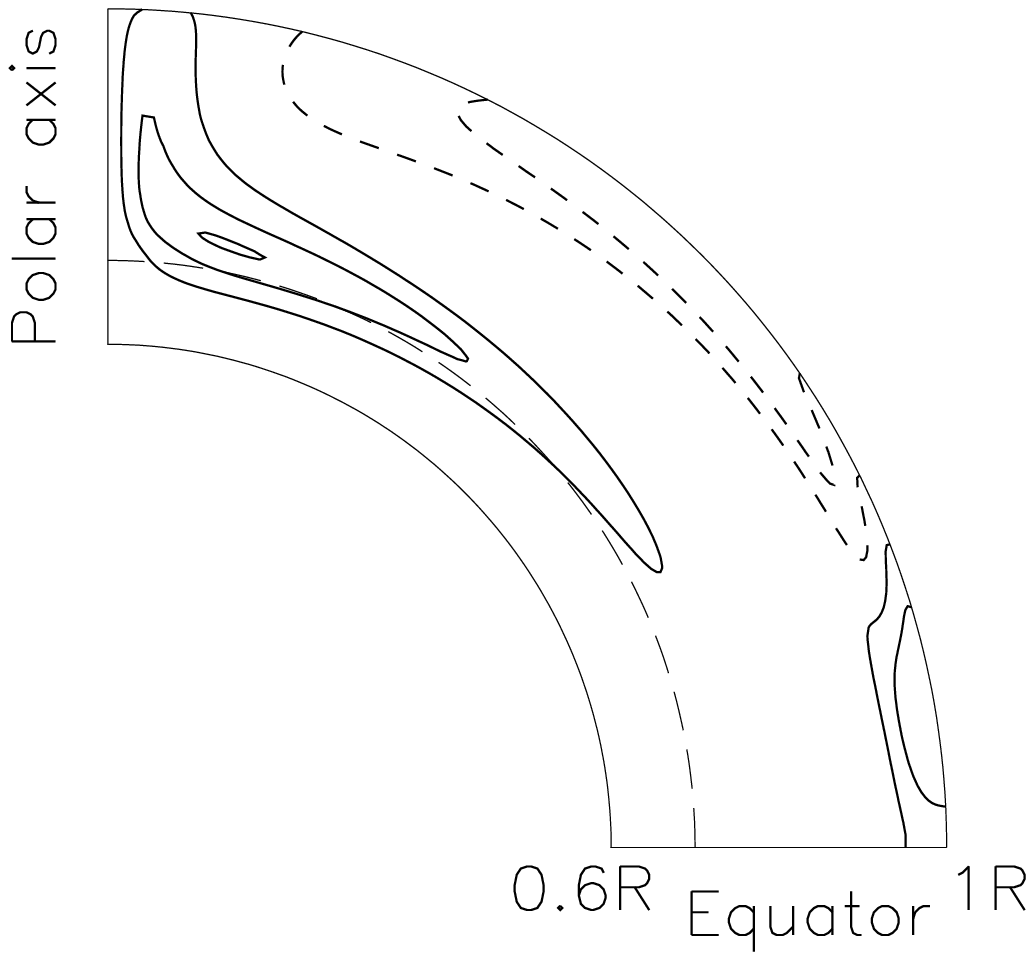,height=4.0cm}
      \psfig{file=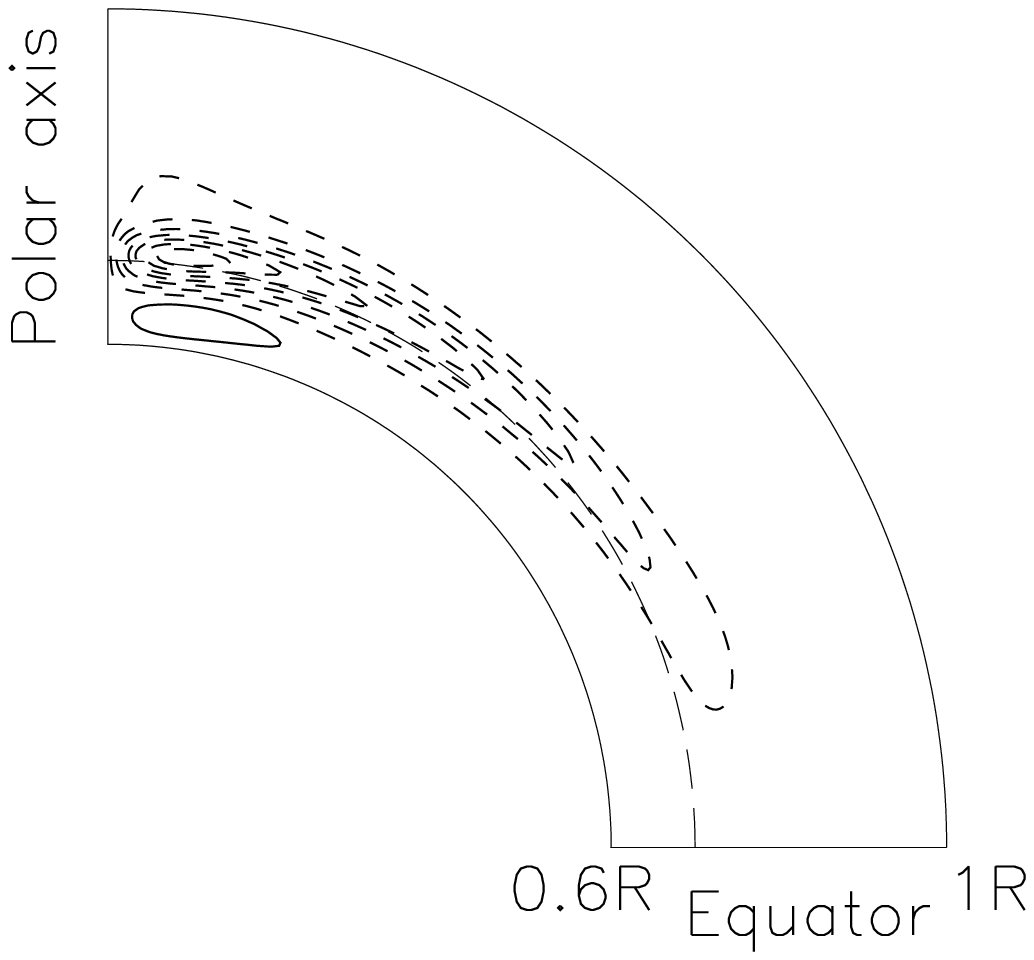,height=4.0cm}
     }
\mbox{(c)
      \psfig{file=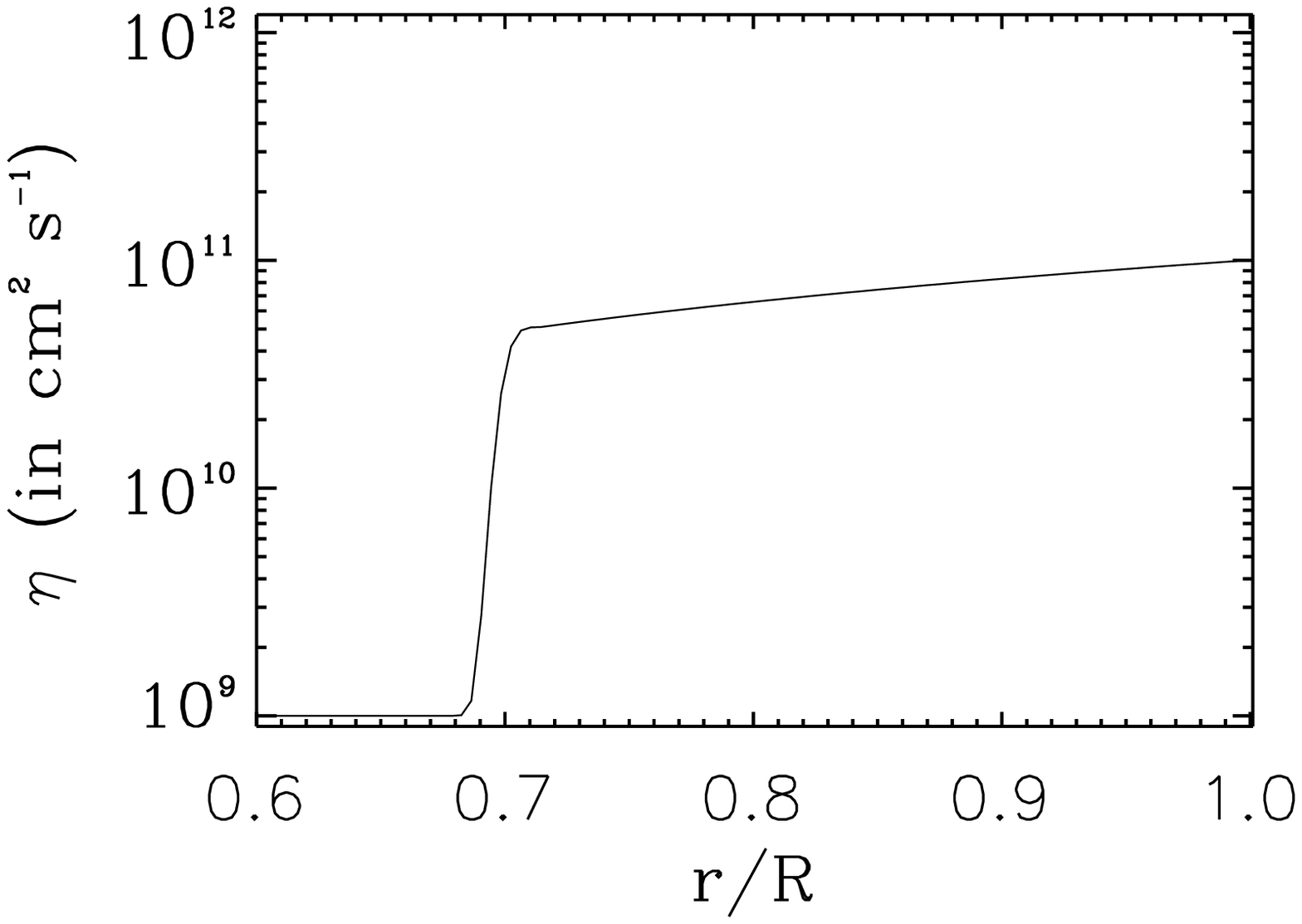,height=4.0cm}
      \psfig{file=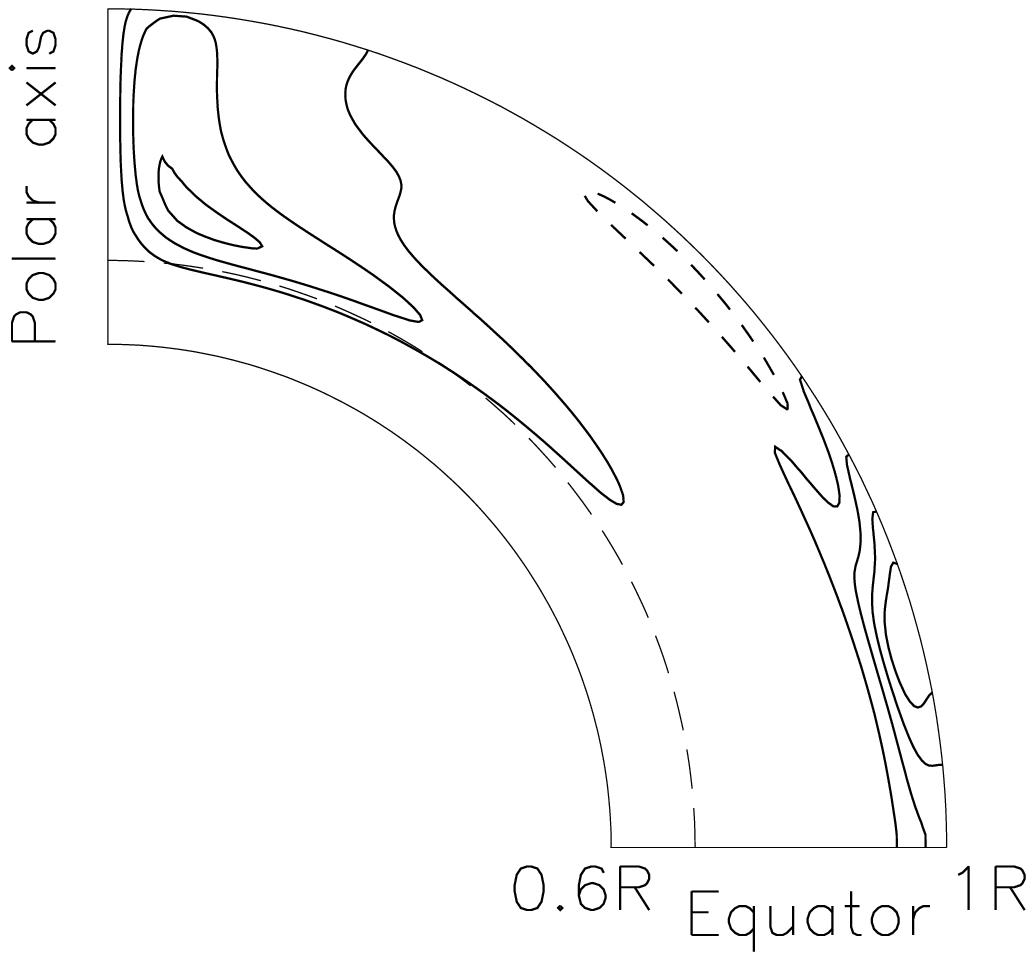,height=4.0cm}
      \psfig{file=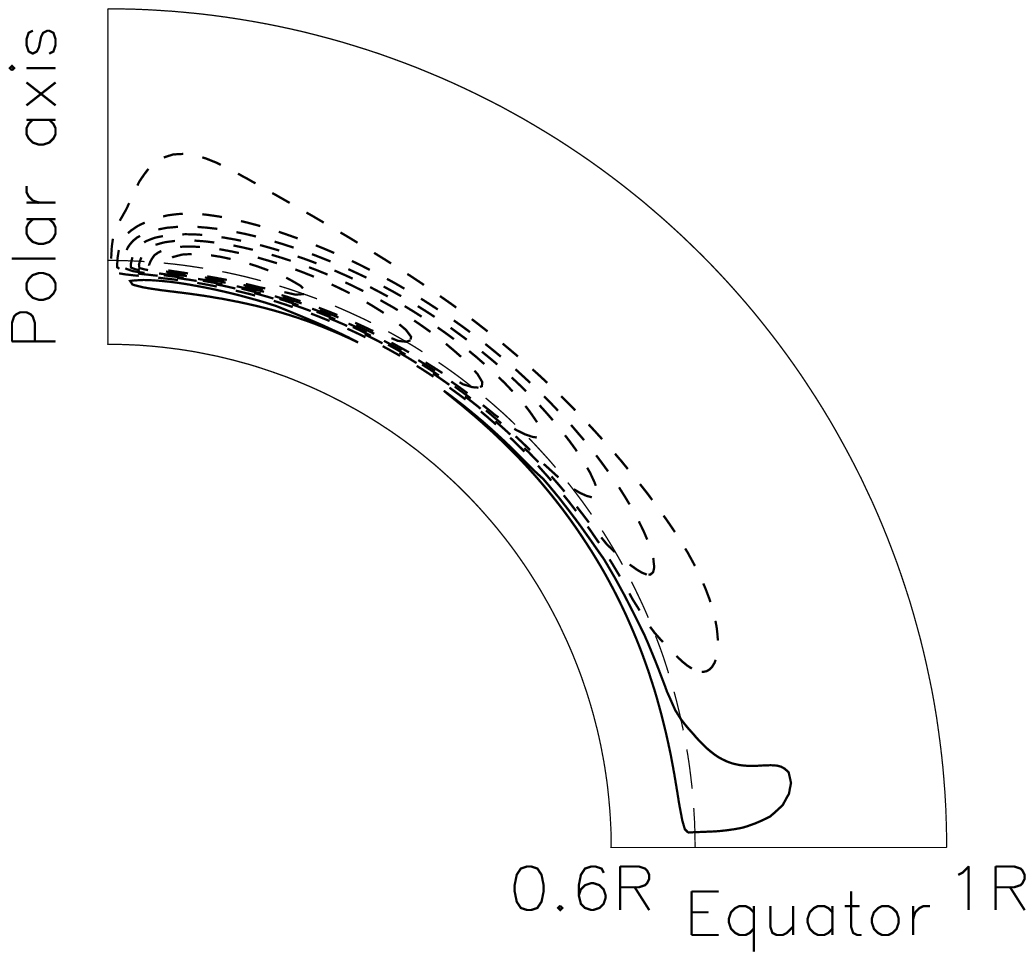,height=4.0cm}
     }
\mbox{(d)
      \psfig{file=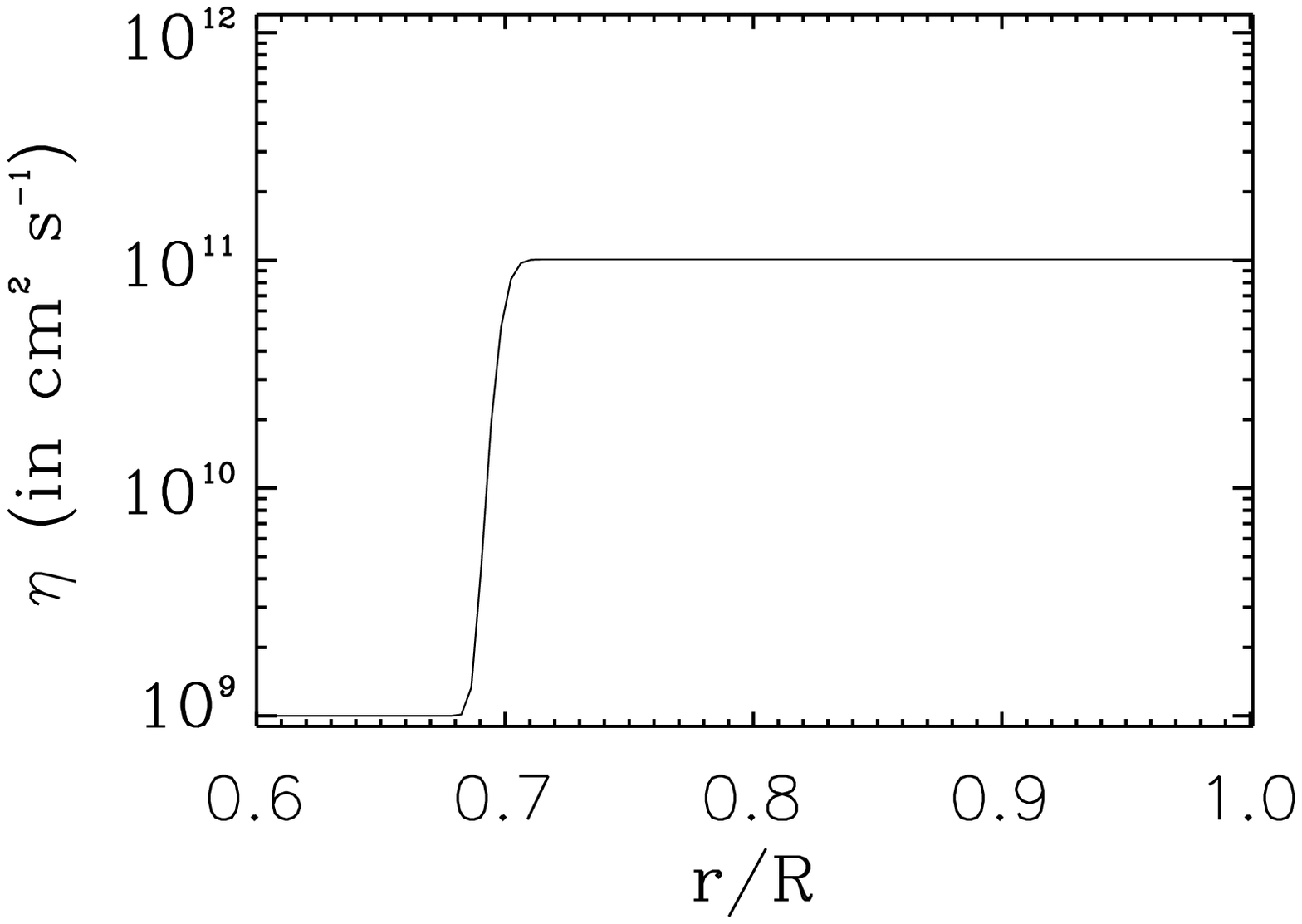,height=4.0cm}
      \psfig{file=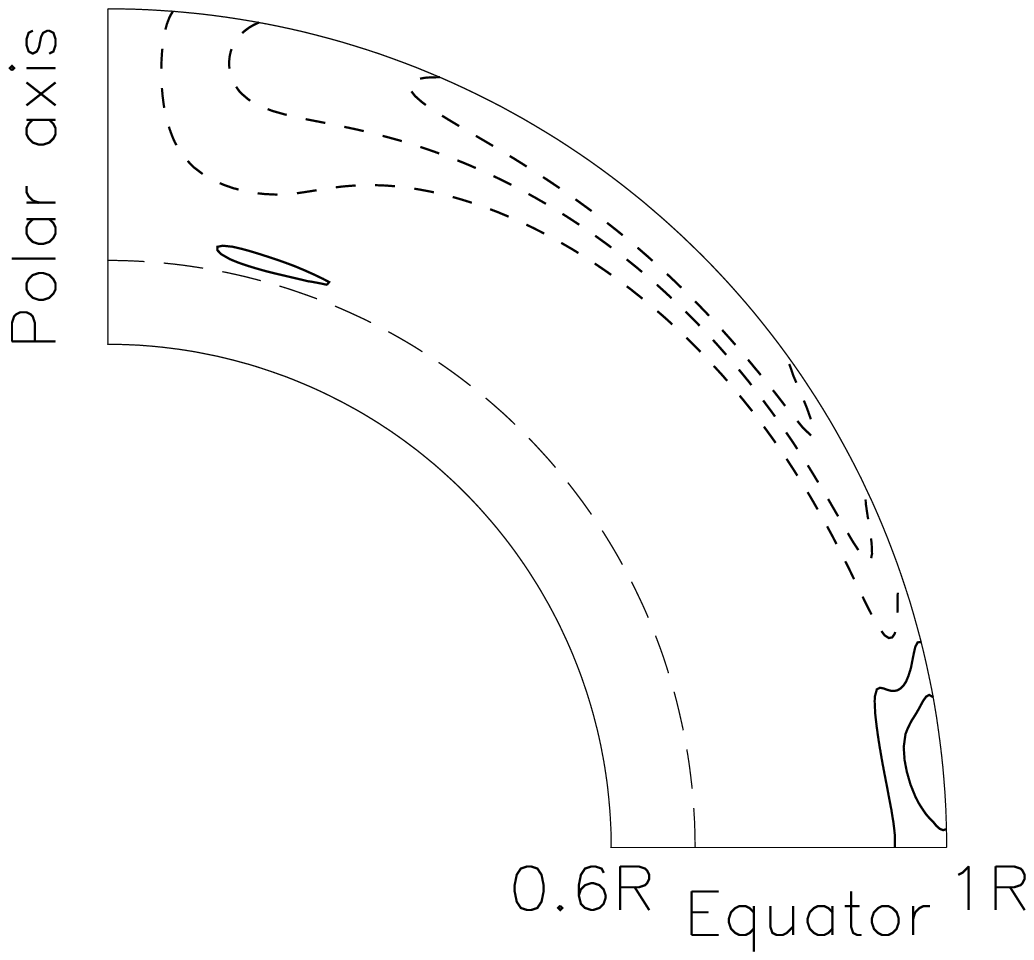,height=4.0cm}
      \psfig{file=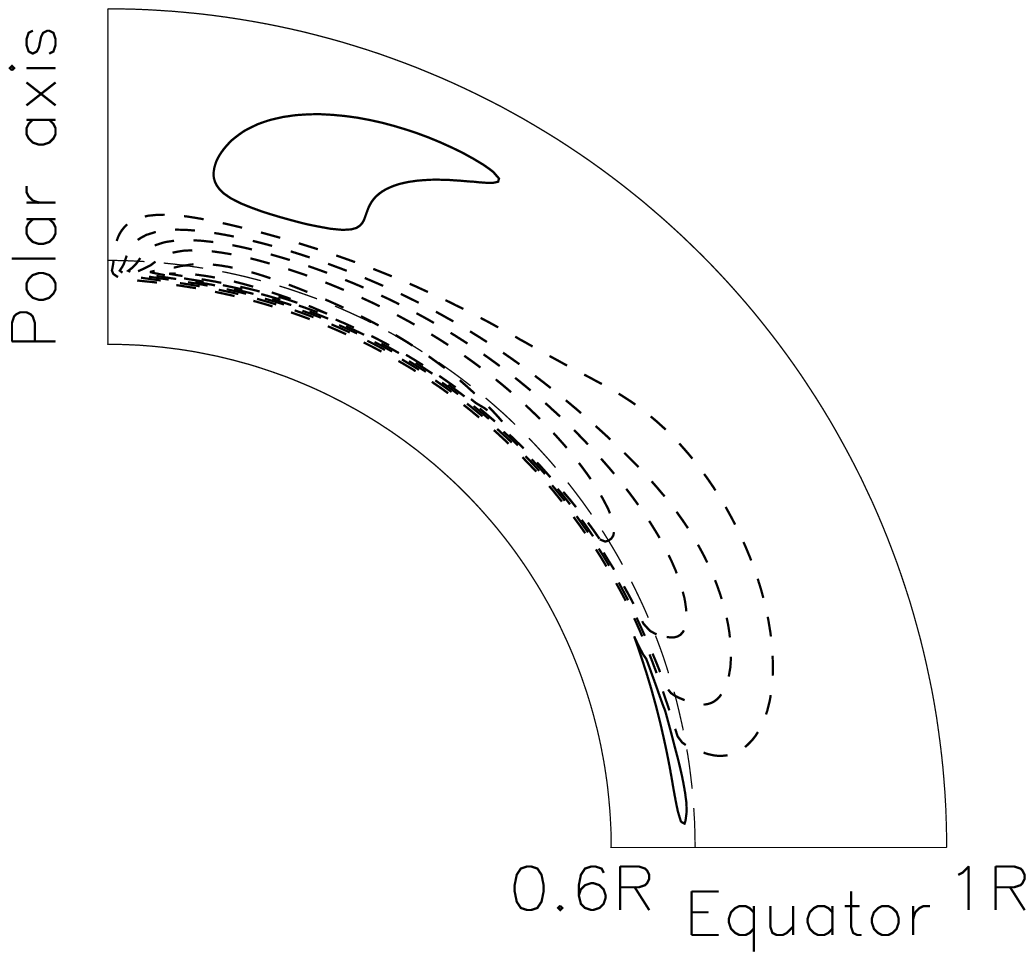,height=4.0cm}
     }
\mbox{(e)
      \psfig{file=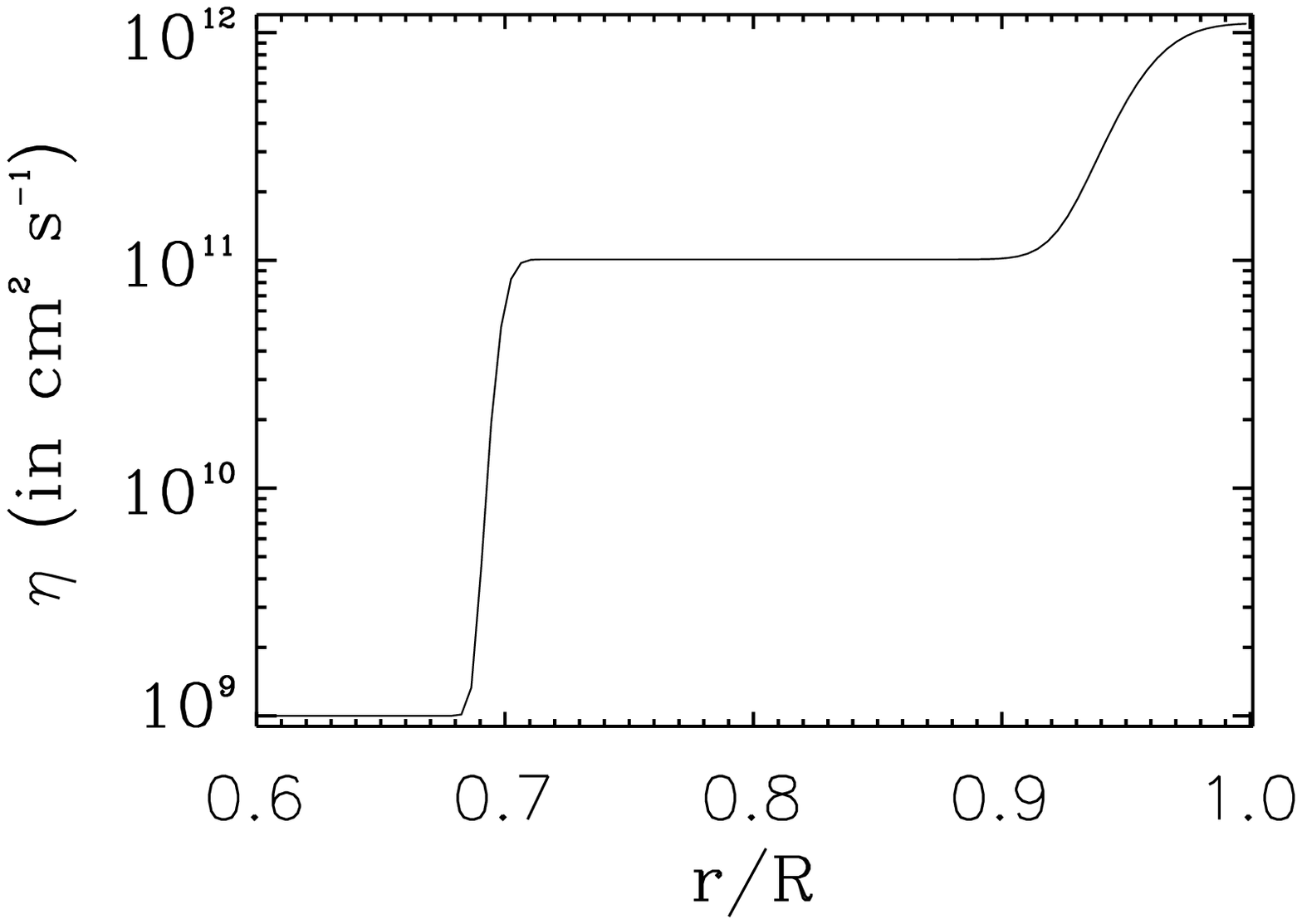,height=4.0cm}
      \psfig{file=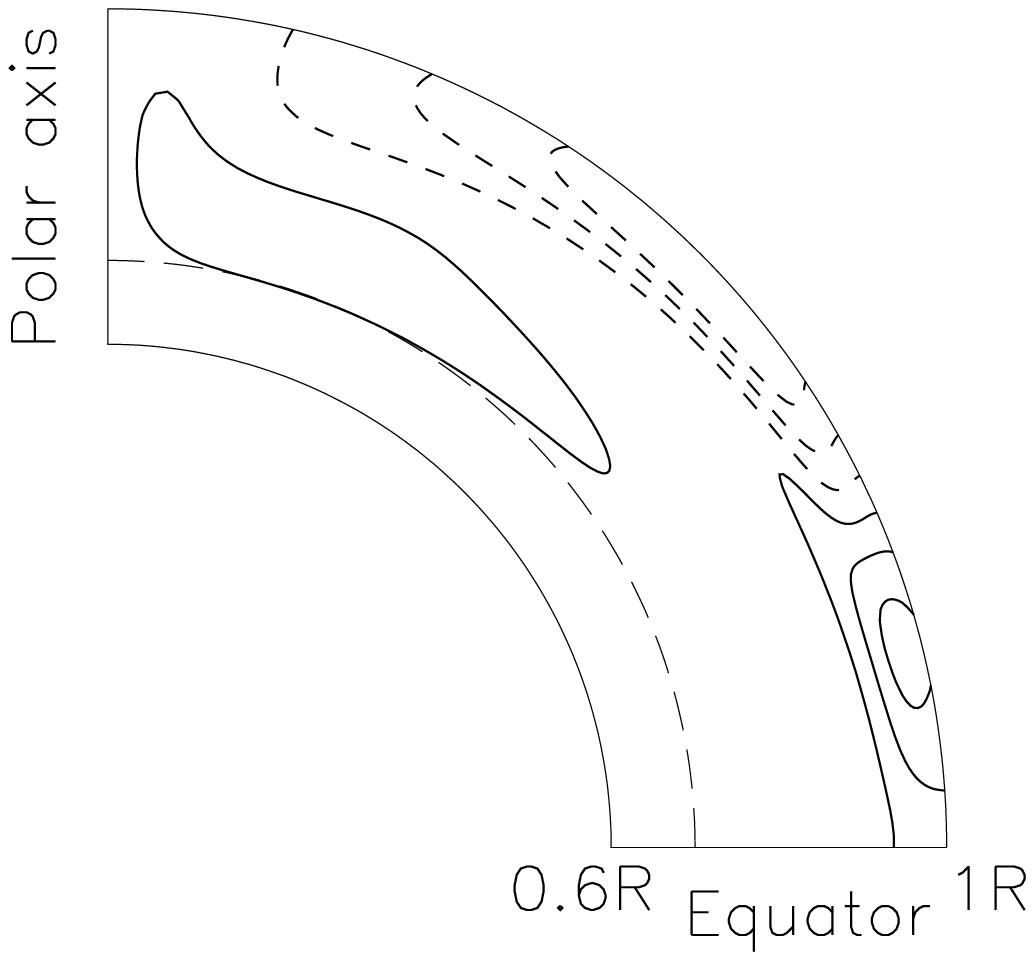,height=4.0cm}
      \psfig{file=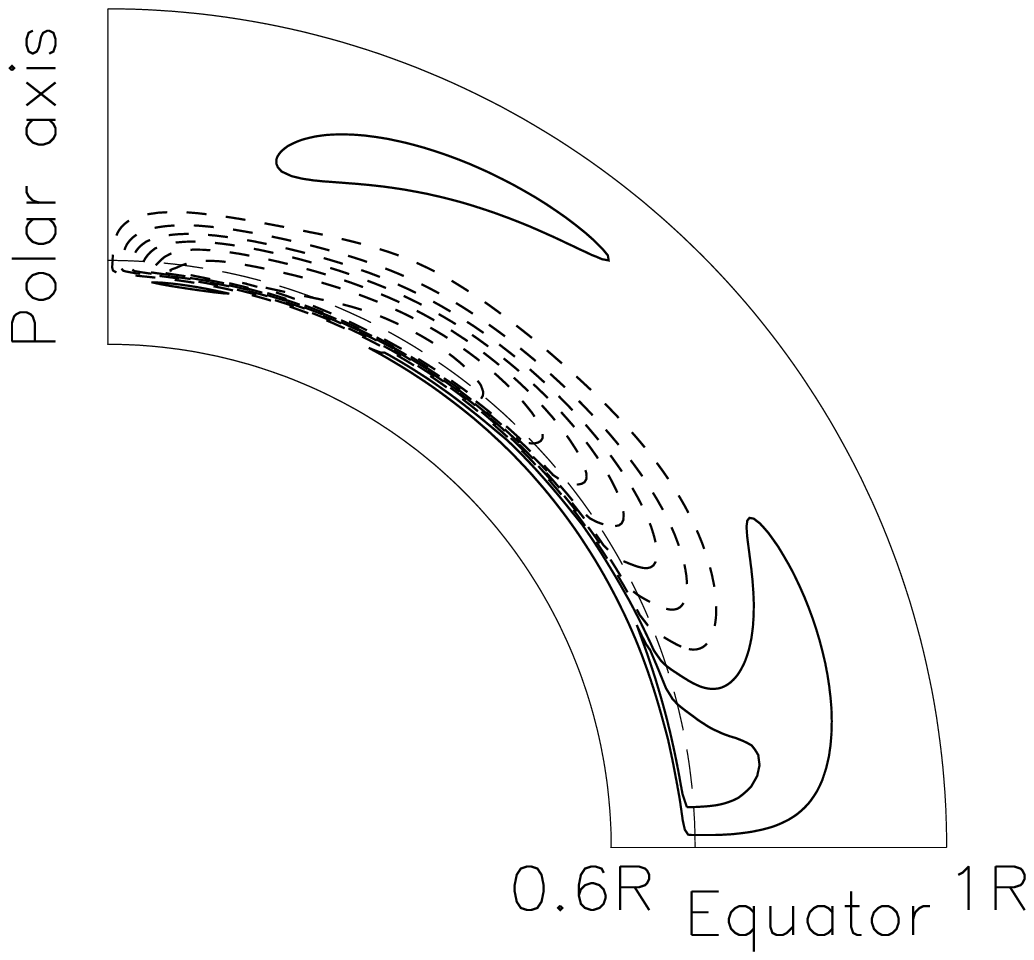,height=4.0cm}
     }
\caption{Experiment I: Diffusivity profiles with 
corresponding poloidal (left) and toroidal (right) field profiles
near solar maximum.
}
\label{bfields1}
\end{figure}
\clearpage

\subsection{Results of Experiment I: Various diffusivity profile shapes}
\label{resultsI}
\vspace{0.2cm}
We compare the output for dynamo runs initialized with 
magnetic diffusivity profiles
illustrated in Figure 1.

While the meridional flow speed primarily governs the dynamo cycle period 
($T$) in flux-transport models, 
we find that differences in the diffusivity profile 
can modify the cycle period.
While our imposed dynamo driver is initially set to a period of 22 years, 
the code responds to different diffusivity profiles 
by yielding actual cycle periods varying from 
about 19 to 30 years.
And
while we set the
poloidal quenching field strength at $B_0=50$ kGauss, 
the maximum toroidal fields produced in the tachocline 
vary with different diffusivity profiles. 
Table I summarizes the cycle period
$T$ and maximum toroidal field strength at the tachocline for each simulation.
%
\vspace{0.1cm}

\begin{tabular}{lcc} \\ \hline
diffusivity profile & $max(B_{\phi})$ & cycle period ($T$)\\ 

\hline
(a) constant       &  56 kG & 23.5 years \\ 
(b) linear         & 103 kG & 24.0 years \\
(c) step-to-linear & 175 kG & 29.7 years \\
(d) single-step    & 168 kG & 28.0 years \\ 
(e) double-step    &  50 kG & 19.0 years \\ 
\hline
\end{tabular}

Table I. Maximum toroidal field strength and dynamo cycle period ($T$) at
the tachocline for
each diffusivity profile in Experiment I (Fig.1).
\vspace{0.2cm}
%
%

\subsubsection{Relations between cycle times and field patterns }
\label{discussI}
\vspace{0.2cm}
By examining the evolution of poloidal field patterns in each case,
we can learn why the dynamo cycles vary for different magnetic
diffusivity profiles.
Consider the three most physically plausible profiles:
(c) step-to-linear, (d) single-step, and (e) double-step.
In Table I, the cycles are longest (28 years or more) in 
cases (c) and (d), and closest to the true solar cycle in cases (e)
and (a).  
With its constant diffusivity profile, 
case (a) is patently non-physical. Its artificially high diffusivity
in the lower half of the convection zone enhances flux diffusion
and suppresses flux storage and amplification near the tachocline.
We will find that the evolution of the magnetic flux contours of the
double-step profile (e) are far more physical,
even though its cycle time and maximum tachocline field resemble
case (a).

We argue that an effective decrease in advective-diffusive 
transport processes near the tachocline region in cases (c) and (d)
causes the dynamo to produce longer cycles, whereas an effective increase
in the transport near the surface due to enhanced diffusivity in 
case (e) is the reason for the shorter dynamo cycles. 
The enhanced near-surface effective transport in case (e) 
also enables poloidal fields to
reach mid-latitudes in the tachocline faster than in the other four
cases. 

\subsubsection{Time evolution of magnetic fields }
\begin{figure}[ht]
\centering
\mbox{
      \psfig{file=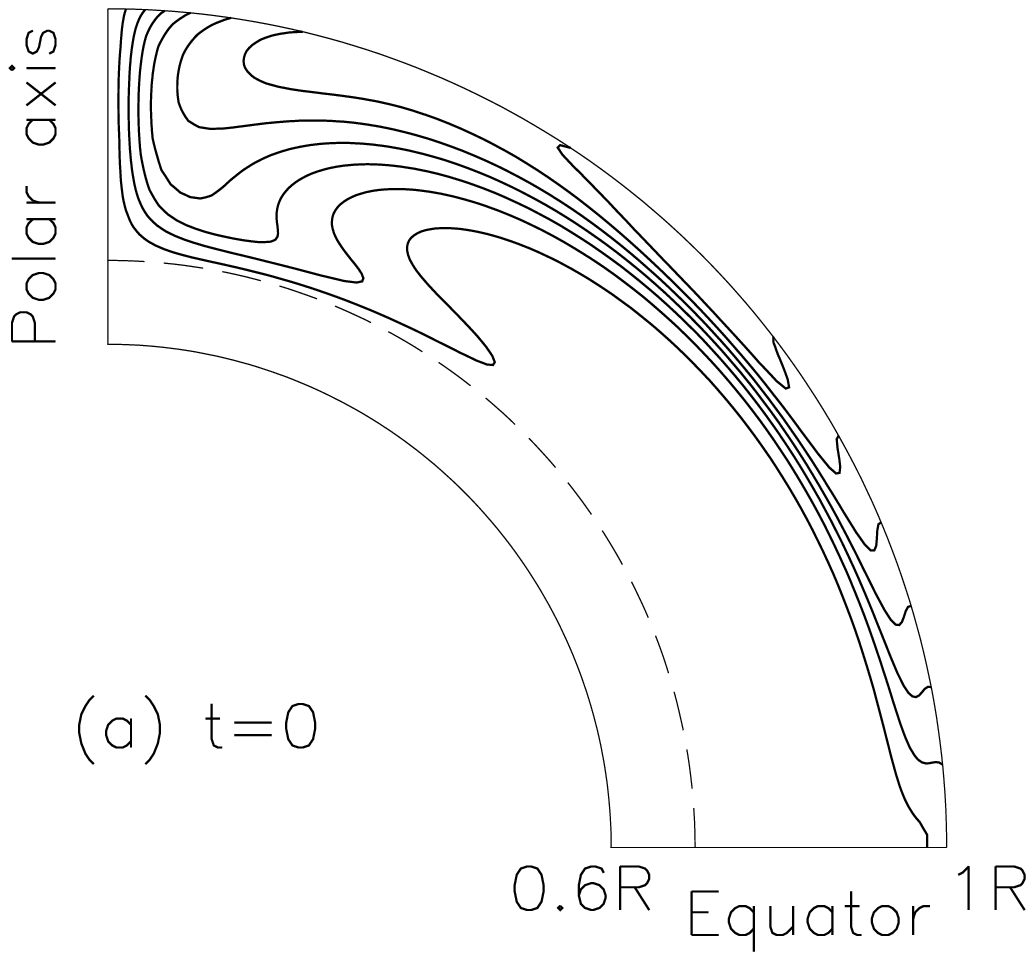,height=4.5cm}
      \psfig{file=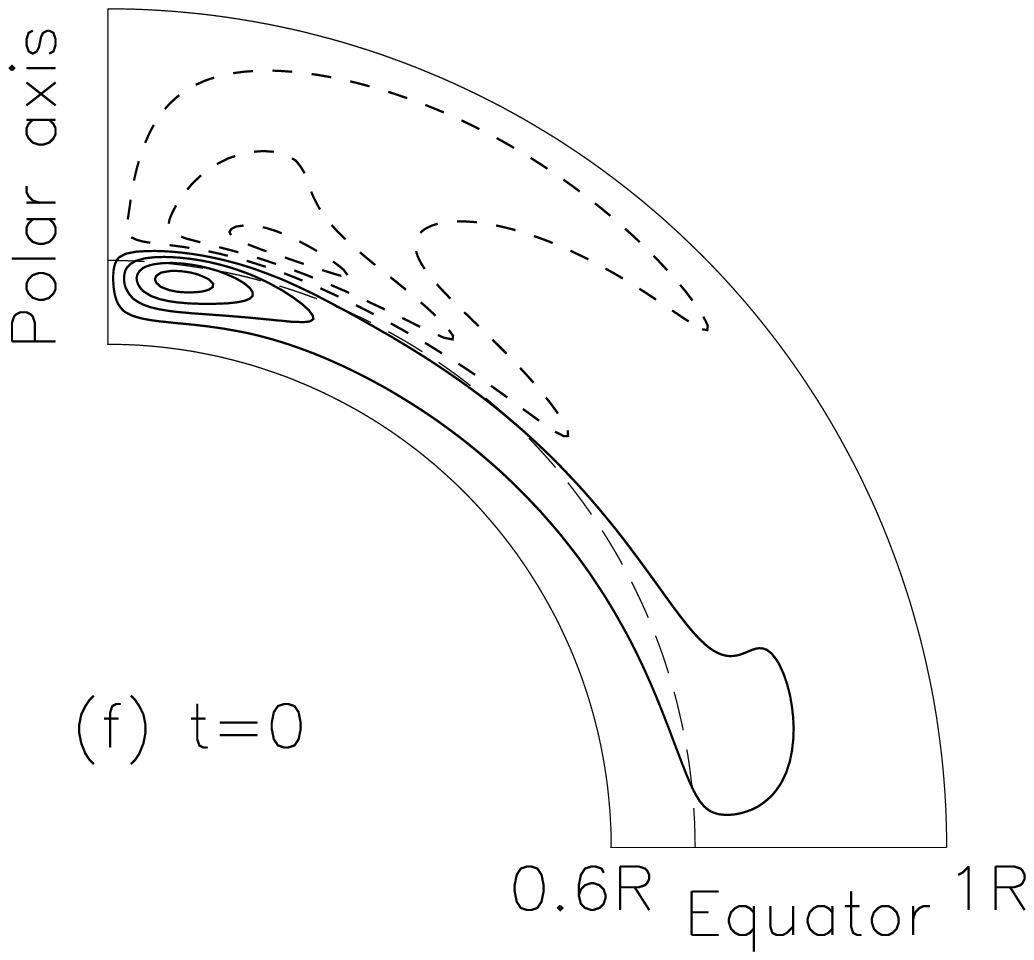,height=4.5cm}
     }
\mbox{
      \psfig{file=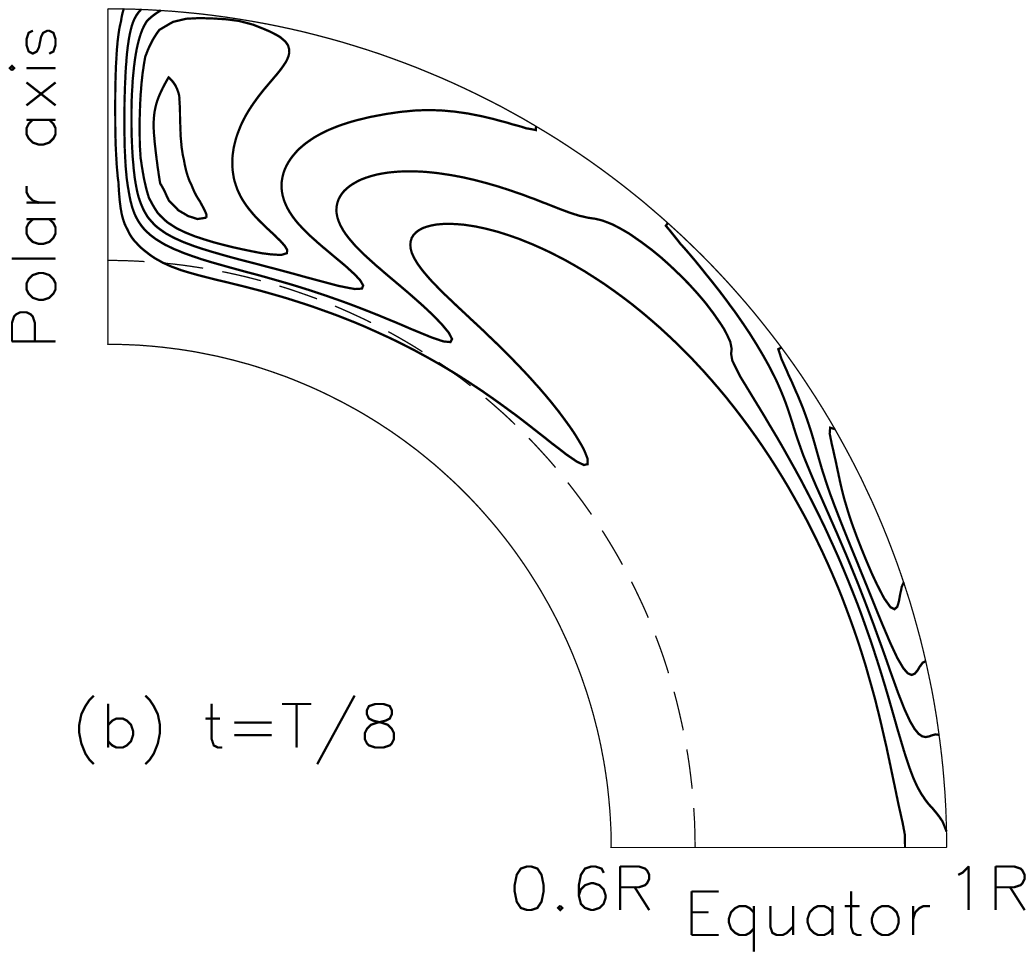,height=4.5cm}
      \psfig{file=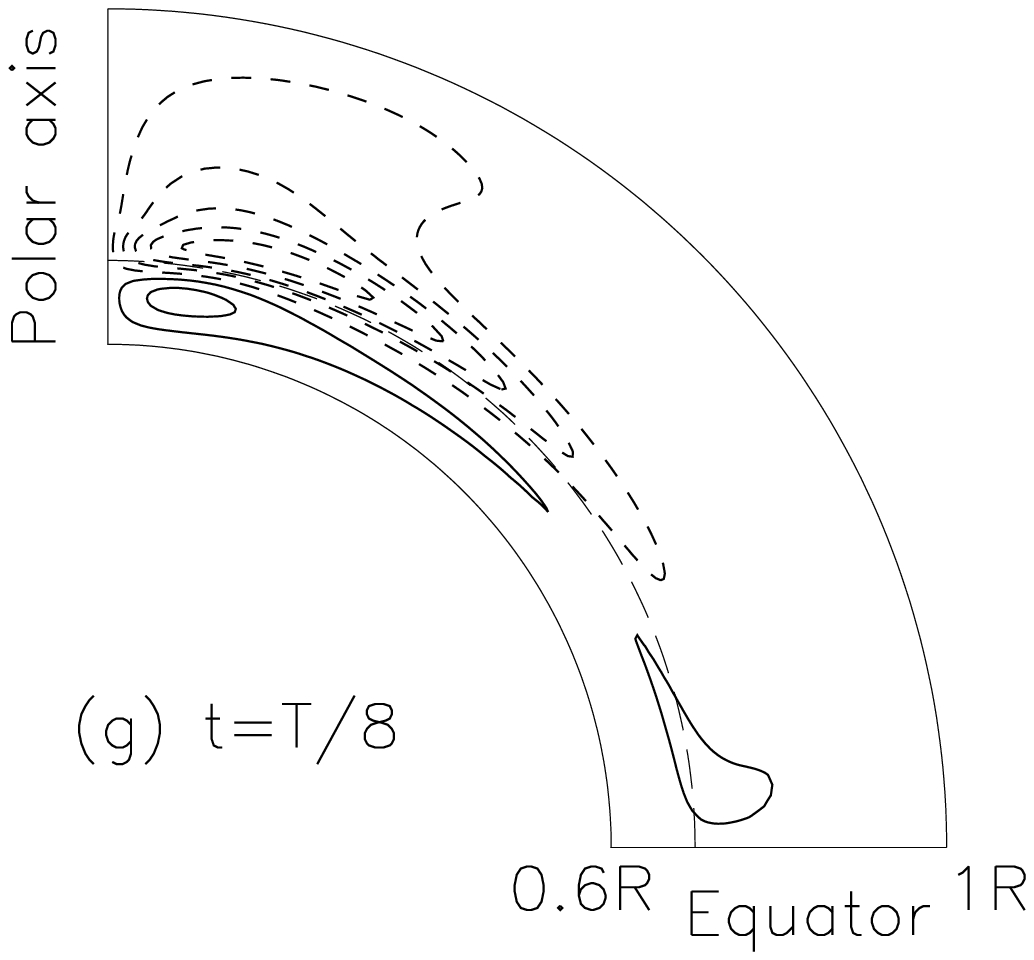,height=4.5cm}
     }
\mbox{
      \psfig{file=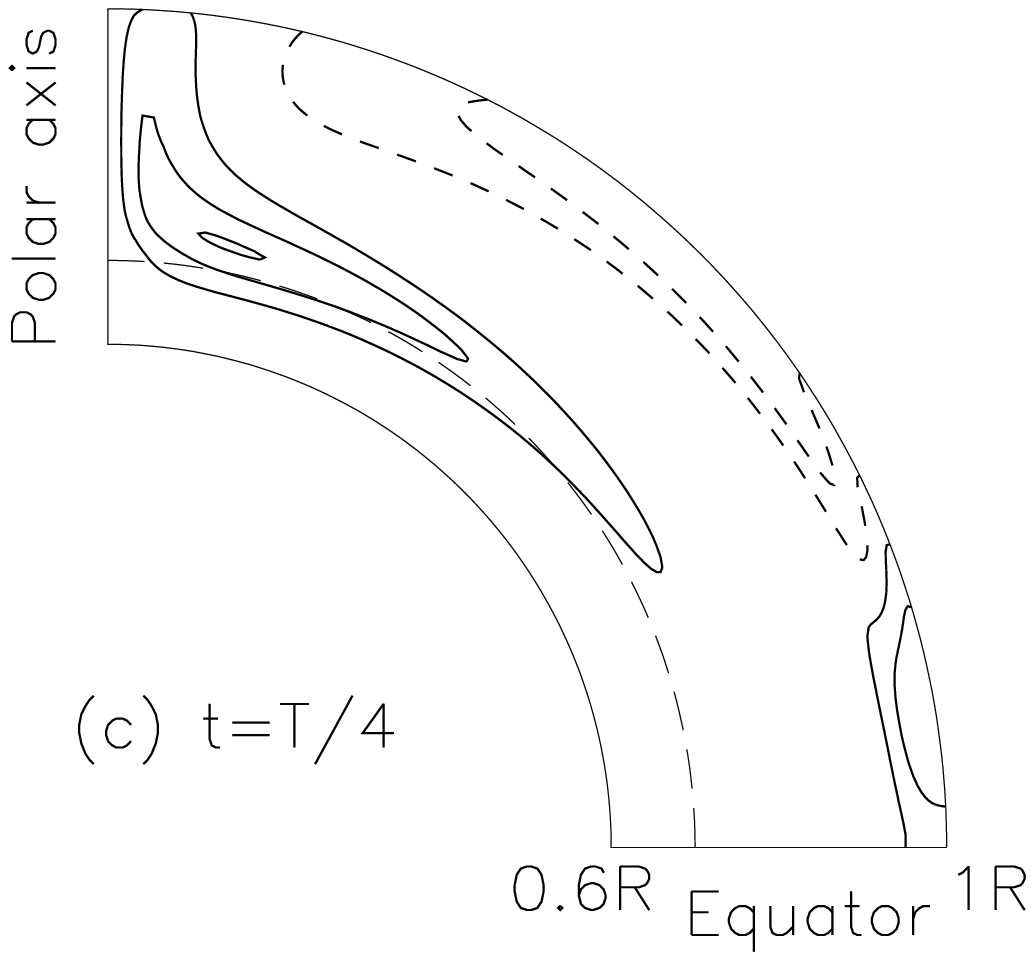,height=4.5cm}
      \psfig{file=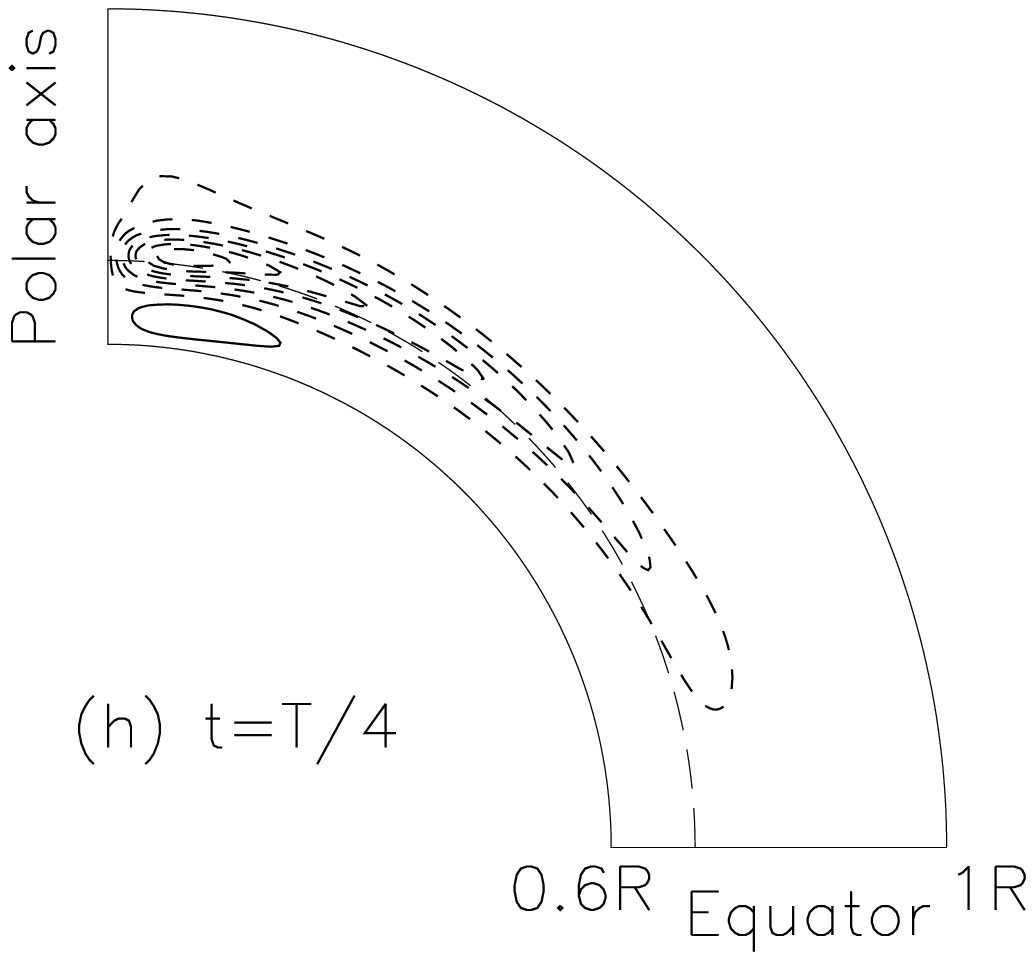,height=4.5cm}
     }
\mbox{
      \psfig{file=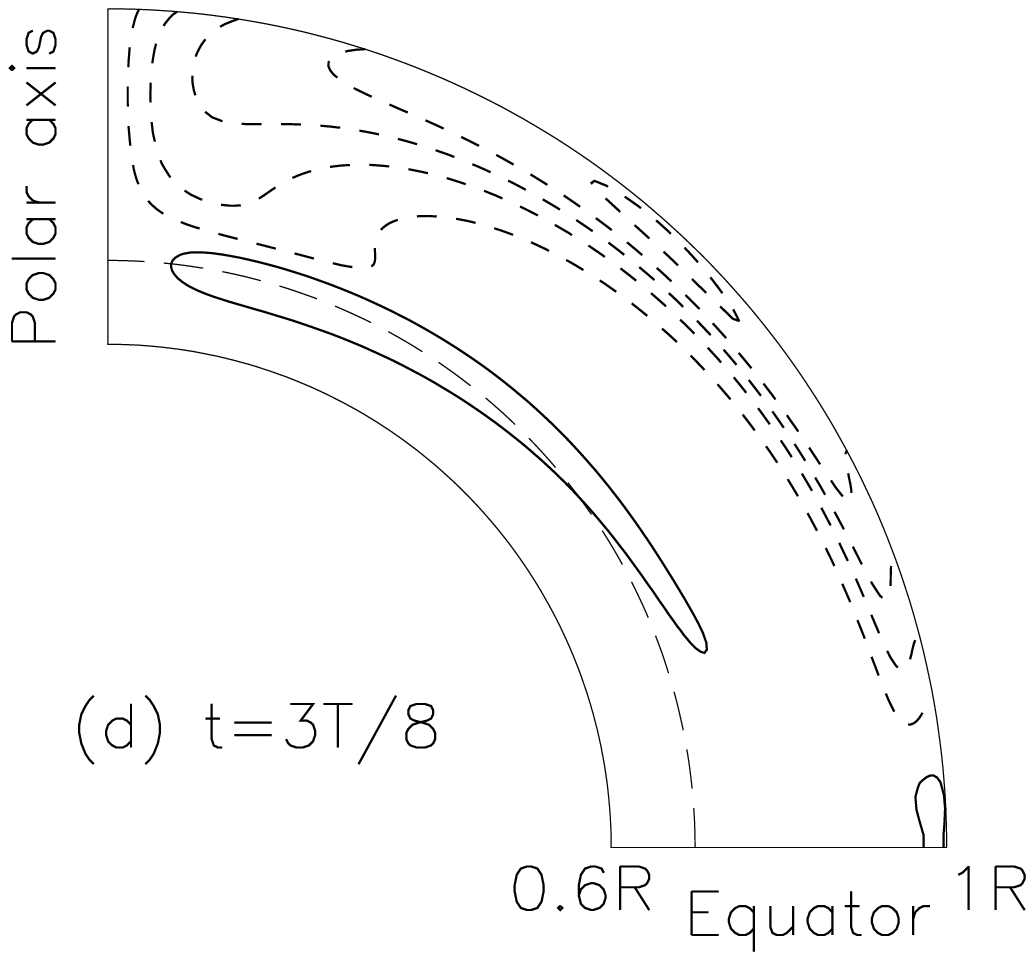,height=4.5cm}
      \psfig{file=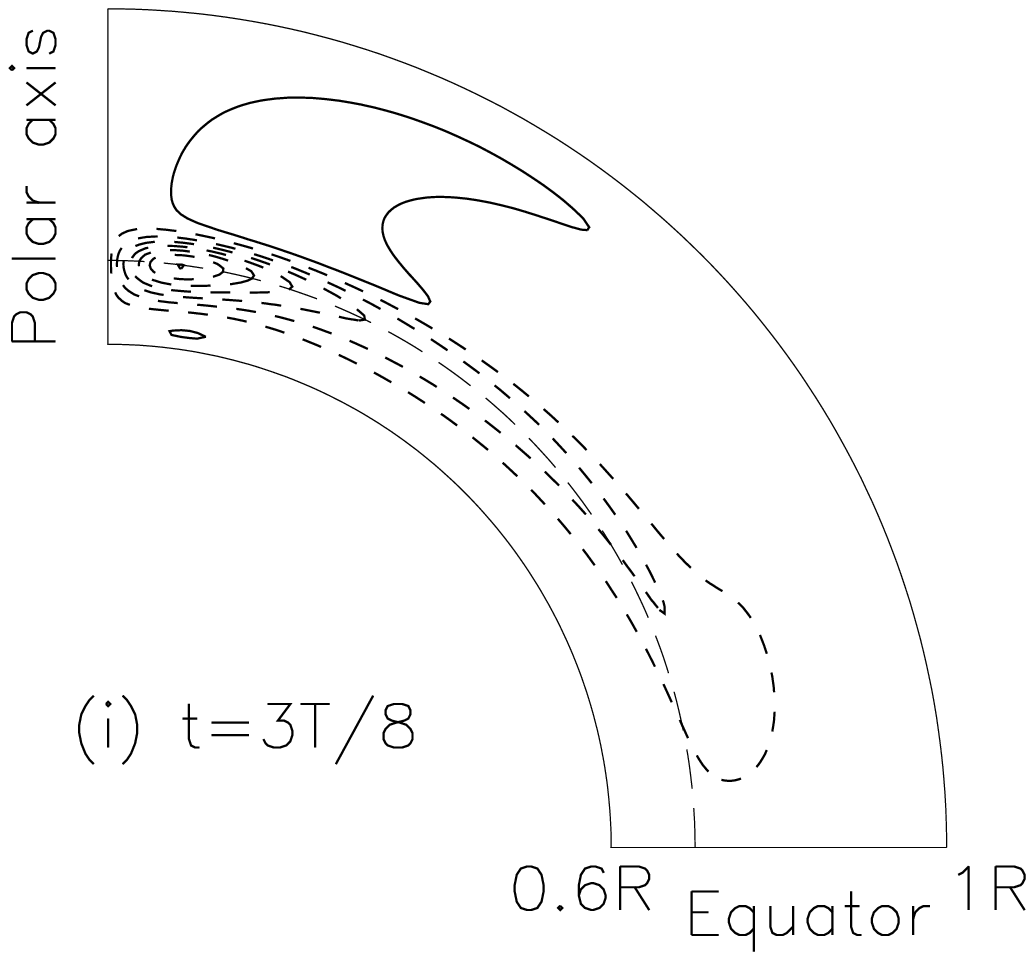,height=4.5cm}
     }
\mbox{
      \psfig{file=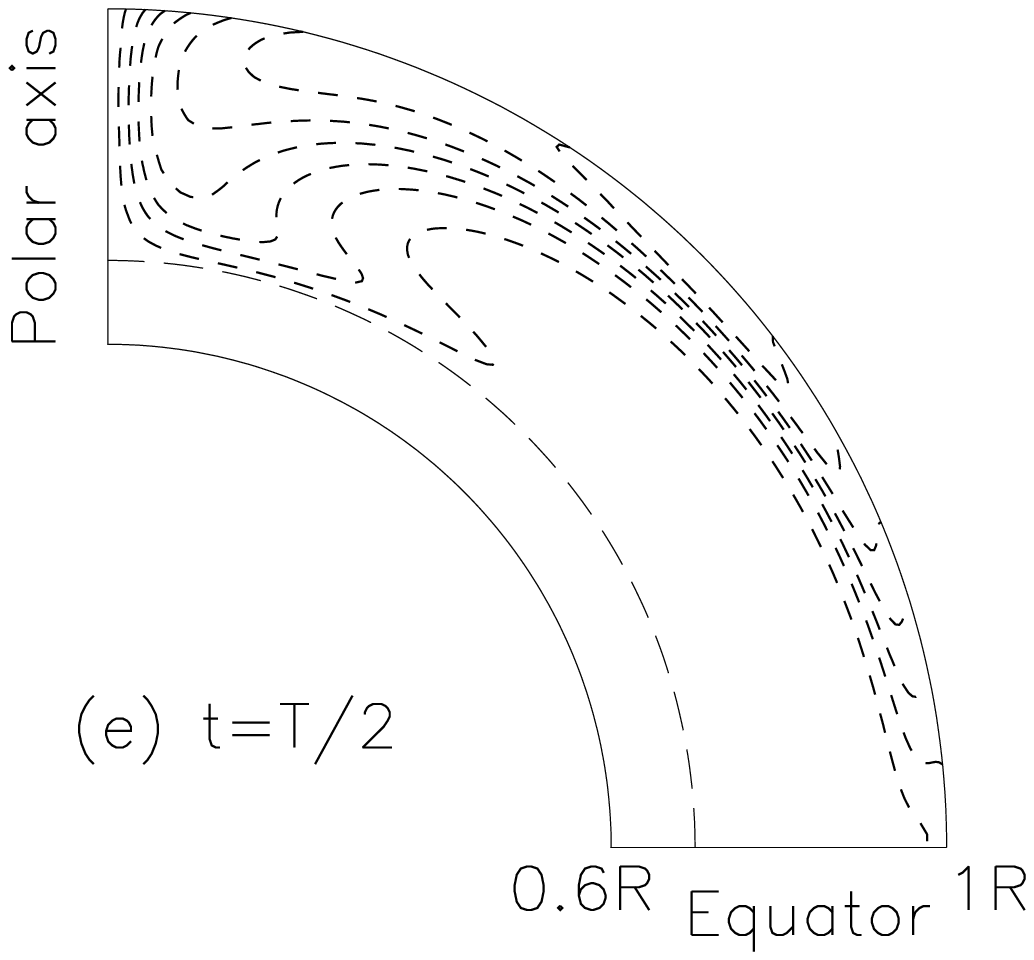,height=4.5cm}
      \psfig{file=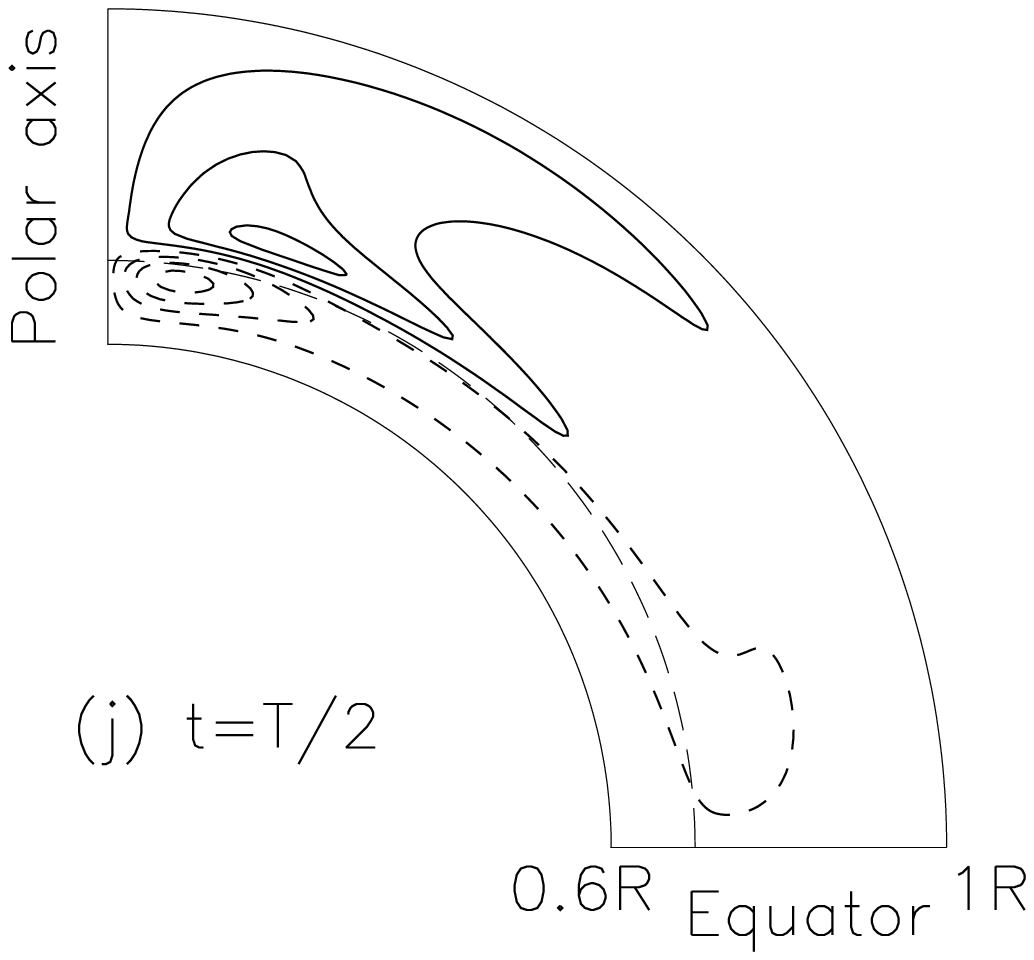,height=4.5cm}
     }
\caption{Linear diffusivity profile from Fig.1b.
Field lines plotted in logarithmic intervals.
Left = Poloidal field, Right = Toroidal field.
Time advances downward.}
\label{b3plots}
\end{figure}

\begin{figure}[ht]
\centering
\mbox{
      \psfig{file=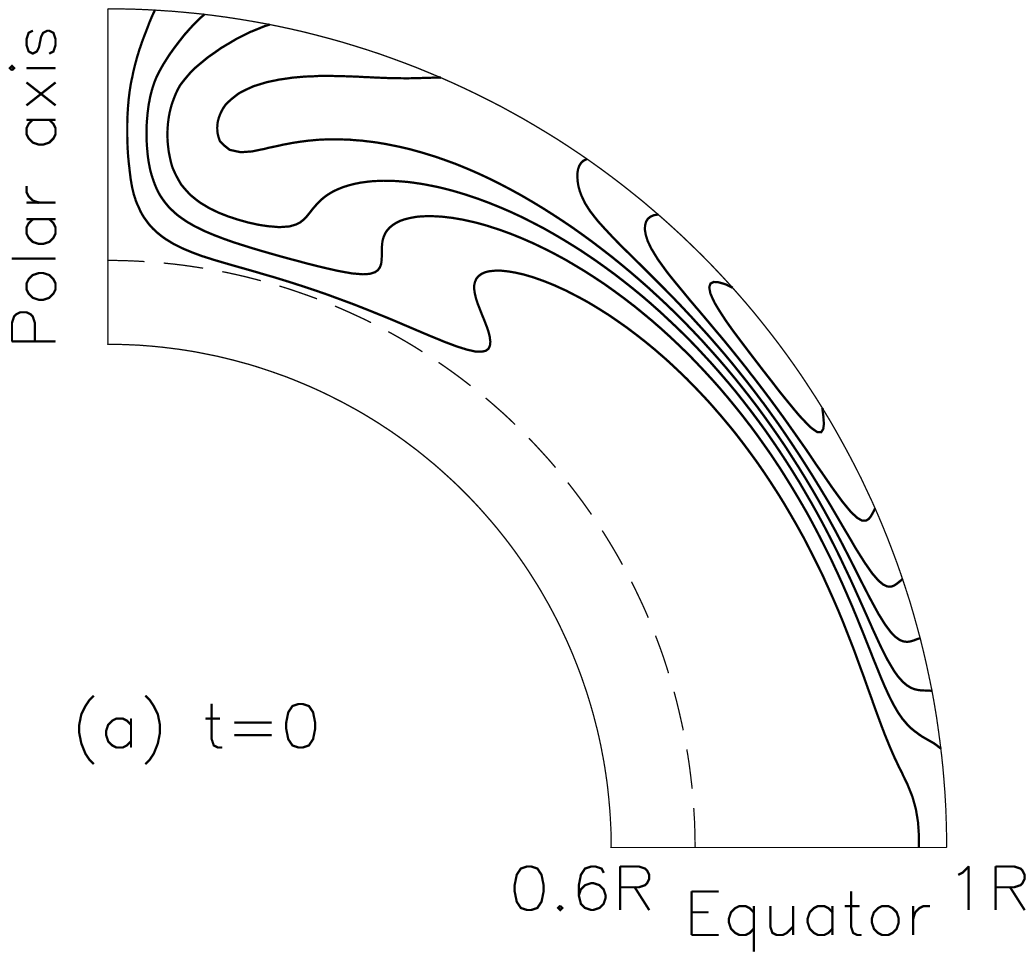,height=4.5cm}
      \psfig{file=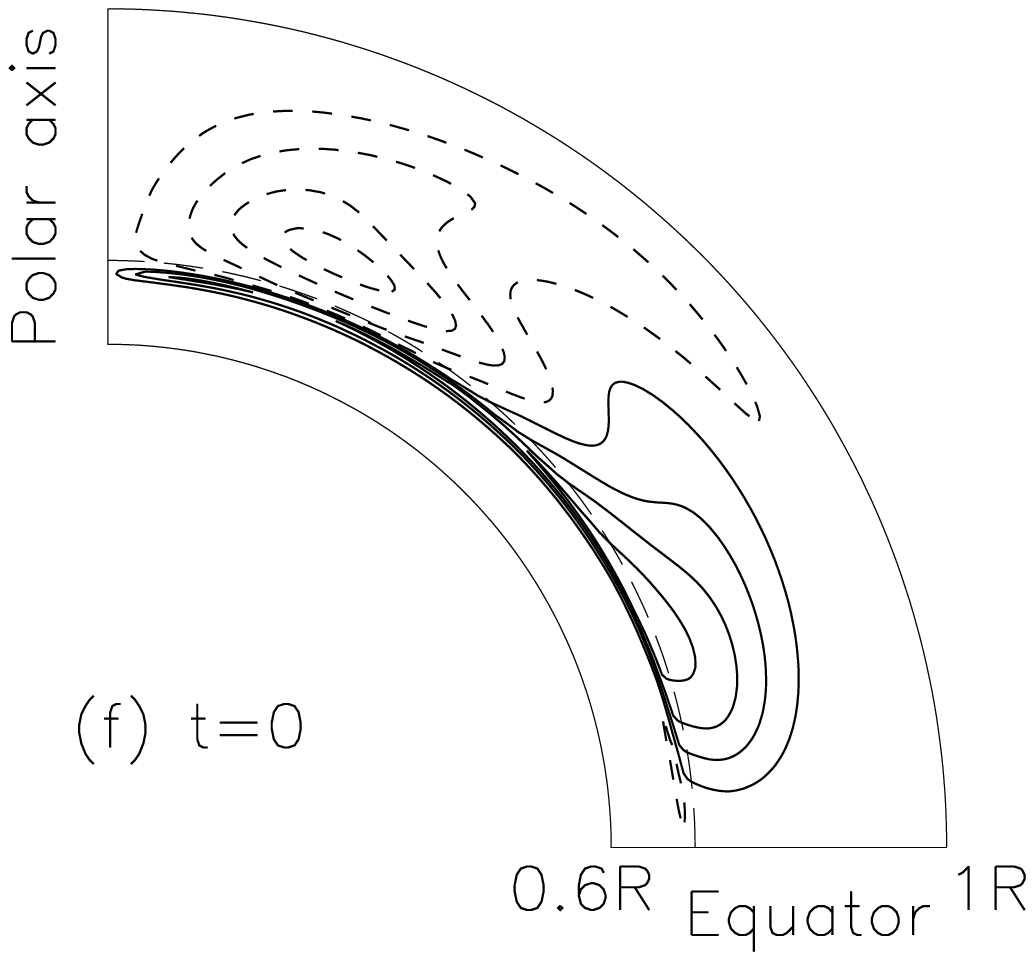,height=4.5cm}
     }
\mbox{
      \psfig{file=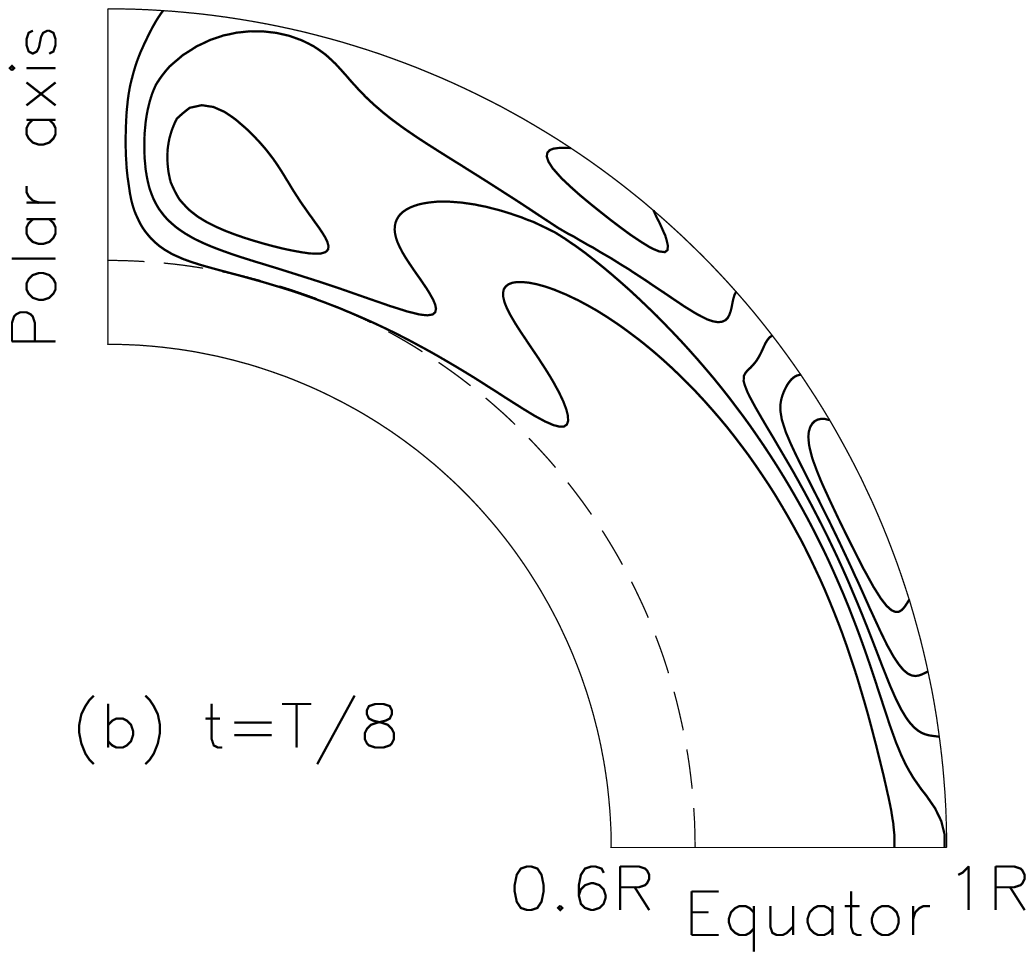,height=4.5cm}
      \psfig{file=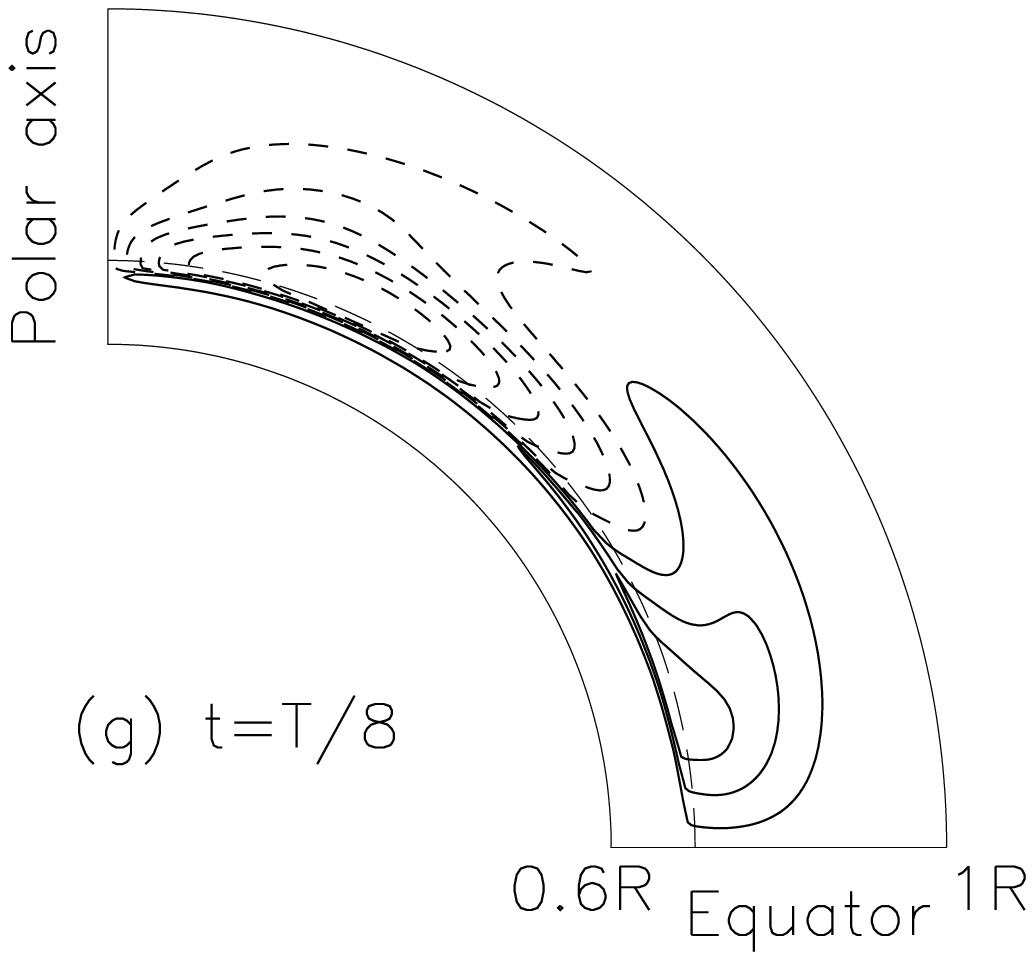,height=4.5cm}
     }
\mbox{
      \psfig{file=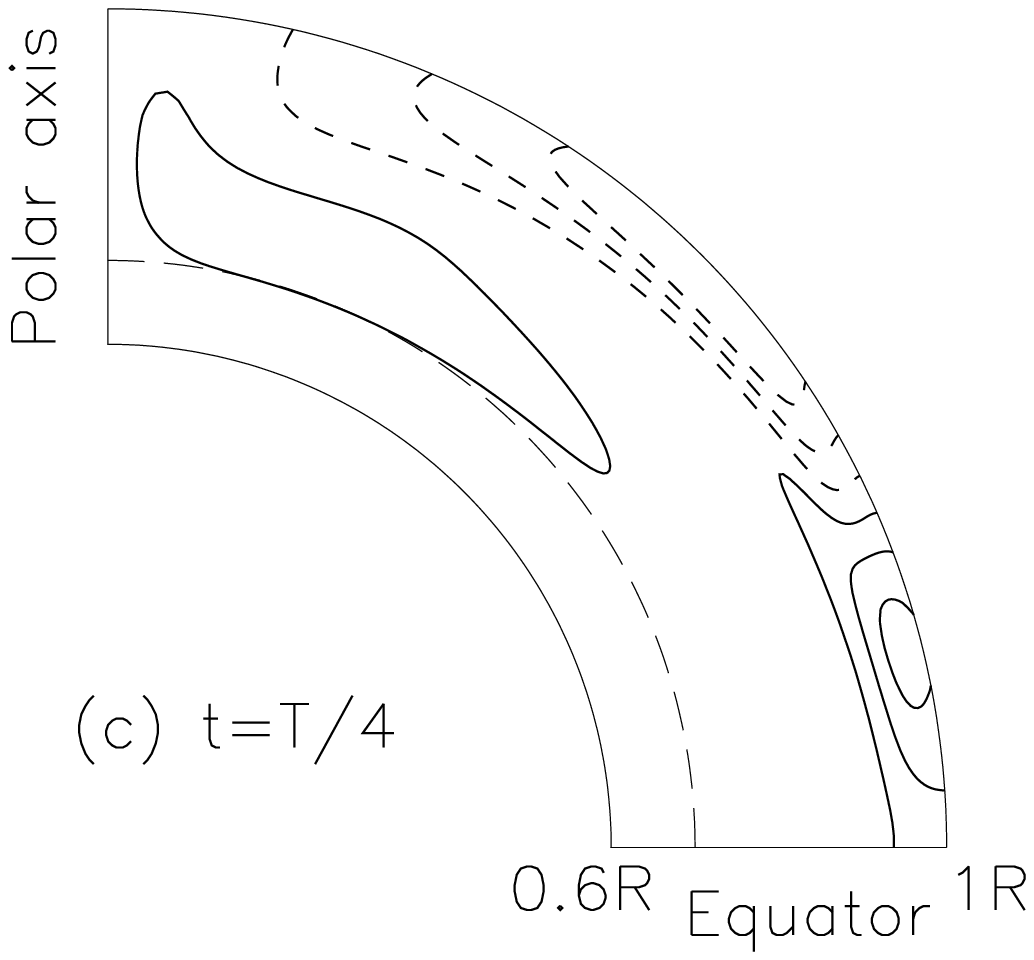,height=4.5cm}
      \psfig{file=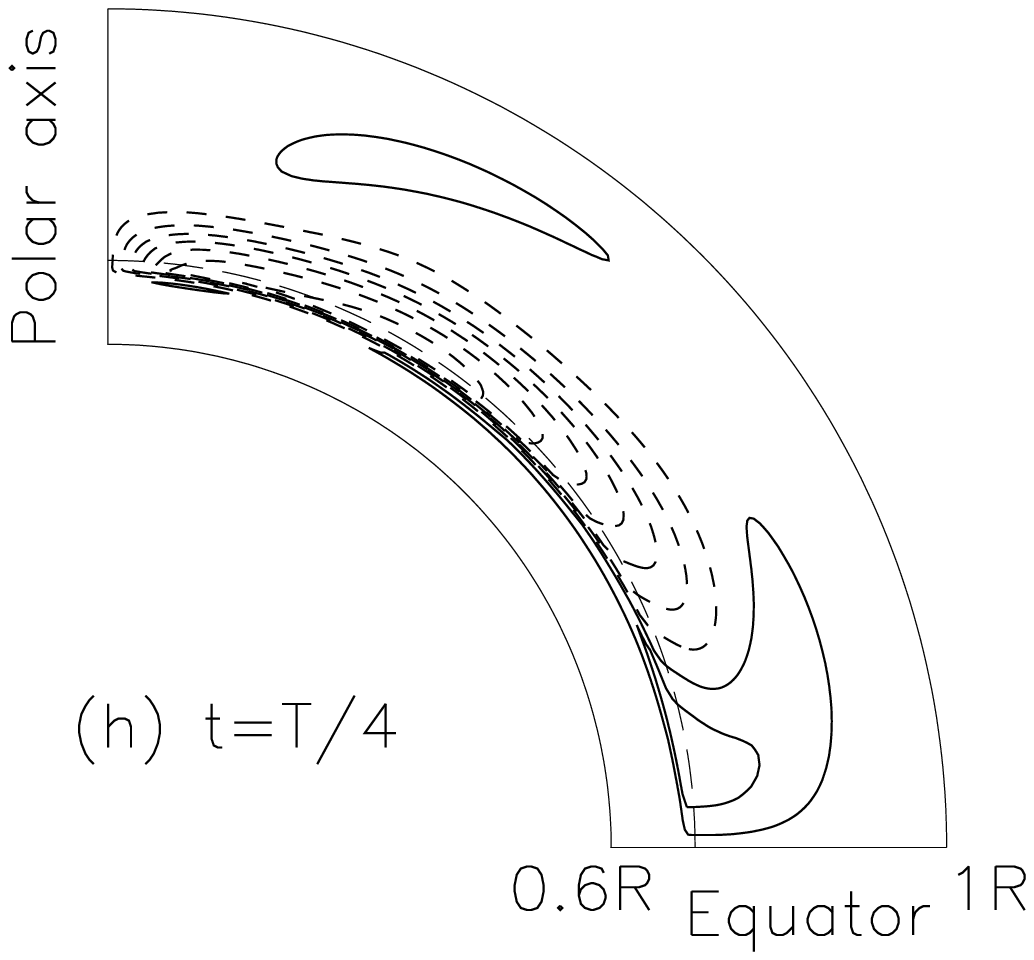,height=4.5cm}
     }
\mbox{
      \psfig{file=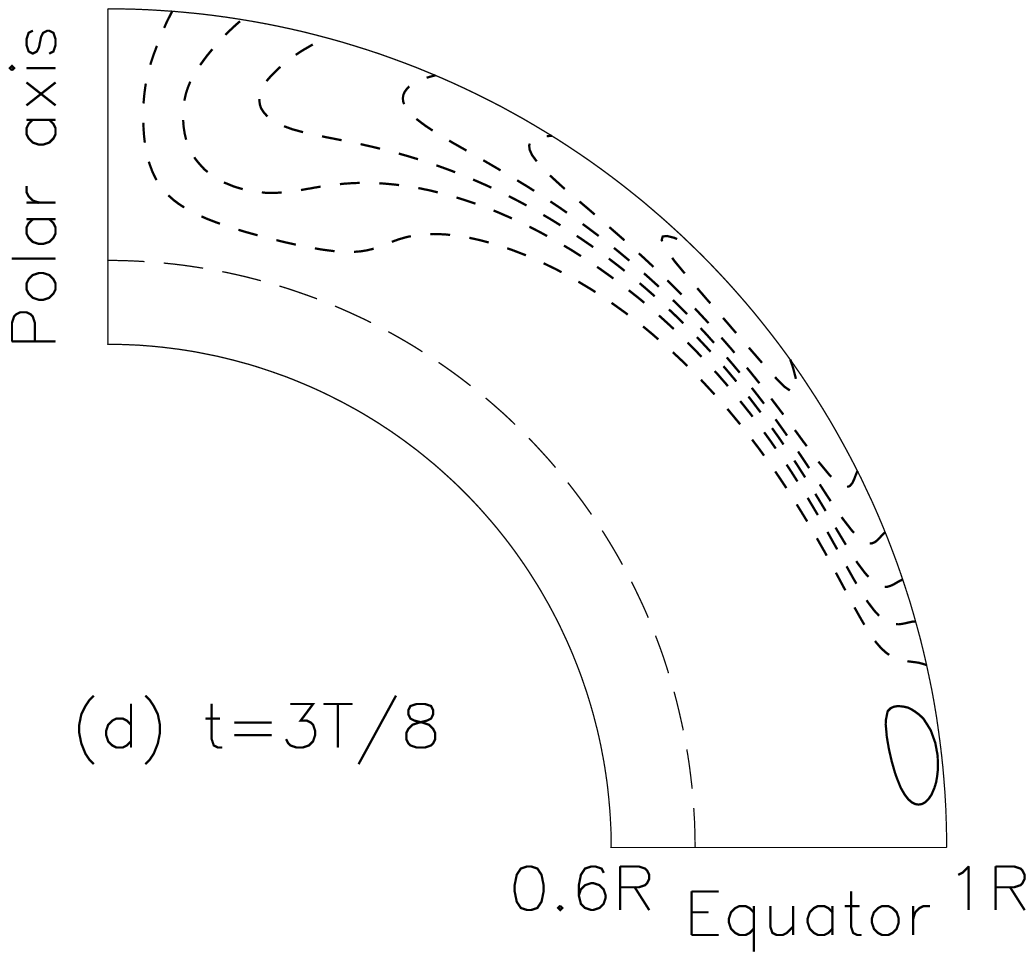,height=4.5cm}
      \psfig{file=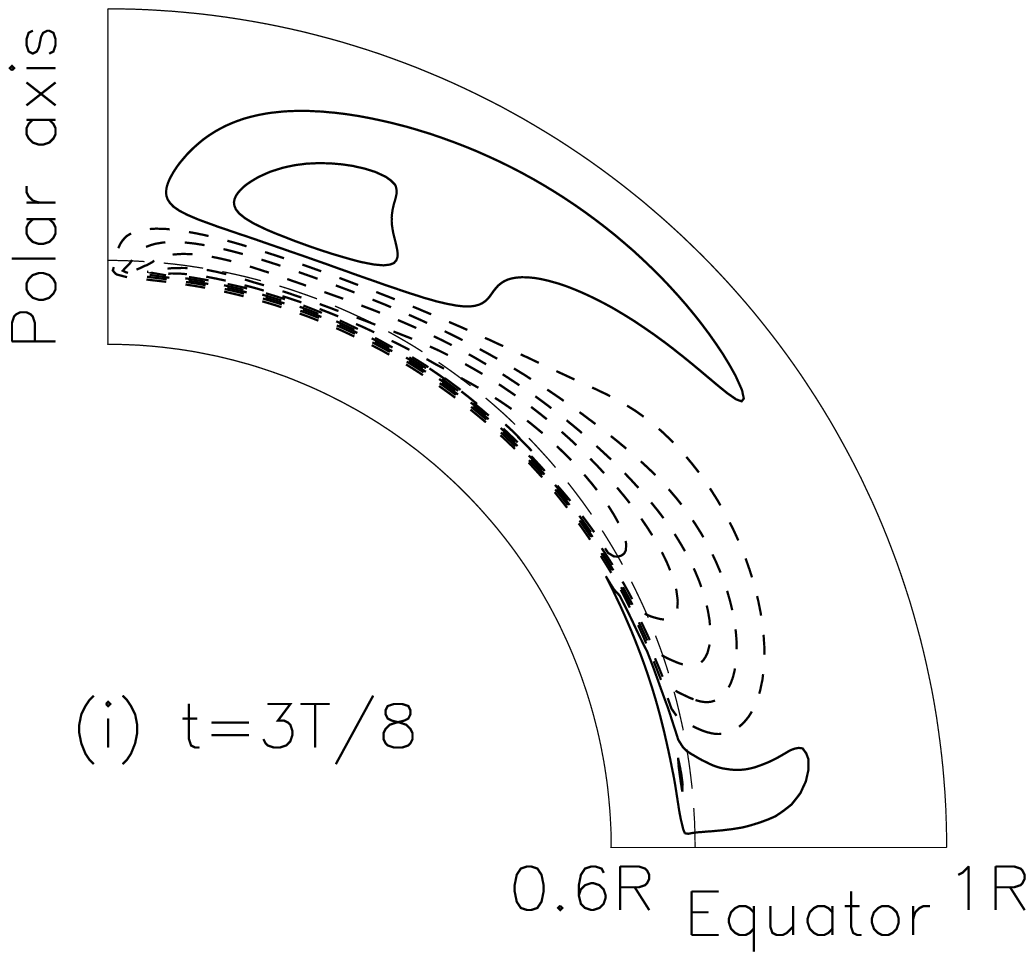,height=4.5cm}
     }
\mbox{
      \psfig{file=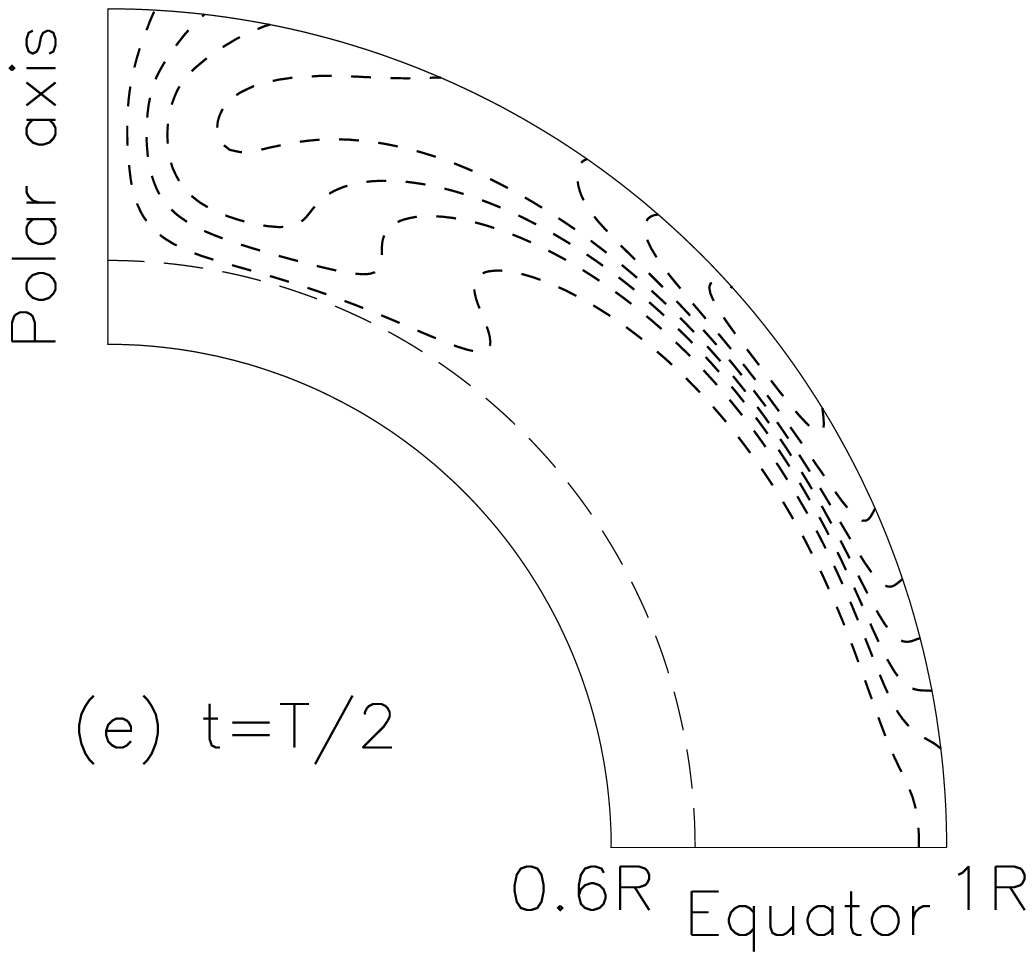,height=4.5cm}
      \psfig{file=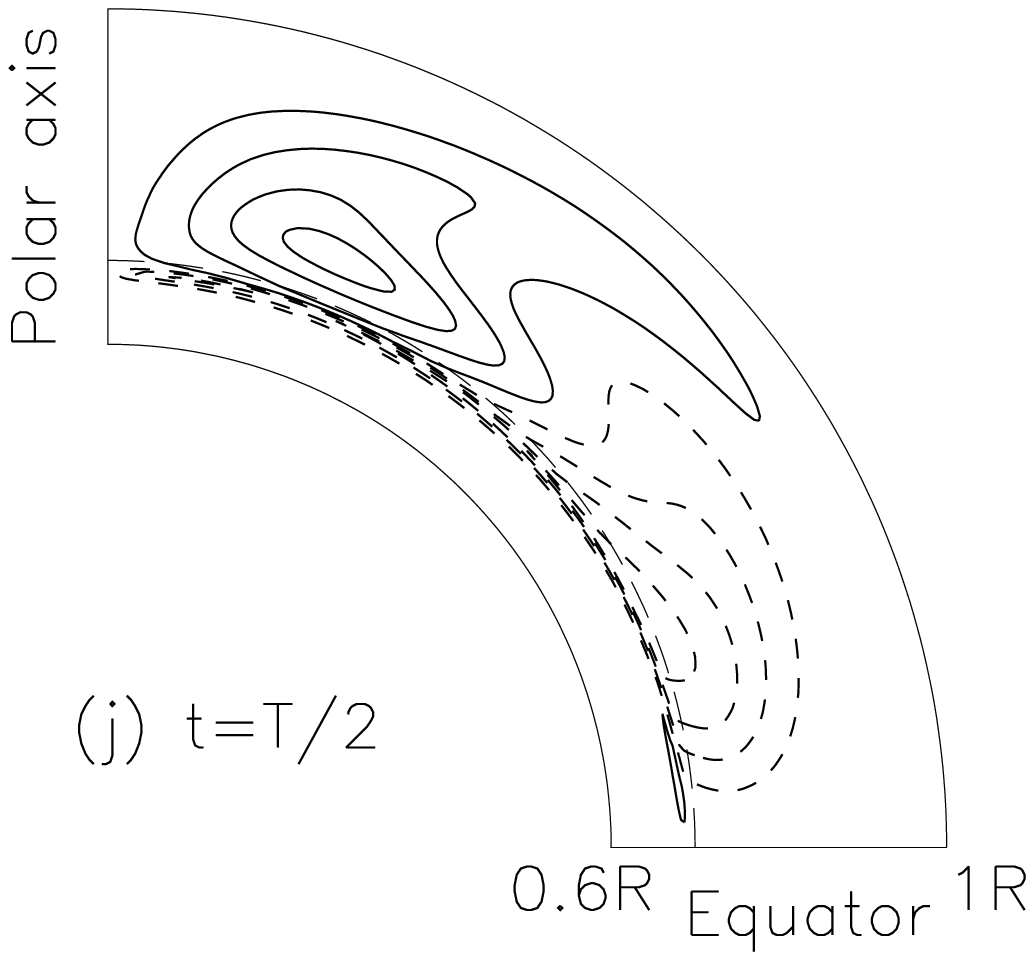,height=4.5cm}
     }
\caption{Double-step diffusivity profile from Fig.1e.
Field lines plotted in logarithmic intervals.
Left = Poloidal field, Right = Toroidal field.
Time advances downward.}
\label{b2plots}
\end{figure}

\vspace{0.2cm}
We can gain more insight by comparing the time evolution of the
poloidal and toroidal fields in the convection zone for 
two very different cases.   
We plot five 
pairs of snapshots of evolving field line contours, 
spanning half a solar cycle, for the linear diffusivity 
profile in Fig.\ref{b3plots} and the
double-step profile in Fig.\ref{b2plots}.
For each case, the left column shows the evolving poloidal 
field lines in logarithmic intervals in meridional planes, and the 
right column shows the toroidal field lines, with time advancing downward.
Solid (dashed) lines represent positive (negative) fields.
Maximum field strengths for each case are listed in Table I, and
each series of plots is scaled to $10 \%$ above the maximum for that case.
The pattern repeats every cycle period $T$; we have chosen
for each diffusivity profile approximately the same phase of a 
cycle to set the $t=0$ snapshot. Thus we can more easily compare 
the time evolution for different cases.

\input{psfig}

\input{psfig}

The linear $\eta(r)$ case (Fig.1b) is almost certainly less physical, 
with unreasonably high resistivity
deeper into the radiative zone.
At $t=0$ for the linear case $\eta(r)$,
positive poloidal fields are distributed along the surface,
concentrated at mid to high latitudes, and strong at the pole
(Fig.\ref{b3plots}a).
Strong negative toroidal field concentrates at the pole at $t=0$, 
and weak positive toroidal field stretches along the tachocline
(Fig.\ref{b3plots}f).
As time progresses, the positive poloidal field sinks from the pole,
stretches along the tachocline, weakens, and disappears (Figs.\ref{b3plots}b-e).
At quarter-cycle (Fig.\ref{b3plots}c), the positive poloidal field is split or bifurcated
into high-latitude poleward and low-latitude equatorward components.
Meanwhile, new poloidal fields of opposite sign (negative) are generated
at midlatitudes at the surface (Fig.\ref{b3plots}c), advance poleward, and 
strengthen, until at midcycle (frames e) they have taken the place of 
the original positive poloidal fields of Fig.\ref{b3plots}a.

For the same linear profile, negative toroidal field is concentrated at 
the pole at $t=0$ (Fig.\ref{b3plots}f) while weaker 
positive toroidal field is distributed along the tachocline.
The negative toroidal field strengthens at the pole, where
its radial extent remains rather broad until reversal is nearly complete
(Fig.\ref{b3plots}f-i).
Meanwhile, positive toroidal field stretches equatorward and weakens
(Fig.\ref{b3plots}g).
As the negative field moves down toward the tachocline, it concentrates, while the
positive field disappears from view (Fig.\ref{b3plots}h).
As the cycle progresses, the negative toroidal field migrates equatorward 
and weakens,
while positive field starts being produced near the surface and strengthens 
as it sinks radially downward. 
Significant field-line features features produced in this broadly-distributed diffusivity 
profile are: (i) in both poloidal and toroidal field patterns, 
one polarity generally dominates over the other, 
with regions of stronger flux oscillating between pole and tachocline
over the cycle, indicating shorter-term memory
(less than a cycle) 
about the past magnetic fields;
(ii) with larger diffusivities
near the tachocline, the skin-depth is
large -- both toroidal and poloidal fields penetrate down to the
bottom boundary.

The constant profile (Fig.1a) gives field evolution results similar to 
the linear profile, but more extreme: (i) very short magnetic memory,
very deep skin depth effect, and (iii) very little field structure
along the meridional flow.

The time evolutions of the magnetic field maps of the
step-to-linear (Fig.1c) and the 
single-step profiles (Fig.1d) 
develop
(i) more evenly distributed polarities, indicating slightly
longer-term memory of at least one sunspot cycle;
(ii) shallower skin depth, due to the diffusivity jump near the tachocline
(iii) more stretching of fieldlines along the meridional flow ``conveyor belt."

Even without seeing time-series of meridional cuts, 
the overall shapes of the fields for each profile are revealing in
Fig.5.  For the flat and linear diffusivity profiles, field lines in
the convection zone are broad and less structured, have greater skin depth,
and show magnetic
memory of only about one cycle.  The three stepped diffusivity profiles 
yield more structured fields, and show more cycles of magnetic memory.

The evolving field lines for the double-step profile 
(Fig.1e) reveal some distinct features in Fig.\ref{b2plots}f-h
compared to all other cases. We can more clearly see
three sets of poloidal fields with 
alternating polarity in Fig.\ref{b2plots}b. Similar toroidal 
field structures of three alternating polarities are also lined up in
the meridional circulation conveyor belt, often in certain phases as in
Figs.\ref{b2plots}h,i. These patterns indicate a memory of magnetic fields
from 1.5-2 previous solar cycles.
This further suggests that the double-step model is the most physical diffusivity profile in our suite.
Next, we should evaluate the effects of varying the location and slope
of this profile's gradient near the tachocline.

\subsection{Results of Experiment II: Variable locations of a fixed gradient}
\label{results2}
\begin{figure}[ht]
\raggedright
\mbox{
      \psfig{file=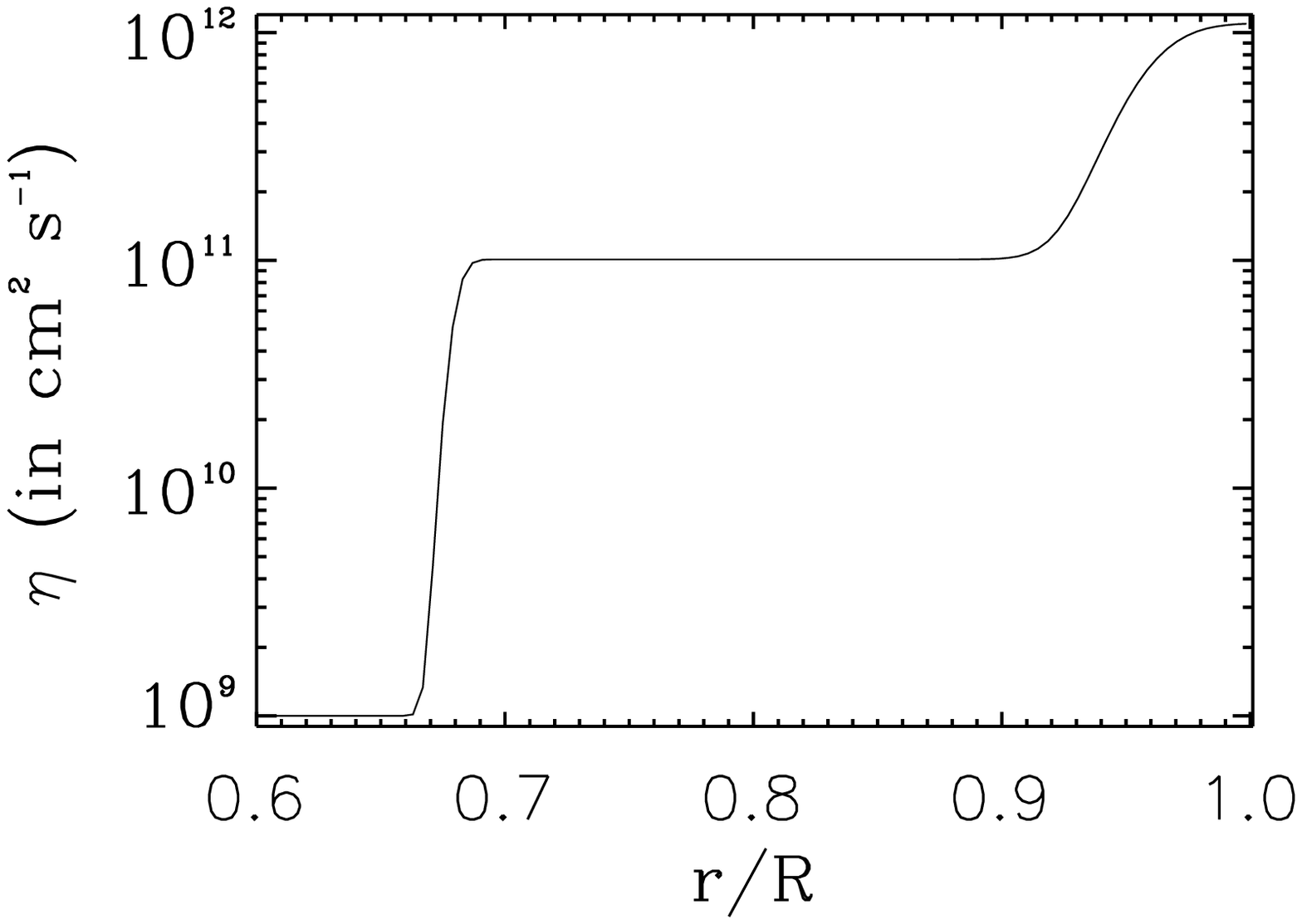,height=4.0cm}
      \psfig{file=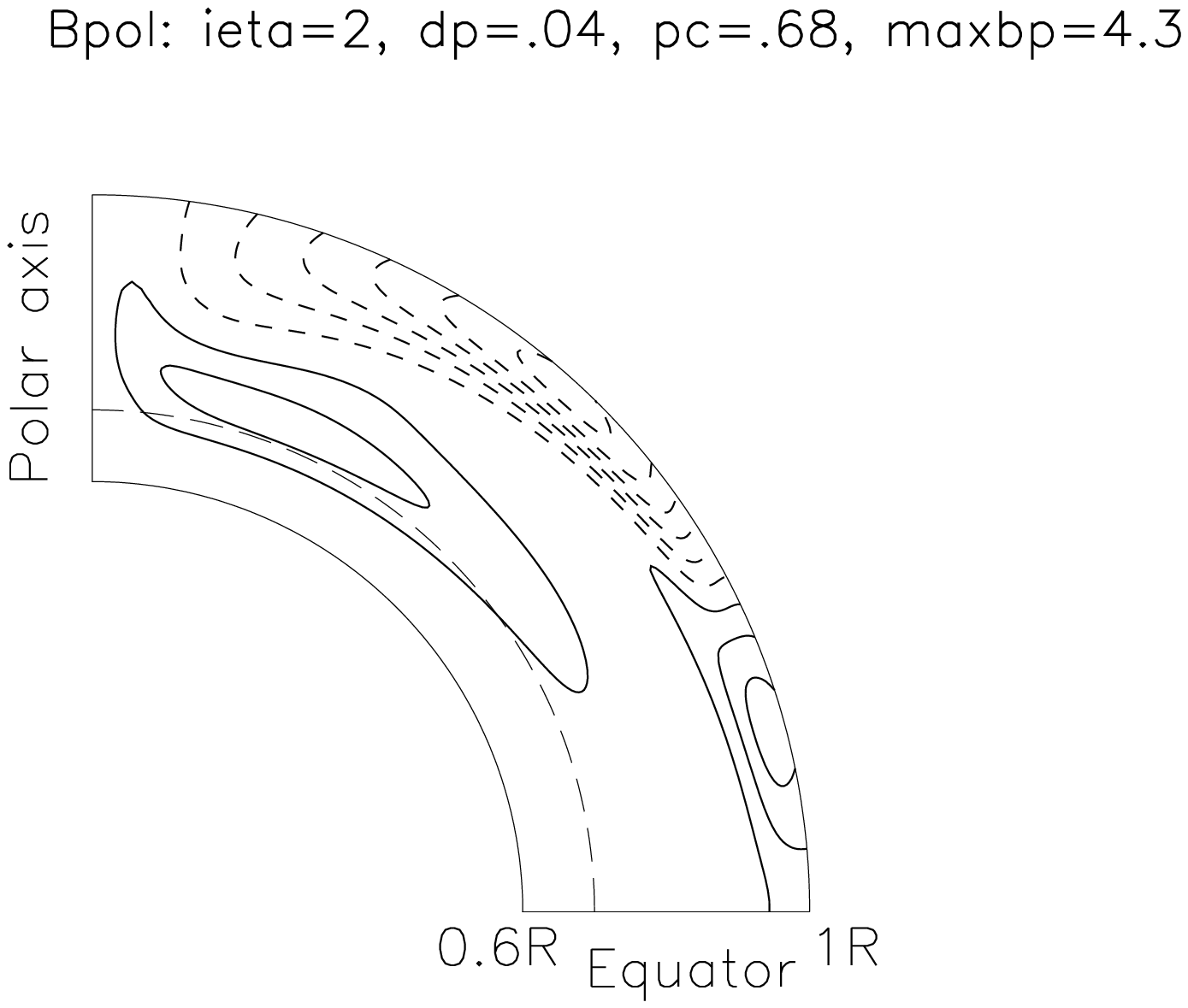,width=5.0cm}
      \psfig{file=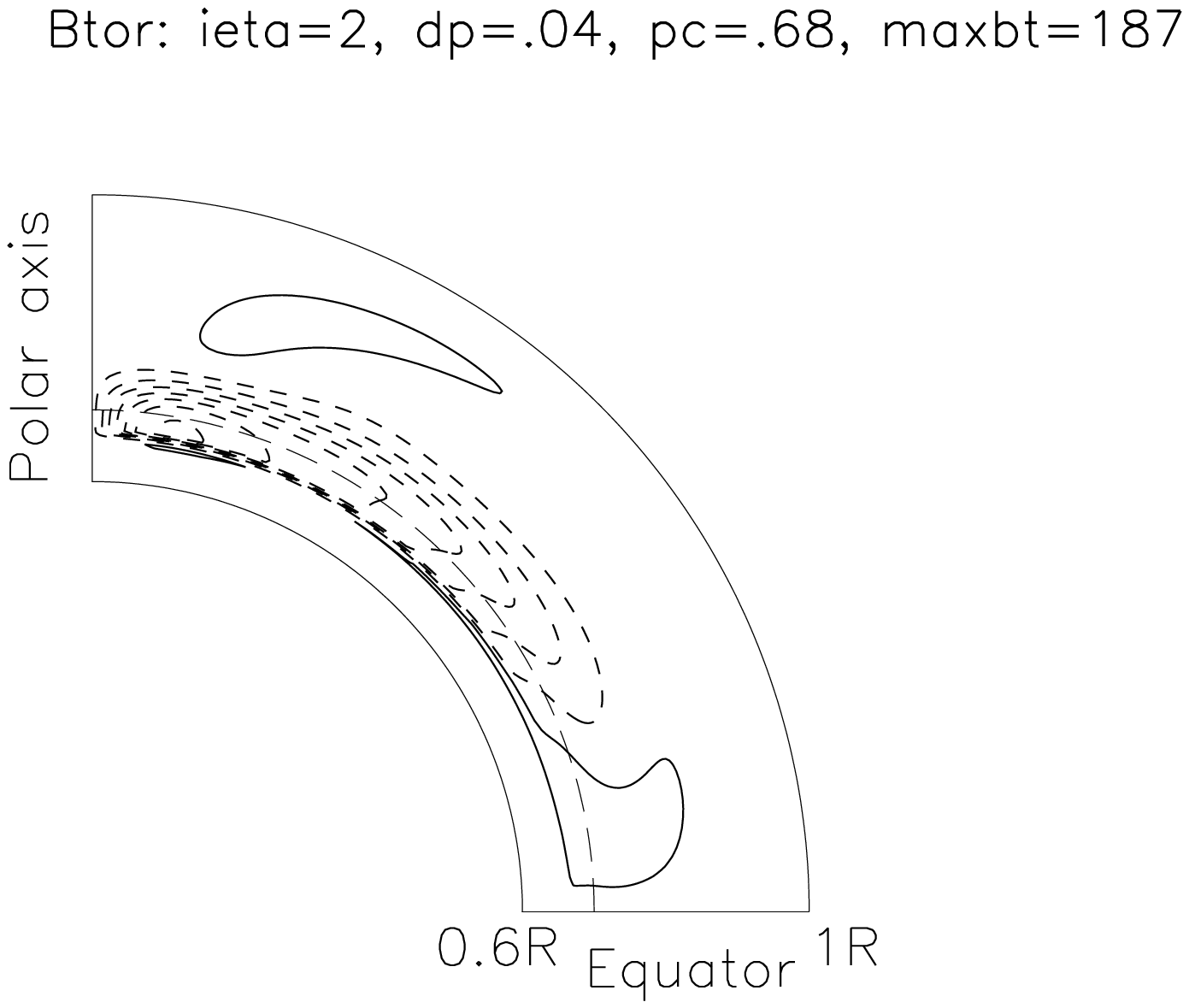,width=5.0cm}
     }
\mbox{(a)}
\mbox{
      \psfig{file=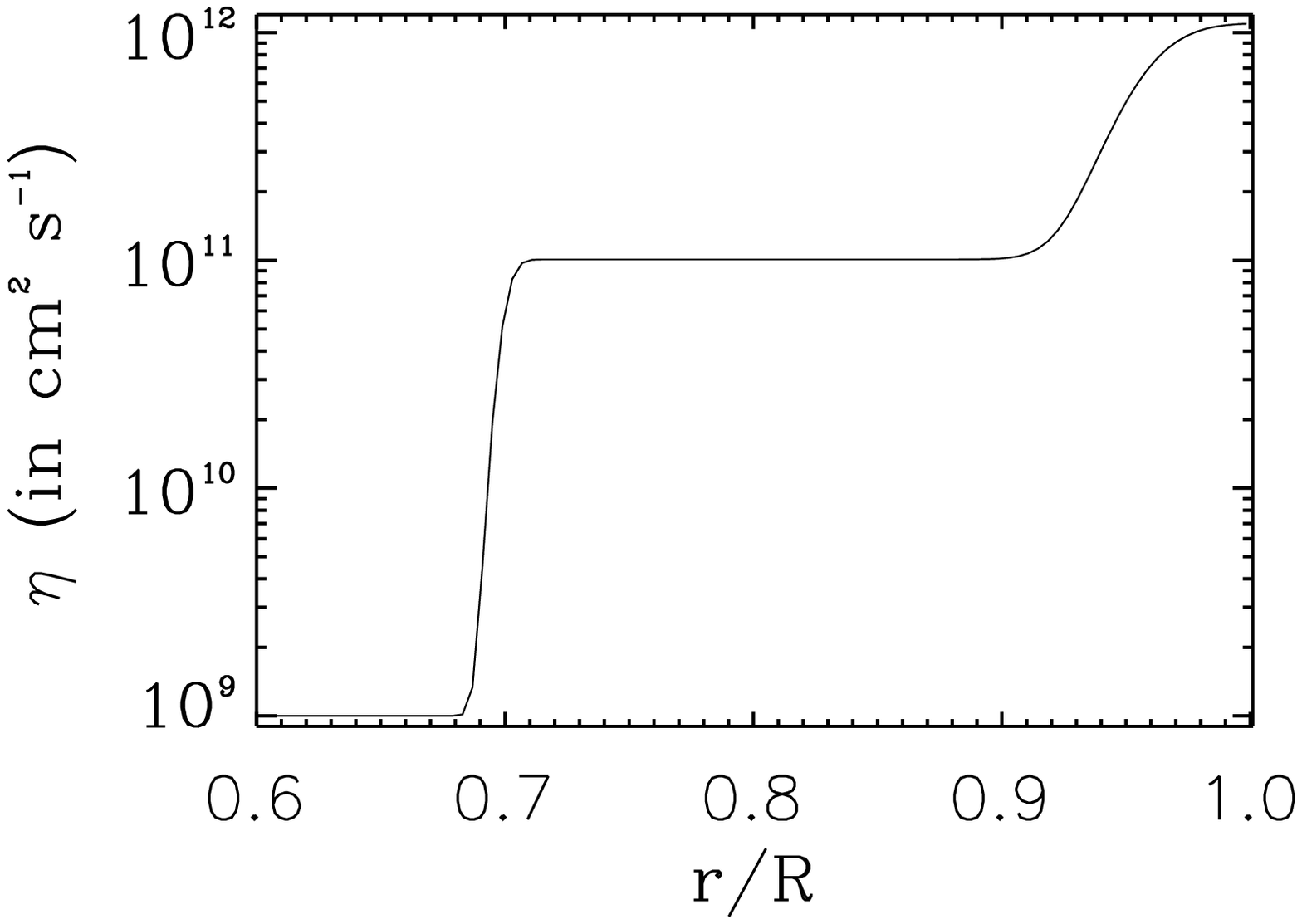,height=4.0cm}
      \psfig{file=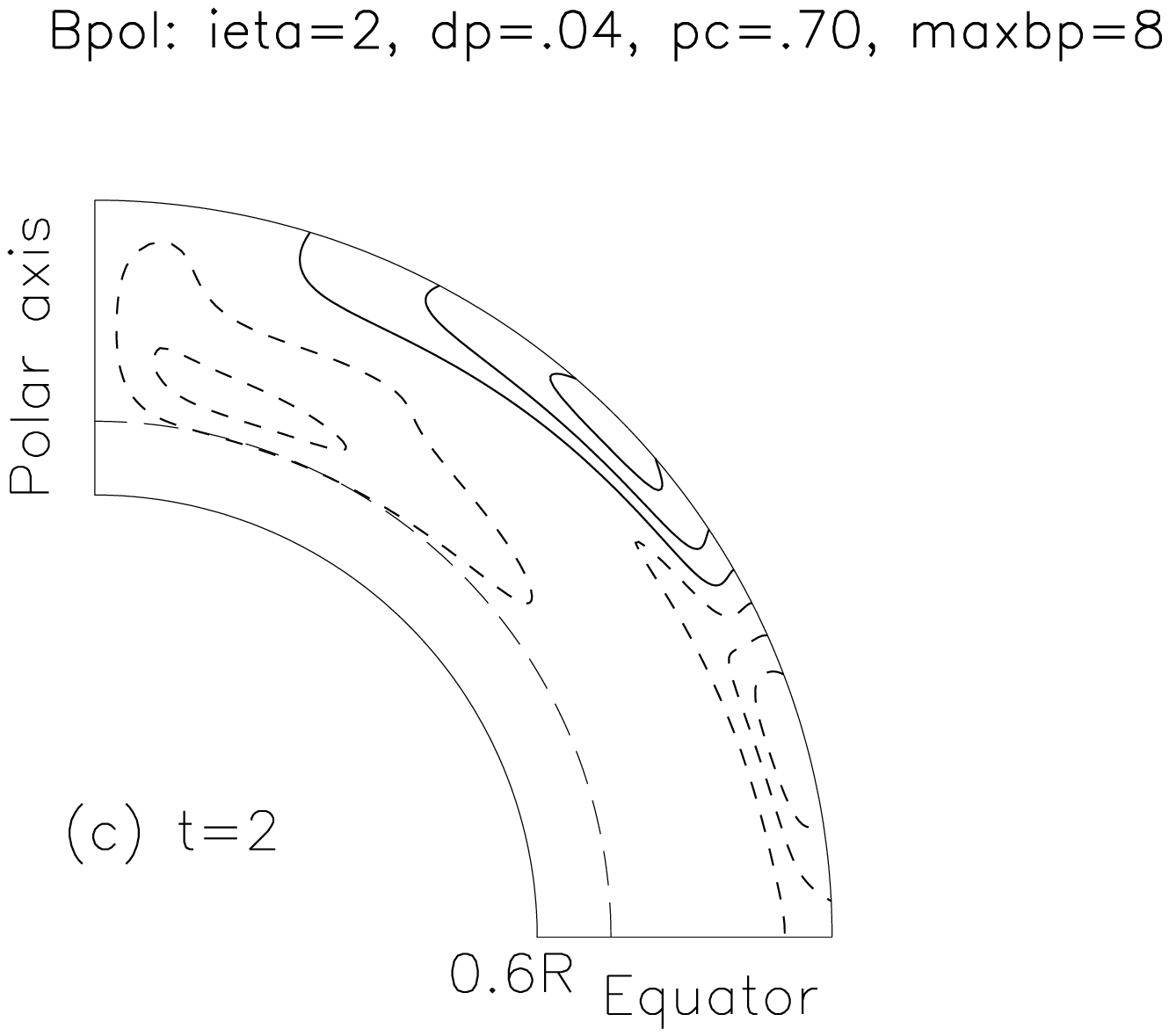,width=5.0cm}
      \psfig{file=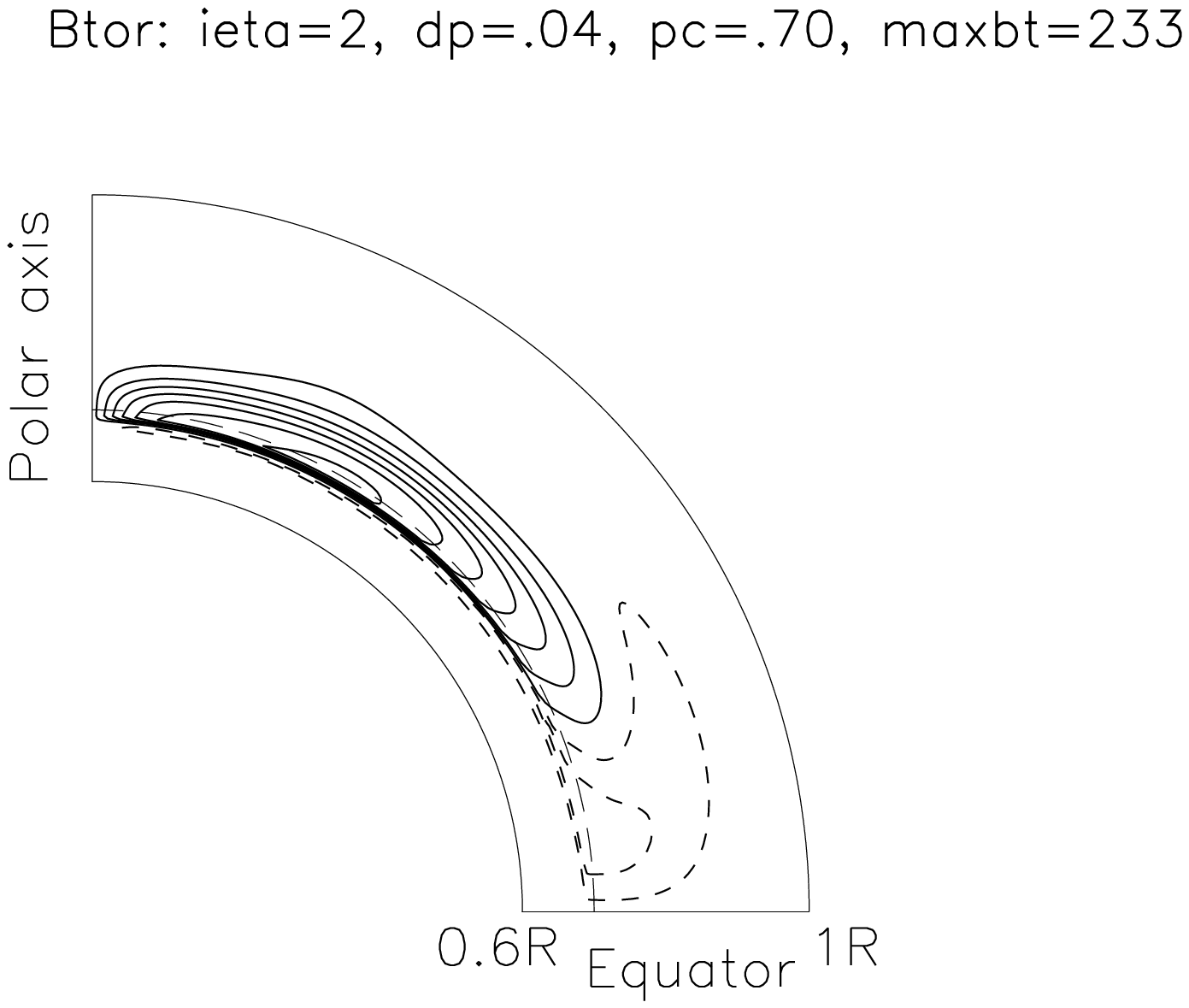,width=5.0cm}
     }
\mbox{(b)}
\mbox{
      \psfig{file=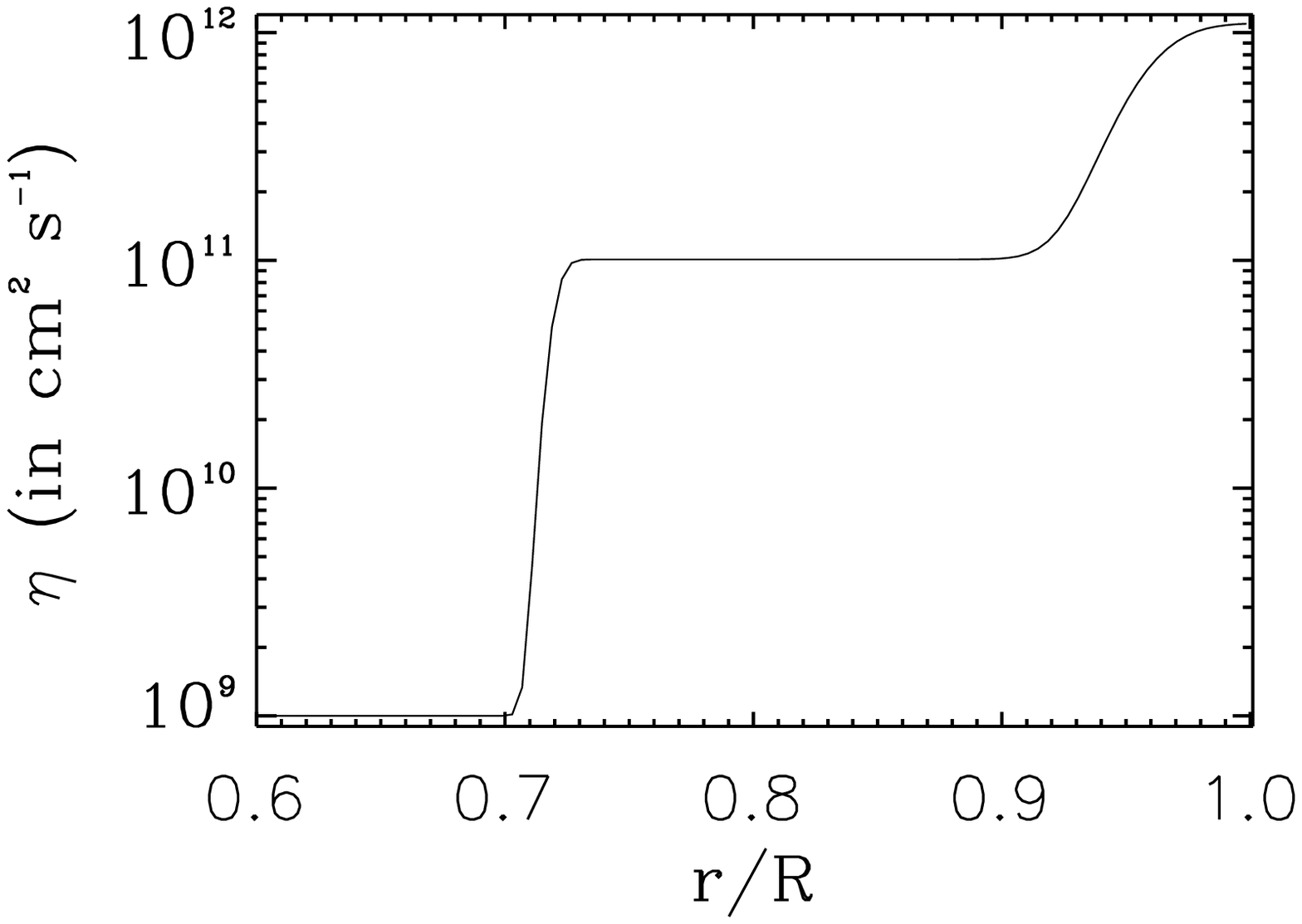,height=4.0cm}
      \psfig{file=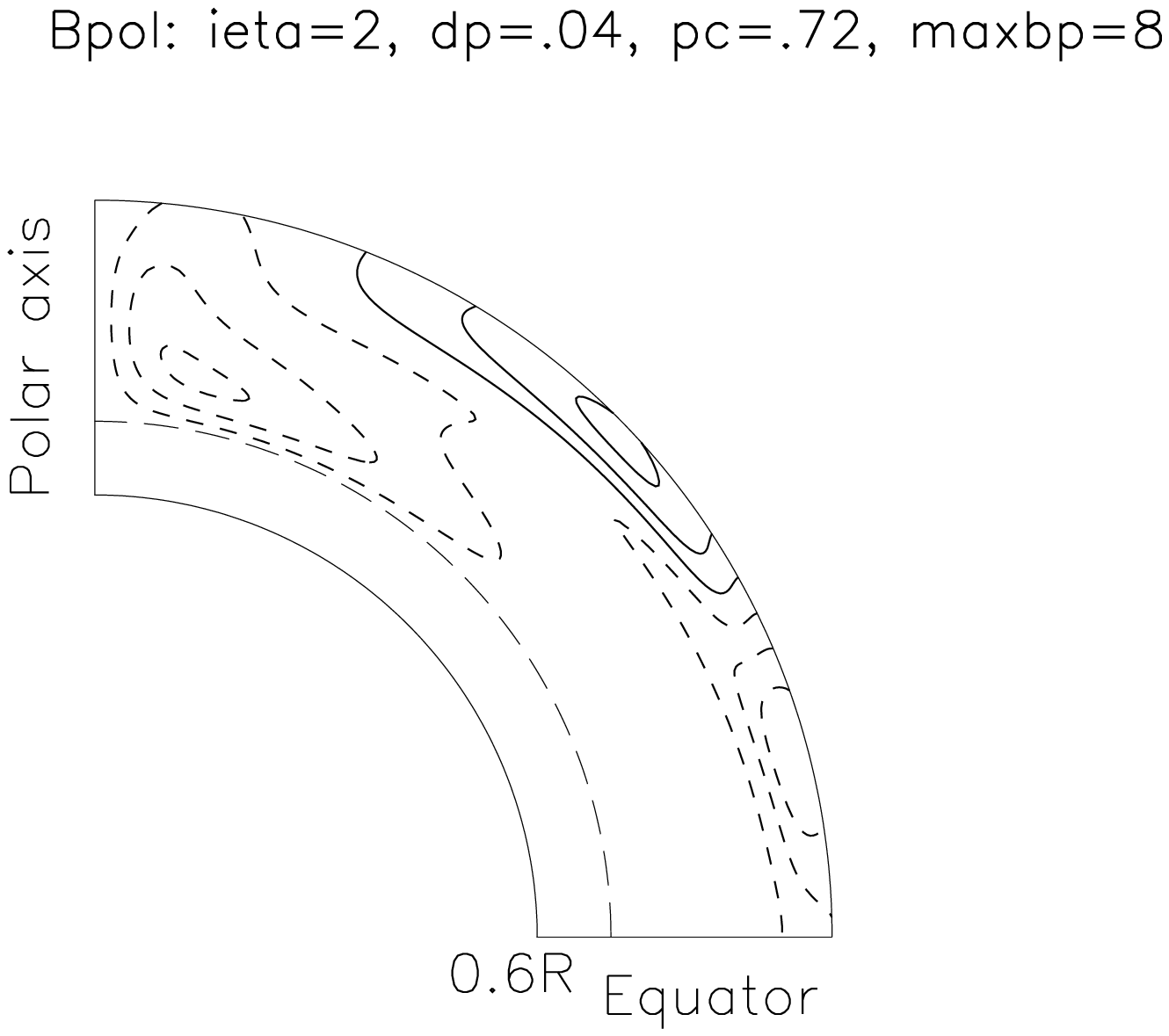,width=5.0cm}
      \psfig{file=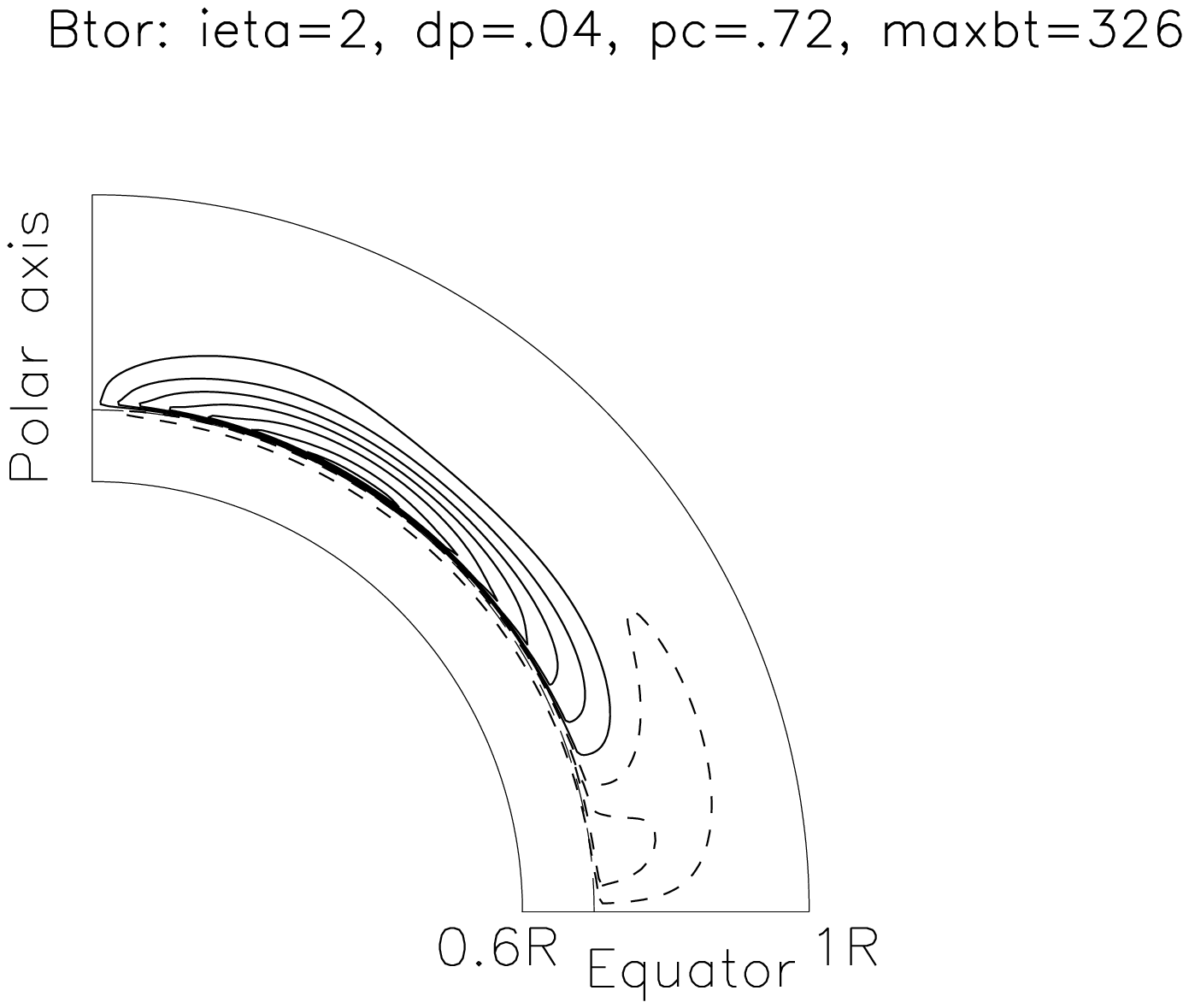,width=5.0cm}
     }
\mbox{(c)}
\mbox{
      \psfig{file=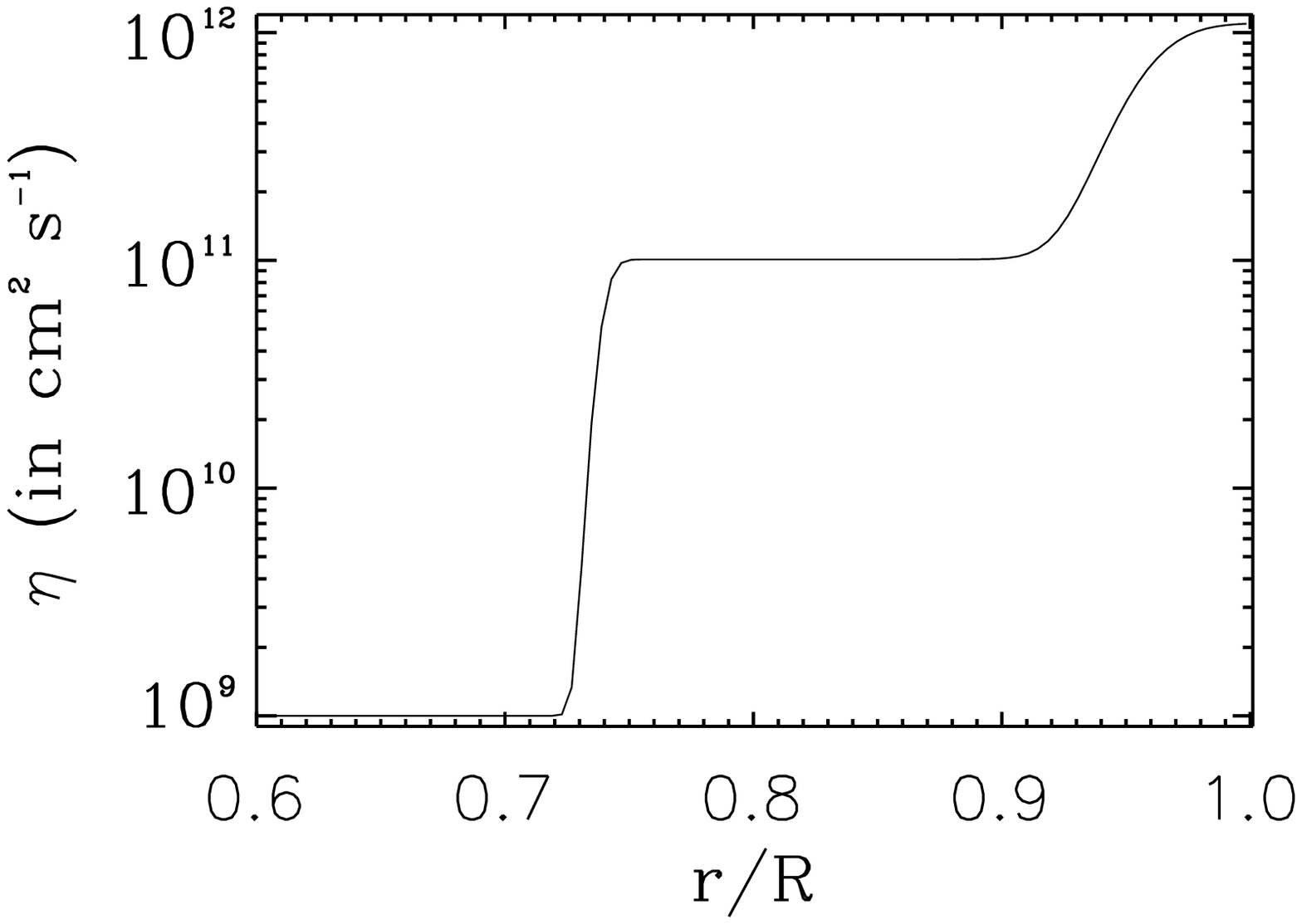,height=4.0cm}
      \psfig{file=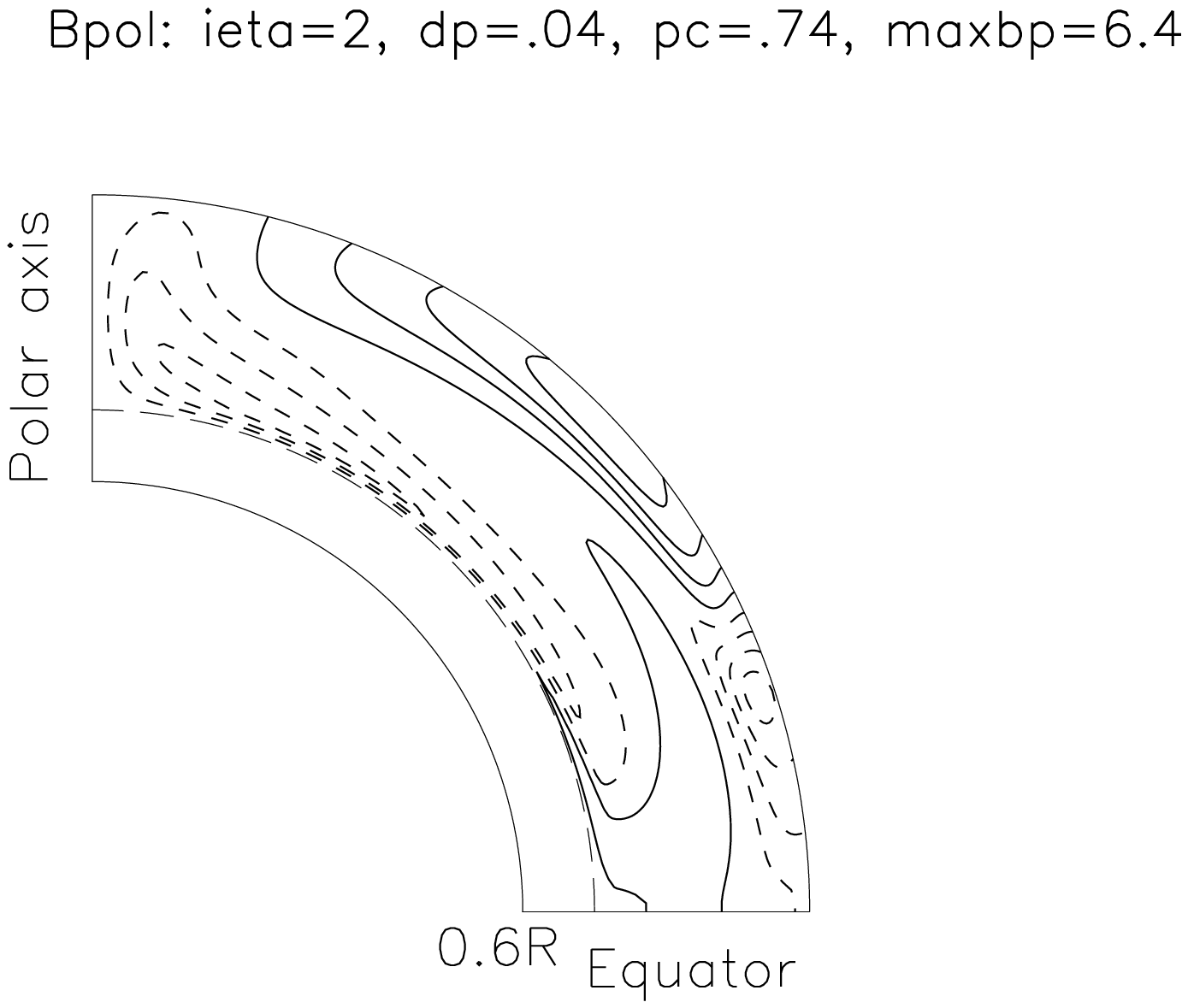,width=5.0cm}
      \psfig{file=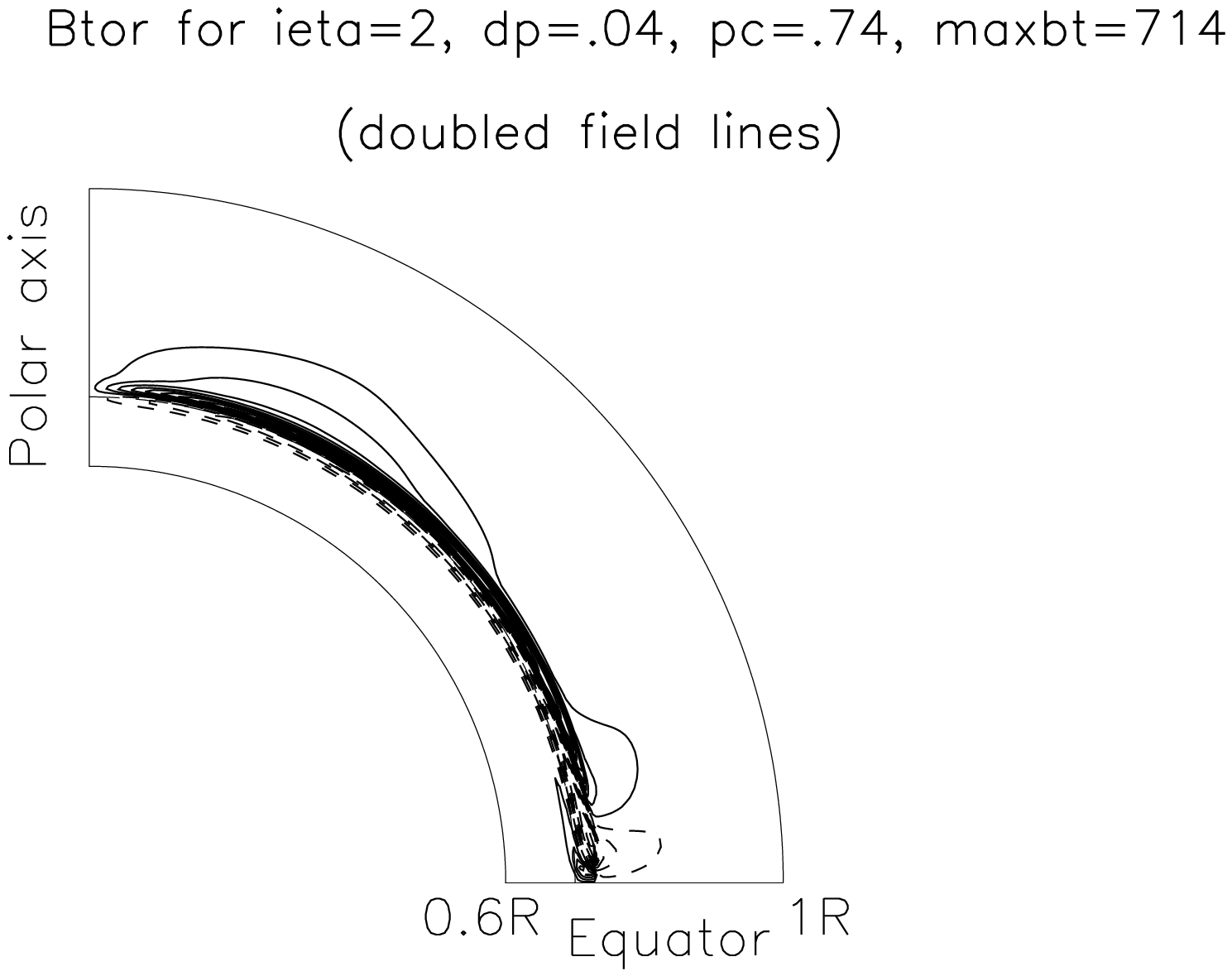,width=5.0cm}
     }
\mbox{(d)}
\caption{Experiment II: Variable locations of a fixed diffusivity gradient (left):
Corresponding poloidal flux contours (center) and
toroidal flux contours (right) in the meridional plane, chosen
near solar max.
}
\label{butter2}
\end{figure}

\vspace{0.2cm}
Greater differences are evident in solar dynamo evolution with identical
diffusivity profiles assigned slight displacements in 
\emph{ gradient location} (Fig.2)
than in the other 
experiments.
This is especially interesting, given recently increasing attention 
to the influence of
detailed \emph{shape} of the diffusivity profile on dynamo evolution
(e.g. Rempel 2006, Guerrero et al. 2008 and 2009, Mu\~noz-Jaramillo et al. 2009).
Fig.\ref{butter2} shows that as long as diffusivity gradients
cross the tachocline (and diffusivity values are within reasonable
boundaries), adequate magnetic flux 
can be transported by the meridional circulation to contribute to a robust
kinematic dynamo (Figs.\ref{butter2}a-c). But when the gradient moves
too far above the tachocline, as in Fig.\ref{etapar}d and Fig.\ref{butter2}ds, 
where $r_{tach} = 0.74$, the diffusivity 
at the base of the tachocline is so low that toroidal flux starts to get
trapped there (see Fig.\ref{butter2}d).  Toroidal field strengths at
the tachocline then become highly amplified (to 266 kG in this case), 
and Table 0 shows that
this tendency worsens as the diffusivity gradient moves
toward the photosphere (cases 10-12).
\vspace{0.1cm}

\begin{tabular}{lcc} \\ \hline
diffusivity profile & $max(B_{\phi})$ & cycle period ($T$)\\ 

\hline
(a) $r_{tach}$ = 0.68       &  20 kG & 20.2 years \\ 
(b) $r_{tach}$ = 0.70       & 50 kG & 19.0 years \\
(c) $r_{tach}$ = 0.72       & 79 kG & 15.9 years \\
(d) $r_{tach}$ = 0.74       & 266 kG & 19.4 years \\ 
\hline
\end{tabular}

Table II. Maximum toroidal field strength and dynamo cycle period ($T$) at
the tachocline for
each diffusivity profile in Experiment II
(Fig.2). Field strength increases as the gradient moves outward, and the
dynamo cycles most rapidly when the gradient straddles the tachocline.
\vspace{0.2cm}

When the diffusivity gradient 
is placed too far above the tachocline,
deeper toroidal flux 
has less freedom to
diffuse. Flux becomes constrained by artificially low diffusivity  
in the tachocline region, and is unable to participate fully in a strong
dynamo process.

The cycle time is faster when the diffusivity gradient straddles
the tachocline (Fig.\ref{etapar}c) and slows when either toroidal flux 
gets trapped by a gradient below the tachocline (d) or the dynamo
proceeds more normally (a,b), with the gradient extending into the
tachocline.

\input{psfig}
\clearpage

\subsection{Results of Experiment III: Diffusivity gradients 
of various slope,
centered at the tachocline}
\label{results3}
\begin{figure}[ht]
\centering
\mbox{(a)
      \psfig{file=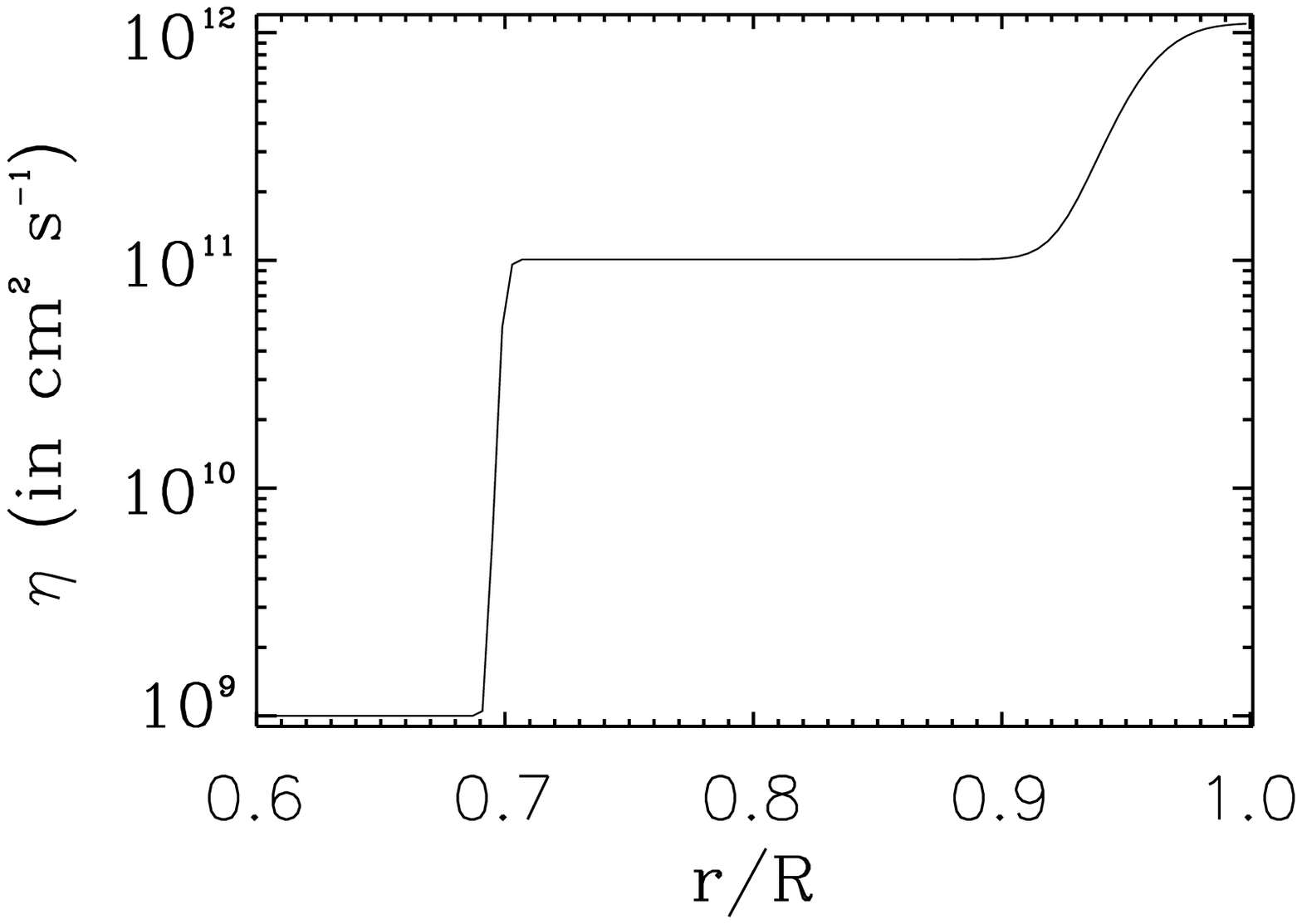,height=4.0cm}
      \psfig{file=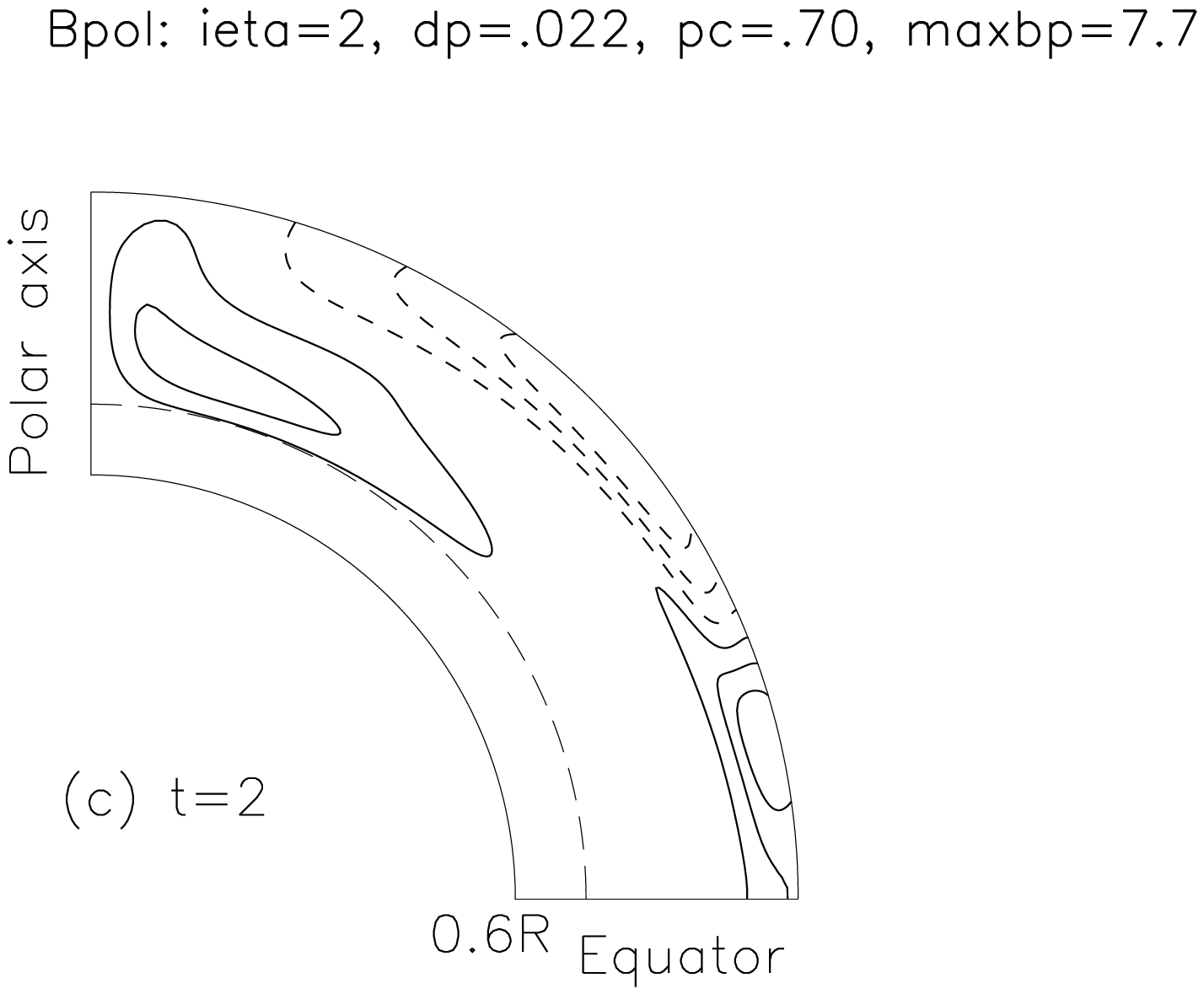,width=5.0cm}
      \psfig{file=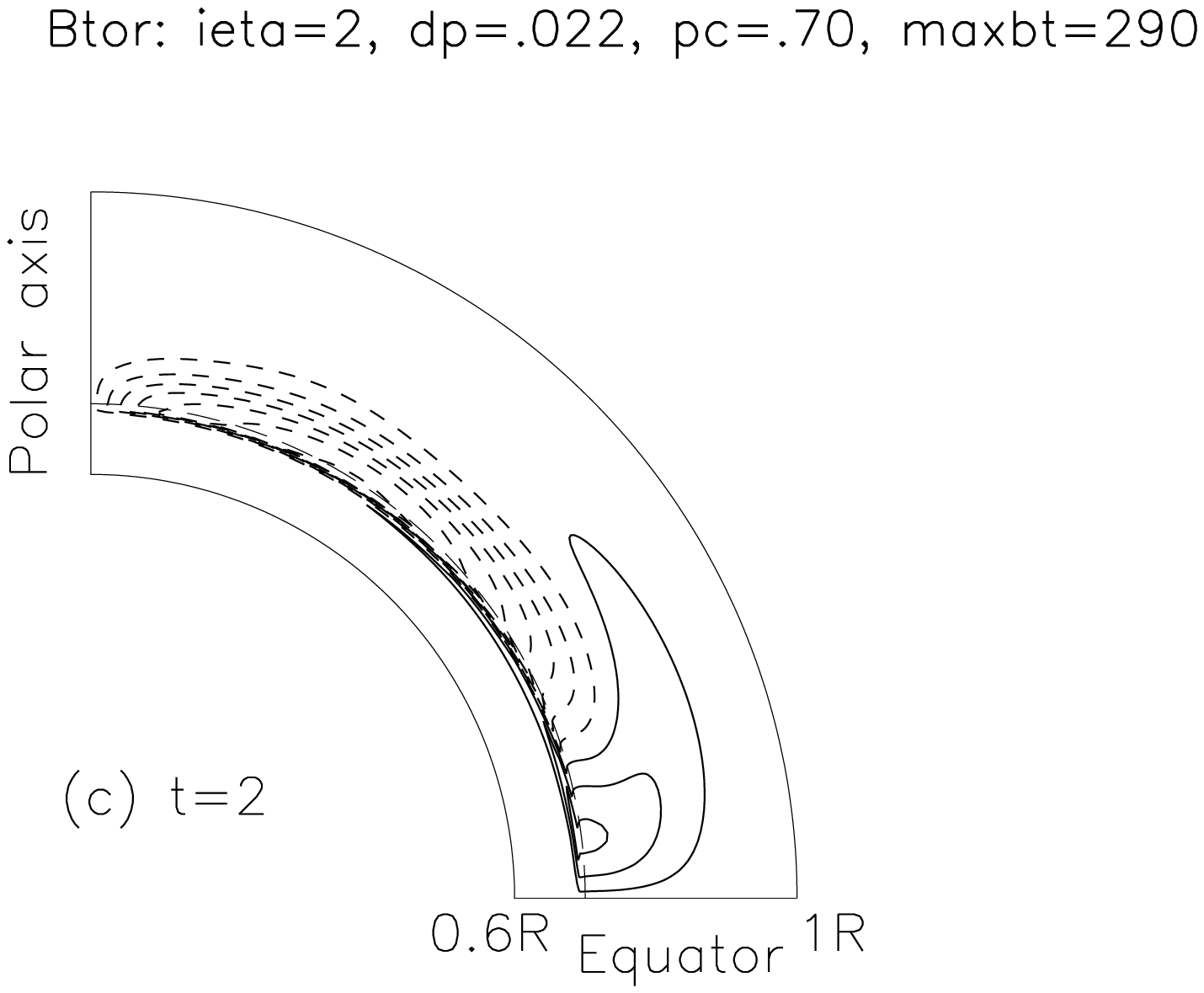,width=5.0cm}
     }
\mbox{(b)
      \psfig{file=etar4b.eps,height=4.0cm}
      \psfig{file=bpol4b2.eps,width=5.0cm}
      \psfig{file=btor4b2.eps,width=5.0cm}
     }
\mbox{(c)
      \psfig{file=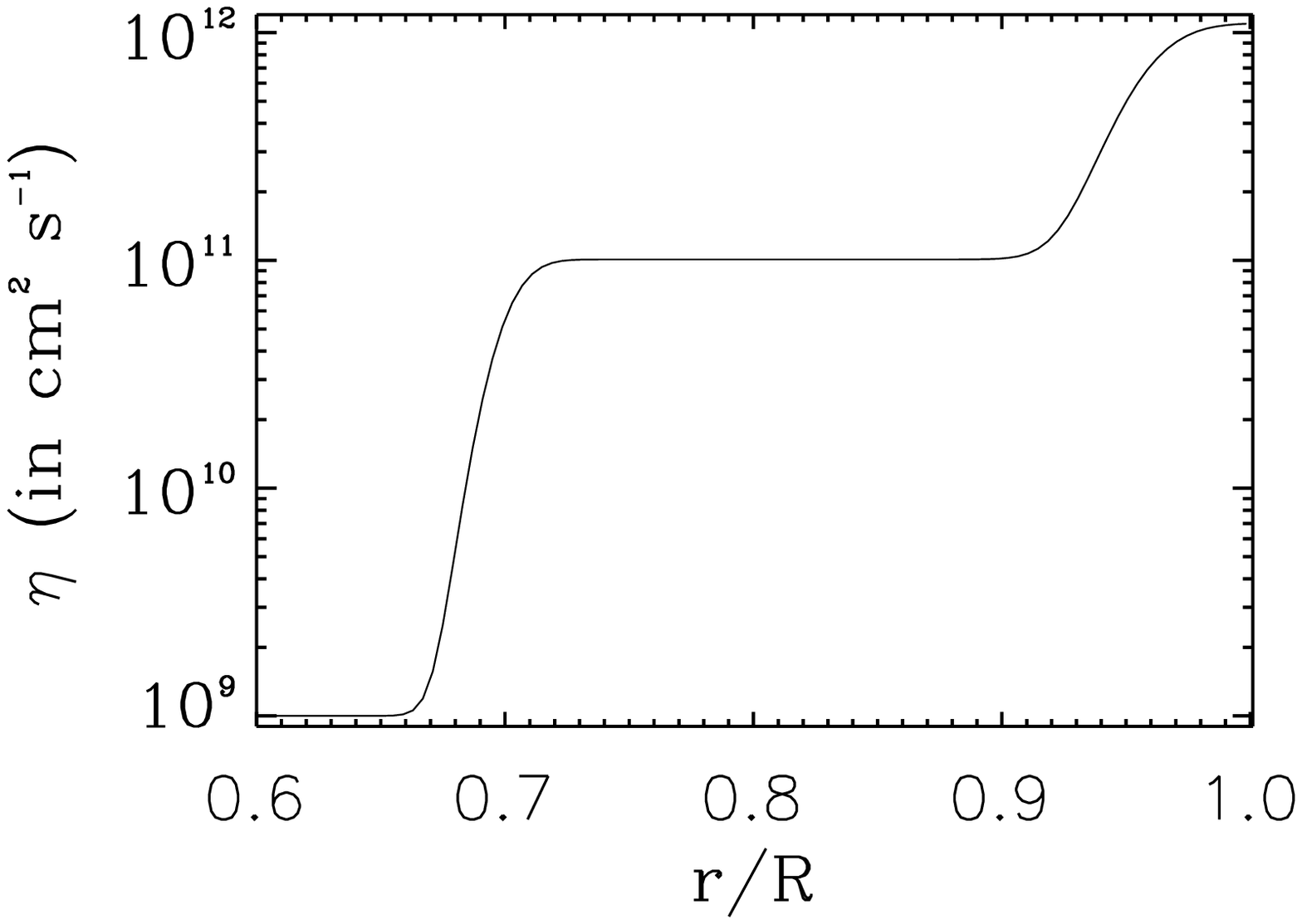,height=4.0cm}
      \psfig{file=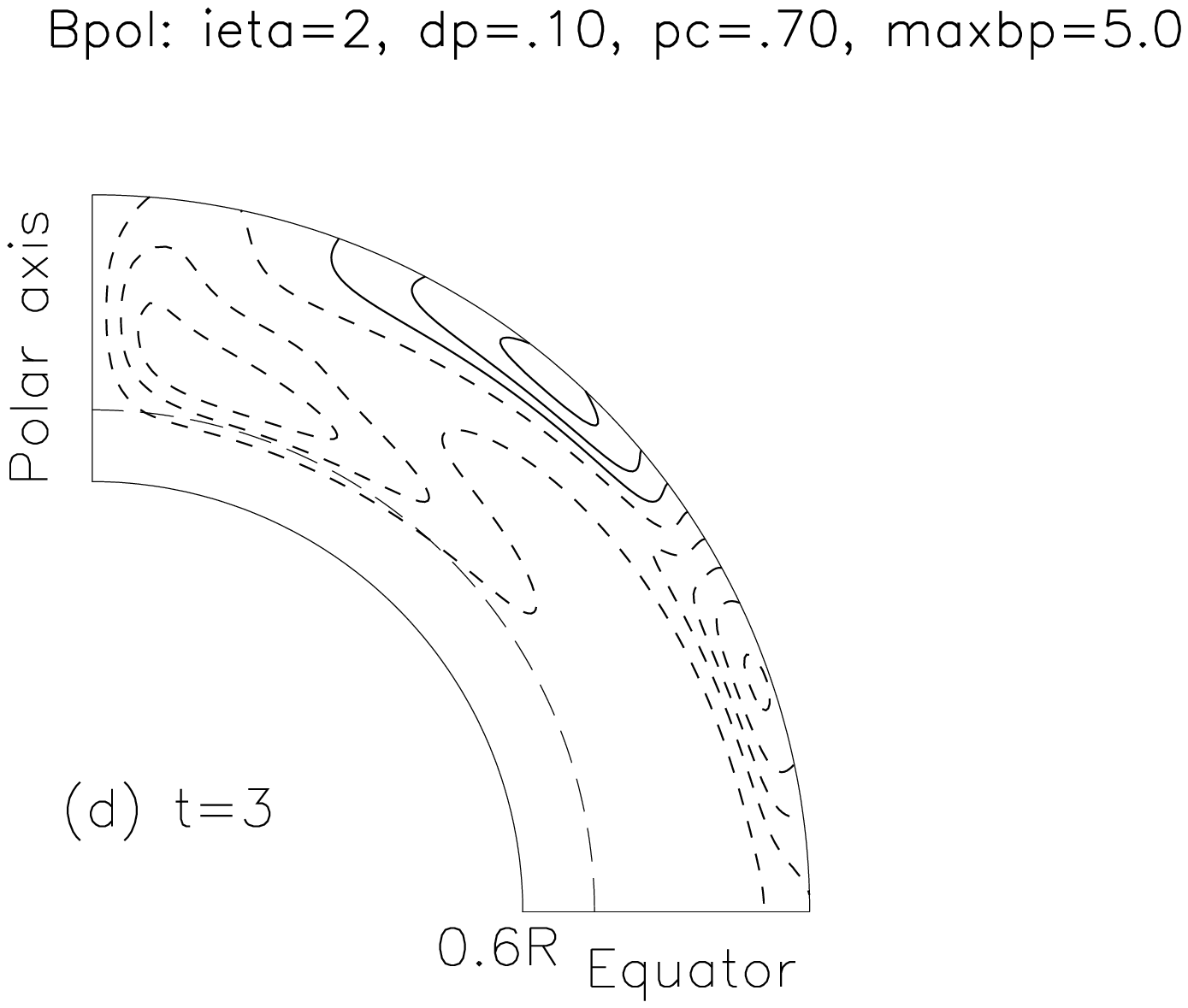,width=5.0cm}
      \psfig{file=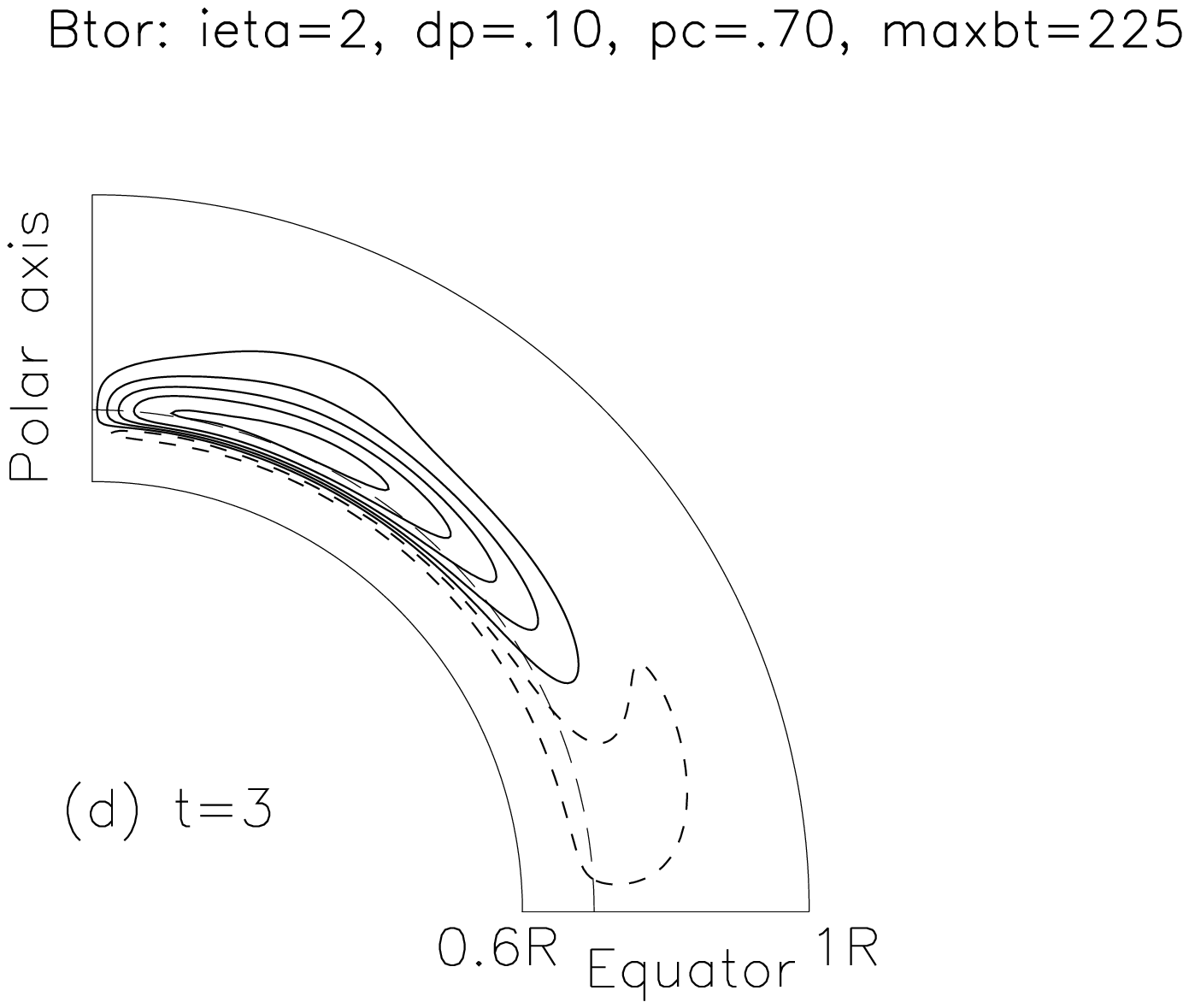,width=5.0cm}
     }
\mbox{(d)
      \psfig{file=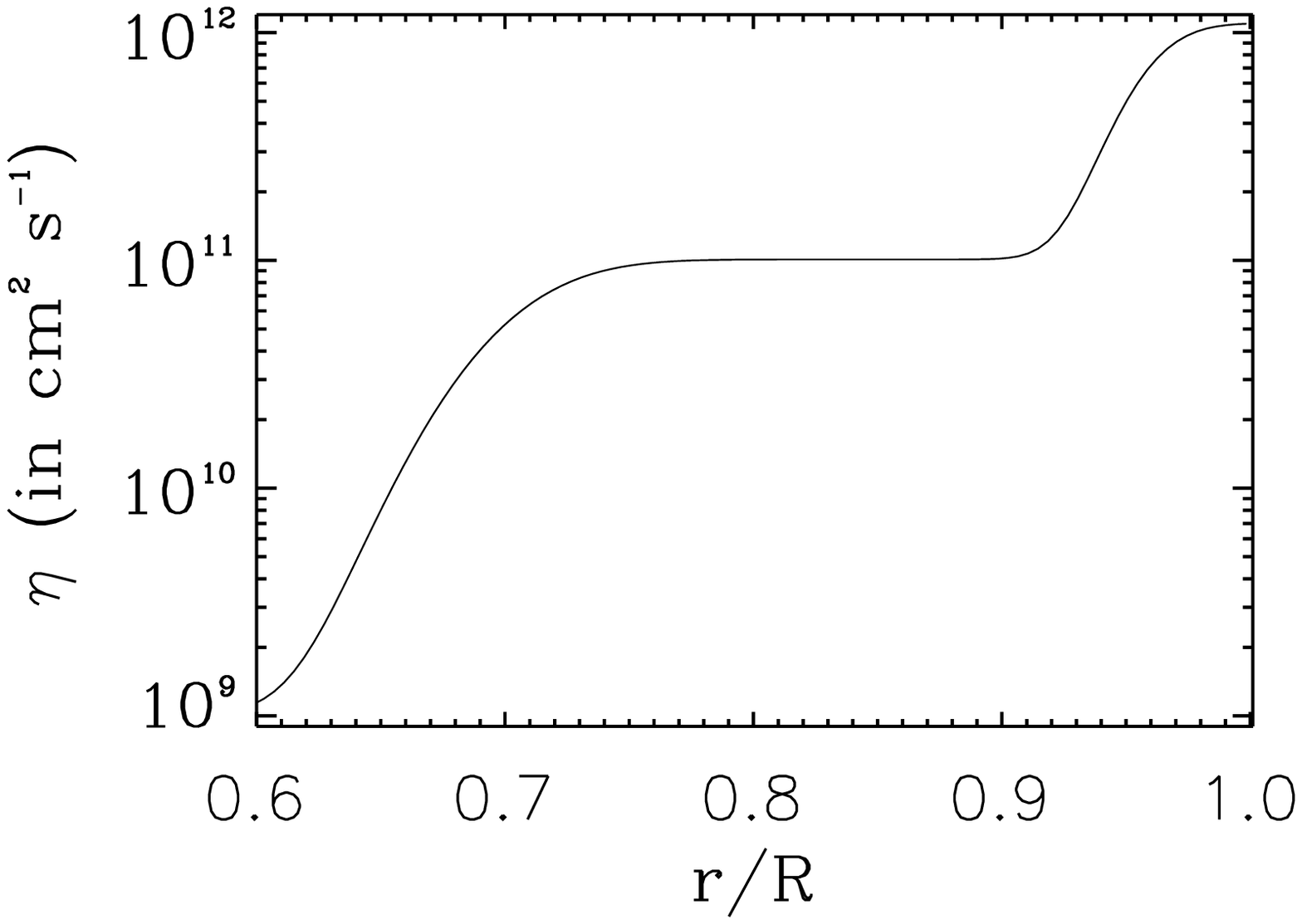,height=4.0cm}
      \psfig{file=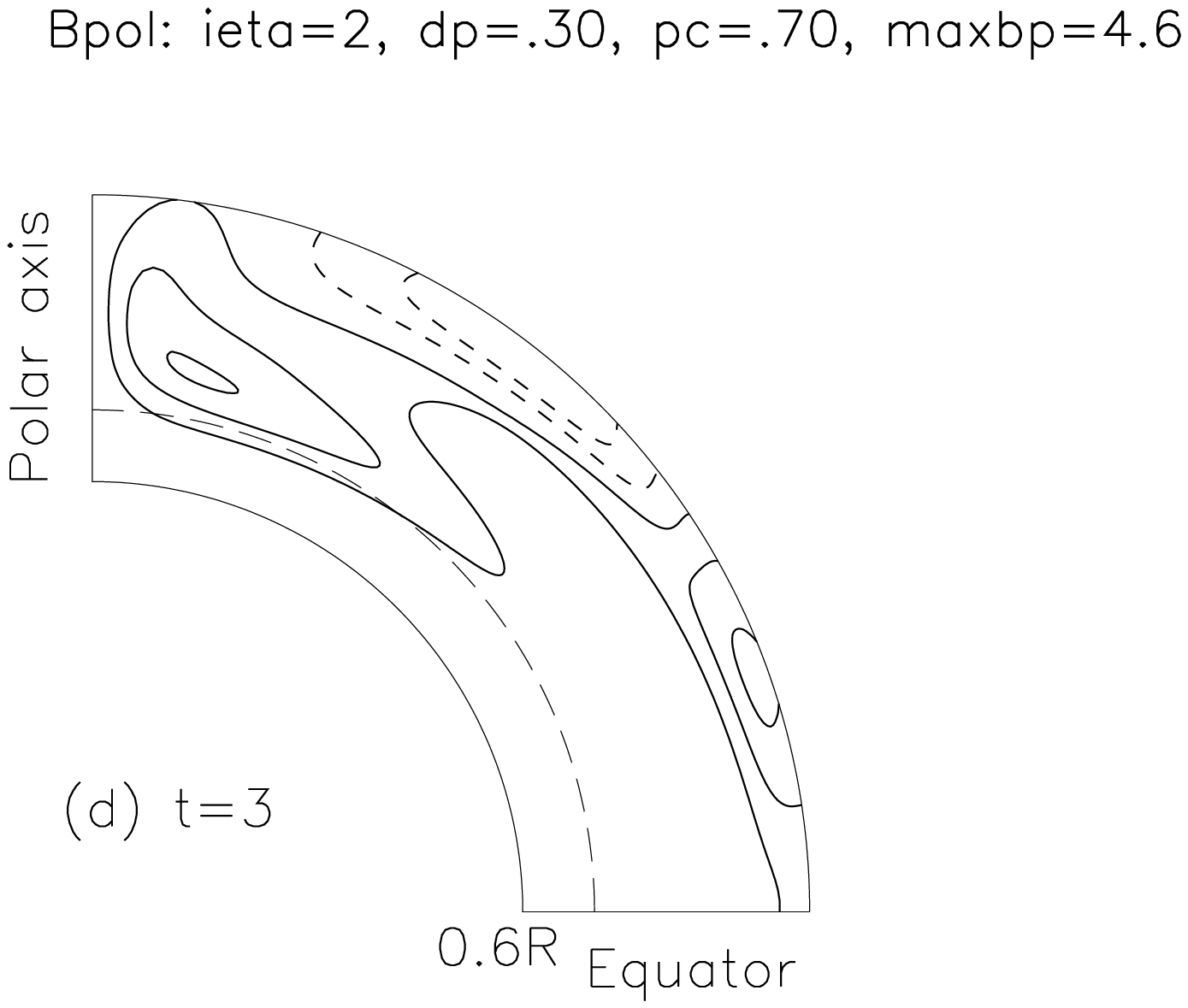,width=5.0cm}
      \psfig{file=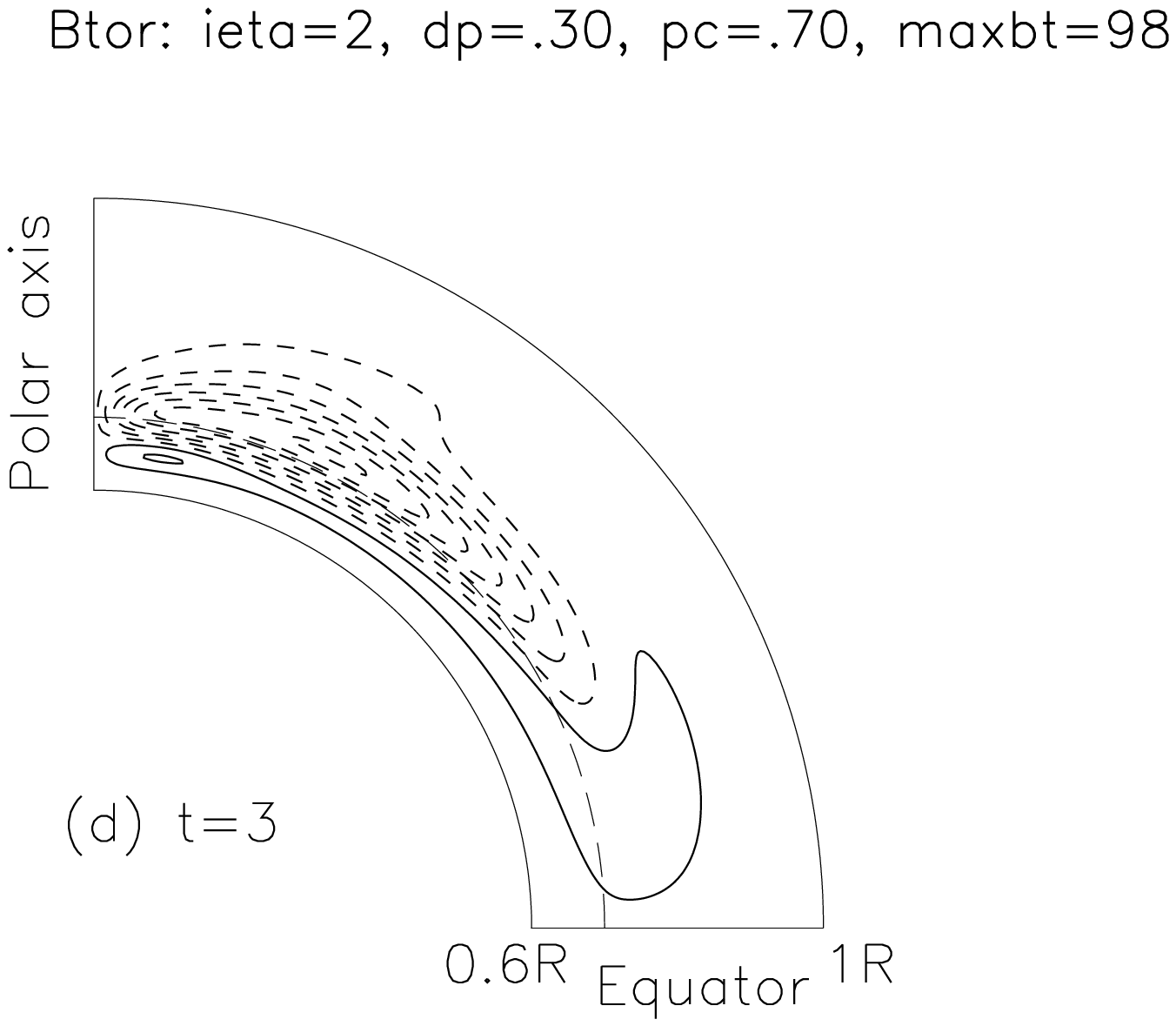,width=5.0cm}
     }
\caption{Experiment III: Diffusivity gradients with various slopes (left): 
Corresponding poloidal flux contours (center) and
toroidal flux contours (right) in the meridional plane, chosen
near solar max.
}
\label{butter3}
\end{figure}

\vspace{0.2cm}
While the diffusivity profiles for this experiment (Fig.3) appear to vary 
far more
than those in Experiment II (Fig.2), in fact we find substantially 
less variation in the
dynamo evolution here. 
Resultant field profiles look similar not
only at solar max (shown in Fig.\ref{butter3}) but throughout their time evolution.  
The broadest profile
does, predictably, show more toroidal flux diffusion below the tachocline, 
and the narrowest profile shows more concentration and therefore flux 
amplification. 
However, the effects of these on poloidal field evolution are 
not very strong.
While Table III shows a smooth increase in tachocline field strength with
diffusivity gradient steepness near the tachocline, 
the cycle time optimizes at 19 years
for our reference case of $\Delta r_{tach}$ = 0.04.  

\vspace{0.1cm}

\begin{tabular}{lcc} \\ \hline
diffusivity profile & $max(B_{\phi})$ & cycle period ($T$)\\ 

\hline
(a) $\Delta r_{tach}$ = 0.022       &  82 kG & 14.9 years \\ 
(b) $\Delta r_{tach}$ = 0.04       & 50 kG & 19.0 years \\
(c) $\Delta r_{tach}$ = 0.10        & 25 kG & 17.9 years \\
(d) $\Delta r_{tach}$ = 0.30        & 12 kG & 15.8 years \\ 
\hline
\end{tabular}

Table III. Maximum toroidal field strength and dynamo cycle period ($T$) at
the tachocline for
each diffusivity profile in Experiment III
(Fig.3). Field strength increases markedly with gradient steepness,
and cycle time is shorter for very broad or very steep gradients.
\vspace{0.2cm}

While Experiment III confirms that a modest diffusivity gradient such as that in our 
reference double-step case (Figs.1e, 2b, and 3b) yields more physical 
dynamo simulations,
(in terms of cycle time, maximum field strength, field structure, and
magnetic memory),
we have also learned that
the kinematic dynamo evolution is {\it not} strongly sensitive to the slope of 
the gradient, so long as it is centered near the
tachocline.

\input{psfig}
\clearpage

\section{Comments and Conclusions}
\label{conclude}

The depth-dependent diffusivity profile $\eta(r)$ is 
a poorly understood ingredient in the mean-field kinematic Babcock-Leighton
flux-transport dynamo model. Mixing-length theory 
and previous numerical investigations
suggest
a value of $\eta$ within the range of 
$10^{11-12} \thinspace {\rm cm}^2 {\rm s}^{-1}$ near the photosphere. Near the
radiative zone, the value of $\eta$ should drop to the molecular
diffusivity value, $10^{3-5} \thinspace {\rm cm}^2 {\rm s}^{-1}$;
computational constraints require us to use higher values at present.
Since we do not have enough observational or theoretical knowledge
to fully specify the diffusivity as function of depth, we have considered
in this paper three relatively plausible $\eta(r)$ profiles
(Figs.1c-e),
two simpler profiles (Figs.1a,b),
and many variations on the most physical diffusivity profile of the five,
a double-stepped profile motivated by a combination of granulation and
supergranulation.
Using a Babcock-Leighton
flux-transport dynamo, we have explored the influence of these 
$\eta(r)$ profiles on solar cycle properties and dynamics.

We performed three numerical experiments to investigate the sensitivity
of the kinematic dynamo model to the (I) \emph{shape}, (II) \emph{gradient location},
and (III) \emph{gradient slope} of the magnetic diffusivity.

In Experiment I, the five $\eta(r)$ profiles we considered are: (a) constant
$\eta$ throughout our computation domain ($0.6R \leq r \leq R$;
$0 \leq \theta \leq \pi/2$), (b) linear,
(c) step-to-linear, (d) single-step and (e) double-step
profiles. 
We have used $\eta_0 = 10^{11} \thinspace {\rm cm}^2 {\rm s}^{-1}$ in the bulk of the convection zone
in all cases, and we raised the surface diffusivity to a maximum of
$10^{12} \thinspace {\rm cm}^2 {\rm s}^{-1}$ in the double-step case.

While our single-step and double-step profiles are particularly
motivated by physical considerations (turbulence and supergranulation),
there may be other
forms of diffusivity profiles that are worth considering
besides those which we studied. Based on our 
investigations, it appears that
synoptic maps alone are unlikely to provide sufficient insight
for the prescription of model diffusivity depth dependence.

The evolutionary patterns
of the toroidal and poloidal fields inside the convection zone show
some distinct features arising from the choice of the depth-dependence
in $\eta(r)$. For example, the field penetration below the center
of the tachocline is greater in cases with
constant and linear profiles than those with
step-to-linear, single-step and double-step profiles. This is because
\textit{higher diffusivity near the tachocline
creates a deeper skin-depth effect}.

Close study of the evolution of flux contours reveals that the 
double-step profile with
$\eta_{core} = {10^{9}} \mathrm{cm^{2} s^{-1}}$,
$\eta_{0} = {10^{11}} \mathrm{cm^{2} s^{-1}}$,
$\eta_{surf} = {10^{12}} \mathrm{cm^{2} s^{-1}}$,
and an inner gradient width of
$\Delta r_{tach} = 0.04$
behaves most consistently with observations and expectations, 
e.g. by showing a magnetic memory of $n=$ 1.5-2 solar cycles,
appropriately structured field evolution,
cycle time of 19 years, and 
$max(B_{\phi})$ of 50 G at the tachocline.

It has been pointed out in some flux-transport dynamos
that the magnetic memory of the model should depend on the
combination of the meridional circulation and the magnetic diffusivity
profiles. Dikpati and Gilman (2006) showed that the increase
in the amplitude of $\eta(r)$ in the bulk of the convection zone from
$3 \times 10^{10}$ to $2 \times 10^{11} \thinspace {\rm cm}^2 {\rm s}^{-1}$
reduces the memory-length of the model from 2 cycles to 1 cycle, and
the memory can be completely lost for 
a diffusivity value of
$2 \times 10^{12} \thinspace {\rm cm}^2 {\rm s}^{-1}$ there. 
They did not explore how a fixed value of $\eta$ in the convection
zone can render different memory-length to the model depending on
its variation with depth. All of our stepped profiles have 
memory-lengths of at least 1.5 cycles, but this reduces to one cycle
for the unphysical constant and linear profiles.  
We find that our double-step case, with surface
$\eta(r) = 10^{12} \thinspace {\rm cm}^2 {\rm s}^{-1}$,
has the most realistic magnetic memory.
\textit{Diffusivity gradients appear to contribute significantly to
storage and recycling of flux from older cycles.}

We find the structure of the toroidal and poloidal magnetic fields
to be more aligned along the meridional flow streamlines
for the stepped profiles, and more broadly distributed in the
cases of constant and linear $\eta$ profiles. \textit{Diffusivity gradients
contribute to more intricate magnetic field structures.} If future helioseismic
observations can help determine the large-scale magnetic field structures
in the bulk of convection zone and tachocline,
this may better constrain theoretical constructions of diffusivity
depth dependence.

In Experiment II, we used the optimal double-step profile from Experiment I,
and shifted the location of its inner gradient with respect to the tachocline.
In two cases (Fig.2a,b), the gradient lay inside the tachocline; in one case
(Fig.2c), it straddled the tachocline; and in one case (Fig.2d), the gradient 
was fixed outside the tachocline. \textit{ We found that the location of the gradient
has strong effects on the dynamo evolution.}    

When the diffusivity gradient lies inside the tachocline, 
as in Fig.2a,b, there is a significant region of high diffusivity subject to
meridional circulation near the tachocline.  Flux within this lower region
appears to make a crucial contribution to the dynamo, for when it is cut off
by lifting the diffusivity gradient above the tachocline, 
too much toroidal flux is trapped and amplified near the 
tachocline.  
\textit{The location of the diffusivity gradient is more important than
the slope of the gradient itself.}

In Experiment III, we modified the slope of the inner gradient of the
same double-step profile, keeping it physically centered just 
below 
the tachocline.
All cases in this experiment generated a strong kinematic
dynamo.
As expected, the broadest gradient, 
with higher diffusivity
deeper below the tachocline,
permits deeper diffusion of magnetic flux, and weaker amplification 
of the toroidal
field at the tachocline.  
The steepest gradient concentrates and amplifies flux.
The two main surprises from this experiment are that the cycle period optimizes
for our original reference choice of diffusivity profile, 
and that \textit{the overall dynamo behavior is 
not very
much changed by our very wide range of diffusivity gradients}.

In general, while Experiment I reveals constraints on the 
evolution of the dynamo and magnetic fields due to
the \emph{shape} of diffusivity profiles; and Experiment III
shows that, for a fixed location of diffusivity gradient with respect to the
tachocline, the 
evolution of the dynamo and magnetic fields are slightly 
sensitive to the \emph{slope} of the gradient; notably Experiment II 
shows that the actual \emph{location} of a given diffusivity gradient can
have profound influences.  In particular, without a gradient close enough
to the tachocline, insufficient toroidal flux may be carried up by the 
meridional circulation to sustain normal dynamo cycles.

Apart from the discovery that \textit{the location of the diffusivity
gradient may be more important than either its slope or the detailed shape of
the diffusivity profile} (given appropriate boundary values),
our results may also suggest that either this kinematic dynamo model 
is so robust that its synoptic diagrams are not
terribly sensitive to the detailed choice of diffusivity profile (within
limits illustrated here; and see Charbonneau 2007) 
and/or that new numerical
diagnostics
are needed to further discern its sensitivity.  

Ongoing observational work (e.g. in helioseismology) 
and numerical work in convection zone 
turbulence can better constrain the values, 
gradients, and gradient locations
of the solar magnetic
diffusivity.  Meanwhile, numerical experiments such as these 
remain among the best tools for learning about
poorly constrained dynamo ingredients such as the diffusivity.

In addition to the depth dependence, $\eta(r)$ probably varies with
latitude and time.
For example, Gilman \& Rempel (2005) describe such
dependence via $\eta$-quenching by strong magnetic fields.
Diffusivity quenching may locally amplify magnetic fields, suppressing both turbulence and turbulent enhancement of diffusivity.
New work such as that of Guerrero et al. (2009) addresses
influences of
$\eta(r,\theta)$-quenching on solar cycle features in a
flux-transport dynamo.
We should also investigate magnetic advection effects due to 
diffusivity gradients (Zita, 2009).
\acknowledgements
We gratefully acknowledge the use of Mausumi Dikpati's kinematic dynamo model,
and critical guidance and helpful feedback from Dr. Dikpati and Dr. Peter Gilman
(HAO/NCAR, Boulder CO 80301),
including a thorough review of an early version of this manuscript. 
We thank Lockheed Martin Solar and Astrophysics Laboratory for 
generously providing facilities during the completion of this work.

This work was partially supported by 
NASA grants
NNH05AB521, NNH06AD51I 
and the NCAR Director's opportunity fund; 
NSF grant 0807651, 
the Visitor Program of the High Altitude Observatory at NCAR, 
and The Evergreeen State College's Sponsored Research program.

The National Center for Atmospheric Research (NCAR) is sponsored by the National
Science Foundation (NSF).


%
\end{article} 
\end{document}